\documentclass[aps,pre,twocolumn,showpacs,floatfix]{revtex4}

\usepackage{amsmath}
\usepackage{amssymb}
\usepackage{graphicx}
\usepackage{bm}
\usepackage{epstopdf}
\usepackage{graphicx,epstopdf}
\usepackage{color}

\newcommand{\la}{\lambda}

\newcommand{\al}{\alpha}

\newcommand{\K}{\mathrm{K}}
\newcommand{\E}{\mathrm{E}}
\newcommand{\prt}{\partial}

\begin{document}

\title{
Undular bore theory for the Gardner equation}

\author{A. M. Kamchatnov$^{1}$}
\author{Y.-H. Kuo$^{2}$}
\author{T.-C. Lin$^{2}$}
\author{T.-L. Horng$^{3}$}
\author{S.-C. Gou$^{4}$}
\author{R. Clift$^{5}$}
\author{G. A. El$^{5}$}
\author{R. H. J. Grimshaw$^{5}$}

\address{
$^1$ Institute of Spectroscopy, Russian Academy of Sciences,
Troitsk, Moscow Region, 142190, Russia\\
$^{2}$ Department of Mathematics, National Taiwan University, Taipei, Taiwan \\
$^{3}$ Department of Applied Mathematics, Feng Chia University, Taichung 40724, Taiwan\\
$^{4}$ Department of Physics, National Changhua University of Education, Changhua 50058, Taiwan \\
$^{5}$ Department of Mathematical Sciences, Loughborough
University, Loughborough LE11 3TU, UK \\
}

\date{\today}

\
\begin{abstract}
We develop modulation theory for undular bores (dispersive shock waves) in the framework of the Gardner, or extended
Korteweg--de Vries, equation, which is
a generic mathematical model for weakly nonlinear and weakly dispersive wave propagation, when effects of higher order
nonlinearity become important.
Using a reduced version of the finite-gap integration method we derive the Gardner-Whitham modulation system in a Riemann
invariant form and show
that it can be mapped onto the well-known modulation system for the Korteweg--de Vries equation. The transformation between
the two counterpart modulation systems is, however, not invertible. As a result, the study of the resolution of an initial discontinuity
for the Gardner equation  reveals a rich phenomenology of solutions  which, along with the KdV type simple undular bores, include nonlinear trigonometric
bores, solibores, rarefaction waves
and composite solutions representing various combinations of the above structures. We construct full parametric maps
of such solutions for both signs of the cubic nonlinear term in the Gardner equation. Our classification is
supported by  numerical simulations.
\end{abstract}


\maketitle

\section{Introduction}

The Gardner equation
\begin{equation}\label{eq1}
    u_t+6uu_x-6\alpha u^2u_x+u_{xxx}=0\, ,
\end{equation}
is a fundamental mathematical model for the description of weakly nonlinear dispersive waves in the situations when
the higher order nonlinearity
effects, described by the cubic term $-6\alpha u^2u_x$,  become important. It first arose as an auxiliary mathematical
tool in the derivation of the infinite set of local conservation laws of the Korteweg--de Vries (KdV) equation
\cite{gardner} but has been shown later to describe nonlinear wave effects in a number of physical contexts including
plasma physics \cite{wat,rud}, stratified fluid flows \cite{grimshaw-2002} and quantum fluid dynamics \cite{dm-2011}.
One of the most important and best known
applications of the Gardner equation is the description of large-amplitude internal waves  (see
\cite{grimshaw-2002,hm-2006,aosl-2007} and references therein).
The coefficient $\alpha$ in (\ref{eq1}) can be positive or negative depending on the physical problem under consideration.
In the context of internal waves, this depends on the stratification, see \cite{grimshaw-2002}. In the particular case of
a two-layer fluid it is always positive \cite{kakyam78}.

When $\alpha=0$ the Gardner equation (\ref{eq1}) reduces to the KdV equation
\begin{equation}\label{kdv}
 u_t+6uu_x+u_{xxx}=0\, .
\end{equation}
Using the change of variables
\begin{equation}\label{transmkdv1}
w=u  -  \frac{1}{2 \alpha}, \qquad x'=x+\frac{3}{2\alpha} t \, .
\end{equation}
one transforms (\ref{eq1}) to the  modified KdV (mKdV) equation
\begin{equation}\label{mkdv1}
w_t-6\alpha w^2w_{x'} + w_{x'x'x'} = 0
\end{equation}
(note the change of boundary conditions at infinity).

The Gardner equation (\ref{eq1}) is invariant with respect to the transformation:
\begin{equation}\label{inv}
  u \to \frac{1}{\alpha} - u \, ,
\end{equation}
which makes  the existence of  solutions of different polarity (e.g. ``bright'' and ``dark'' solitons) possible {\it for the  same system}, depending on the initial conditions.
This is markedly different from the properties of the KdV equation, which admits, for a given set of coefficients, solitary wave solutions of a  fixed polarity,
independently on the initial conditions.

The soliton solutions of the Gardner equation for both signs of $\alpha$ are well-known and,
along with the usual KdV-type ``bright'' and  ``dark'' solitons, include table-top solitons, breathers,
algebraic solitons and kinks (solibores). Much less is known about  dynamics of undular bores described by the Gardner equation.
This problem is of significant theoretical and applied interest and is of particular importance in oceanography, where undular bores play
a key role in the evolution of the internal tide (see \cite{grimshaw-2002,holloway99}).

Undular bores  are nonlinear expanding wave trains connecting two different basic flow states and exhibiting solitary waves
near one of the edges. They are usually formed as a result of dispersive resolution of a shock or an initial discontinuity
in fluid depth and/or velocity (see e.g. \cite{sh88}, \cite{egs-06}) or due to a resonant interaction of a fluid flow with
localized topography (see e.g. \cite{gs86},  \cite{baines}). Formation of  undular bores (also often called dispersive
shock waves) is a generic physical phenomenon which has been observed not only in classical fluids but also in
collisionless plasmas, Bose-Einstein condensates and nonlinear optical media (see \cite{scholarpedia} and references therein).

The analytical description of undular bores is usually made in the framework of the Whitham modulation theory
\cite{whitham1,whitham2} in which the asymptotic solution for the bore is sought in the form of a slowly varying
periodic solution of the governing dispersive equation. The slow evolution of the modulation parameters
(such as mean value, amplitude, wavenumber etc.) is then governed by a hydrodynamic-type system of averaged equations,
called the Whitham equations.  The modulation description of the KdV undular bore was first constructed in the celebrated
paper by Gurevich and Pitaevskii \cite{GP1} and was later generalised to other dispersive systems both integrable
(see e.g. \cite{kamch2000,egp-01} and references therein) and non-integrable (\cite{el-2005,egs-06,egkk07,ep-11}).

The modulation system for the KdV equation can be represented in the Riemann invariant form \cite{whitham1,whitham2}
which  plays the key role in the Gurevich-Pitaevskii analytical construction of the KdV undular bore. For the mKdV equation (\ref{mkdv1})
(both defocusing ($\alpha>0$) and focusing ($\alpha<0$) cases) the modulation system in Riemann invariants was
derived in \cite{dron76} using direct averaging of conservation laws and non-trivial algebraic manipulations leading to the diagonal structure.
The spectral (finite-gap) approach to the derivation of the defocusing mKdV modulation system in the Riemann form was used in \cite{ksk-04}, where the modulation solution was obtained for the undular bore resolving the ``cubic'' wave breaking singularity.  The  undular bore theory for
the focusing ($\alpha <0$) mKdV equation (\ref{mkdv1}) was constructed in \cite{march2008}. It was shown in \cite{march2008} that,
along with the KdV-type cnoidal undular bores, in which the elliptic modulus $m$  varies together with the wave amplitude
$a$ from $m=0$, $a=0$ at the trailing edge to $m=1$, $a=a^+$  at the leading edge, $a^+ > 0$ being the amplitude of
the lead solitary wave,  the focusing mKdV equation supports another type of modulated solutions, termed trigonometric bores,
in which $m=0$ throughout the whole wave train but the amplitude $a \ne 0$ and vanishes only at the trailing edge.
It was also shown in \cite{march2008} that the trigonometric bore is usually realised as part of a composite solution:
either a combined cnoidal-trigonometric bore or a combination of a trigonometric bore and a simple rarefaction wave.
Similar composite solutions were constructed in \cite{kod2008} for the complex mKdV equation (which is related to the
defocusing nonlinear Schr\"odinger (NLS), rather than the KdV, equation). We stress that neither trigonometric bores nor
composite modulation solutions exist in the KdV and NLS modulation theories. These  new patterns owe their existence
to the fact  that the mKdV  modulation systems, unlike the KdV and NLS modulation systems, are neither strictly hyperbolic
nor genuinely nonlinear \cite{kod2008}. Of course, the latter is not surprising  if one remembers that the KdV and mKdV
modulation systems are related by a non-invertible quadratic mapping \cite{dron76}, a modulation counterpart of the
Miura transformation.

The Gardner equation, similar to the KdV and mKdV equations, is a completely integrable system,  which implies that
the Riemann invariants are in principle available for the associated modulation system. However, we are not aware
of any publications containing a consistent and complete derivation of the Gardner modulation system in Riemann
invariant form (we note that some particular results for the Riemann invariants via the mapping between the KdV
and Gardner spectral problems can be found in \cite{pav95}). Consequently, the full theory of the Gardner undular
bores has not been constructed. Some analytical progress has only been made for the case when the coefficient
$\alpha$ is sufficiently small so that the Gardner equation can be asymptotically reduced to the KdV equation
via a near-identity transformation \cite{ms-1990}. The undular bore solutions in this case are qualitatively
similar to their KdV equation counterparts provided initial discontinuity is not very large. An interesting phenomenology
of the Gardner undular bore solutions, beyond the  KdV paradigm, was revealed in the numerical simulations in
\cite{mh87,gcc-2002}, where the problem of the transcritical flow of a stratified fluid was considered in the
framework of the forced defocusing Gardner equation for a broad range of values for $\alpha<0$ and for the external
forcing amplitude.

In this paper we derive the modulation system for the Gardner equation in the Riemann invariant form and construct
a full classification of the asymptotic ($t \gg 1$) solutions to the Gardner equation (\ref{eq1}) with the initial
conditions in the form of a step
\begin{equation}\label{step}
    u(x,0)=\left\{
    \begin{array}{cc}
    {u}^-, & \quad x<0,\\
    {u}^+, & \quad x>0.
    \end{array}
    \right.
\end{equation}
We consider both signs of the coefficient $\alpha$ for the cubic nonlinear term.

In the KdV equation (\ref{kdv}) theory, the resolution of the step (\ref{step}) occurs via the generation of an undular
bore if $u^->u^+$ or a rarefaction wave if $u^-<u^+$.
For the Gardner equation we show that, due to the form of the nonlinear term in (\ref{eq1}), the structure of the solutions
to the initial value problem (\ref{eq1}),
(\ref{step}) also depends on the positions of the initial step parameters $u^+$, $u^-$
relative to the  turning point $u=1/2\alpha$ of the characteristic velocity $ 6u(1-\alpha u)$ of the dispersionless
limit of the Gardner equation.
The full classification encompasses 16 possible cases (8 for each sign of $\alpha$). The wave patterns encountered
include normal (``bright'') and reversed (``dark'') cnoidal undular bores, rarefaction waves, solibores (kinks),
nonlinear trigonometric bores and various combinations of the above patterns.

The structure of the paper is as follows. In Section II we undertake the derivation of two families of periodic solutions of the Gardner equation (\ref{eq1})
corresponding to two signs of $\alpha$.  The solutions are derived in the ``natural'' parametrization by considering
the travelling wave ansatz $u=u(x-Vt)$ in (\ref{eq1}) and reducing it to an ordinary differential equation $u_\xi^2=Q(u)$, $Q(u)$ being a
polynomial of the fourth degree having (generally) four distinct roots $u_1 \le u_2 \le u_3 \le u_4$, only three of
which are independent. The ordinary differential equation is then integrated in terms of Jacobi elliptic functions and the  harmonic ($m \to 0$)
and  soliton ($m  \to 1$)  limits for both families solutions  are then invesigated, $m$ being the modulus of the elliptic solution.
Section III is devoted to the derivation of the Whitham modulation equations in Riemann invariant form. For that,
we take advantage of the reduced version of the finite-gap integration method \cite{kamch2000} to derive the ``spectral''
representation of the periodic solutions obtained in the previous section. The outcome is the set of relationships between
the spectral parameters $r_1, r_2, r_3$ and the dependent set $u_1,u_2,u_3,u_4$ characterising the periodic solution
(two possible sets of relationships ${\bf u}(\bf r)$ are derived for each sign of $\alpha$ --- this is a consequence of the invariance of the Gardner equation with respect to the transformation (\ref{inv})). The Whitham modulation equations are then derived for which $r_j$'s are the Riemann invariants. In Section IV, based on the results obtained in Sections II and III, we construct the full classifications of the solutions to the evolution of an initial discontinuity problem for the Gardner equation with $\alpha>0$ and $\alpha<0$.
In Section V we draw conclusions from our analysis and outline possible applications of the obtained solutions.

\section{Periodic solution of the Gardner equation}

We start with a direct derivation of the periodic travelling wave solution
of the Gardner equation (\ref{eq1}). General expressions for such solutions can be found in \cite{vas11}.
Here we need a more detailed description suitable for our subsequent development of the undular bore theory.
Introducing the substitution
\begin{equation}\label{eq2}
    u=u(\xi),\qquad \xi=x-Vt,
\end{equation}
and integrating twice we arrive at a nonlinear oscillator equation
\begin{equation}\label{eq3}
    u_{\xi}^2=\al u^4-2u^3-Vu^2+Au+B  \equiv  Q(u),
\end{equation}
where $A$ and $B$ are the integration constants. We shall sometimes refer to the polynomial $Q(u)$ in the right-hand side of (\ref{eq3}) as a ``potential curve''
for the nonlinear oscillator described by (\ref{eq3}).
Let $Q(u)$ have four real roots
\begin{equation}\label{eq4}
    u_1\leq u_2\leq u_3\leq u_4
\end{equation}
(the case when $Q(u)$ has two real and two complex conjugate roots corresponds to modulationally unstable solutions \cite{dron76} so we do not consider it here).
The roots (\ref{eq4}) are obviously related by the condition
\begin{equation}\label{11-1}
    \sum_{i=1}^4u_i=\frac2{\al} \, ,
\end{equation}
and hence only three of them are independent. It is still convenient to keep all four $u_j$'s in the
subsequent formulae to preserve the symmetry of the expressions.

\smallskip

We should distinguish between two qualitatively different cases.

\smallskip

\begin{figure}[h]
\begin{center}$
\begin{array}{cc}
\includegraphics[width=5.3cm, clip]{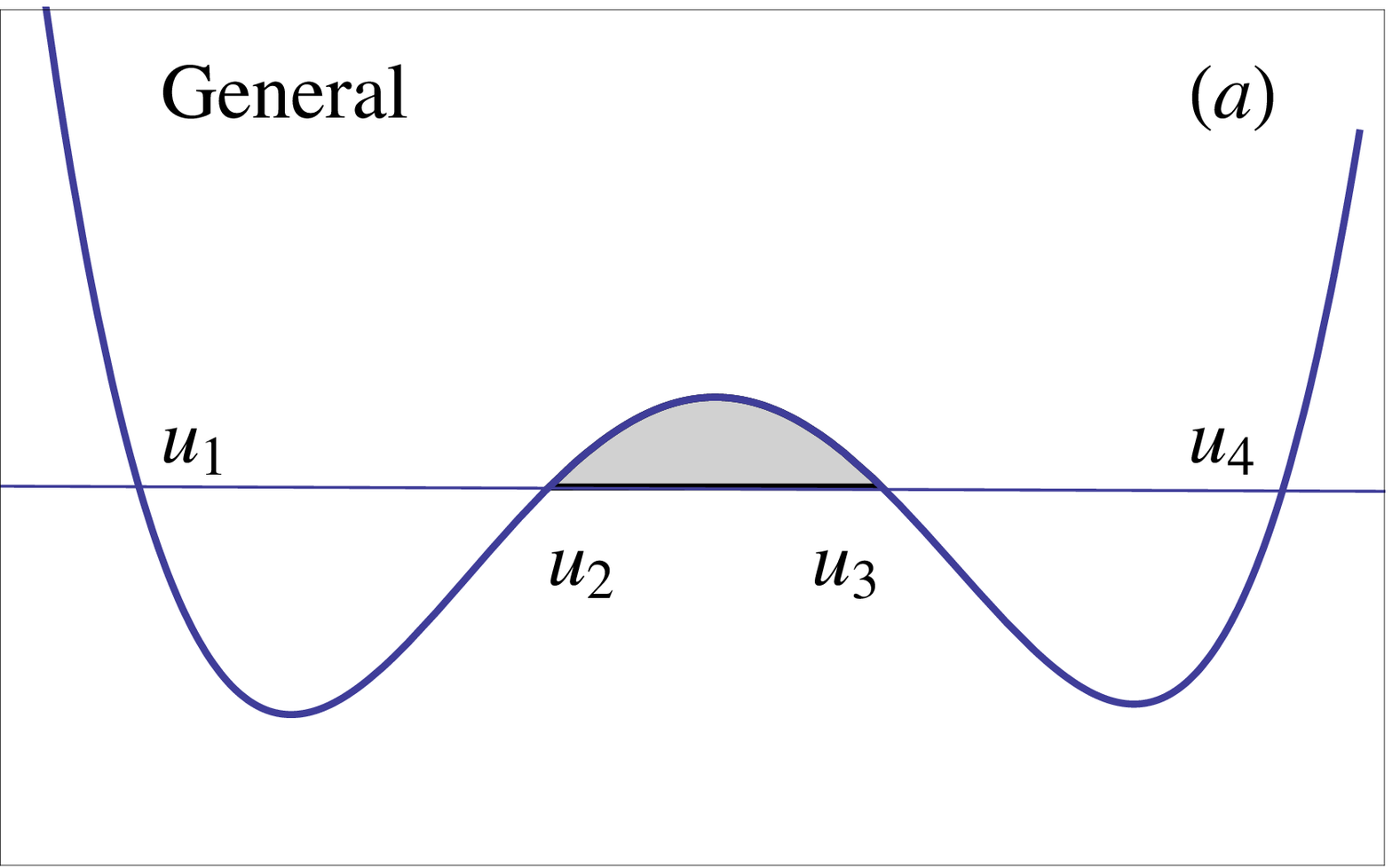}  \\
\includegraphics[width=5.3cm, clip]{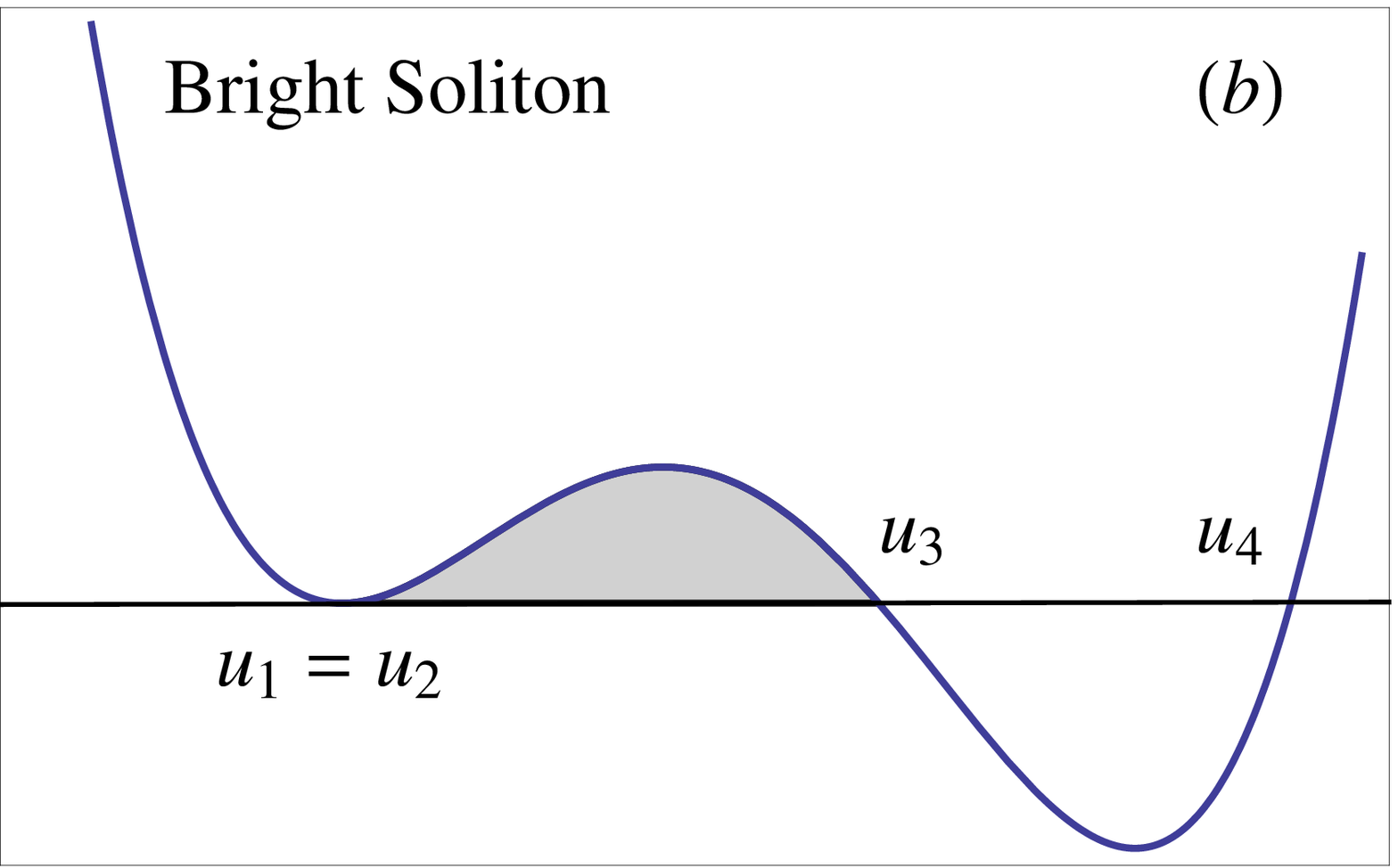} \\
\includegraphics[width=5.3cm, clip]{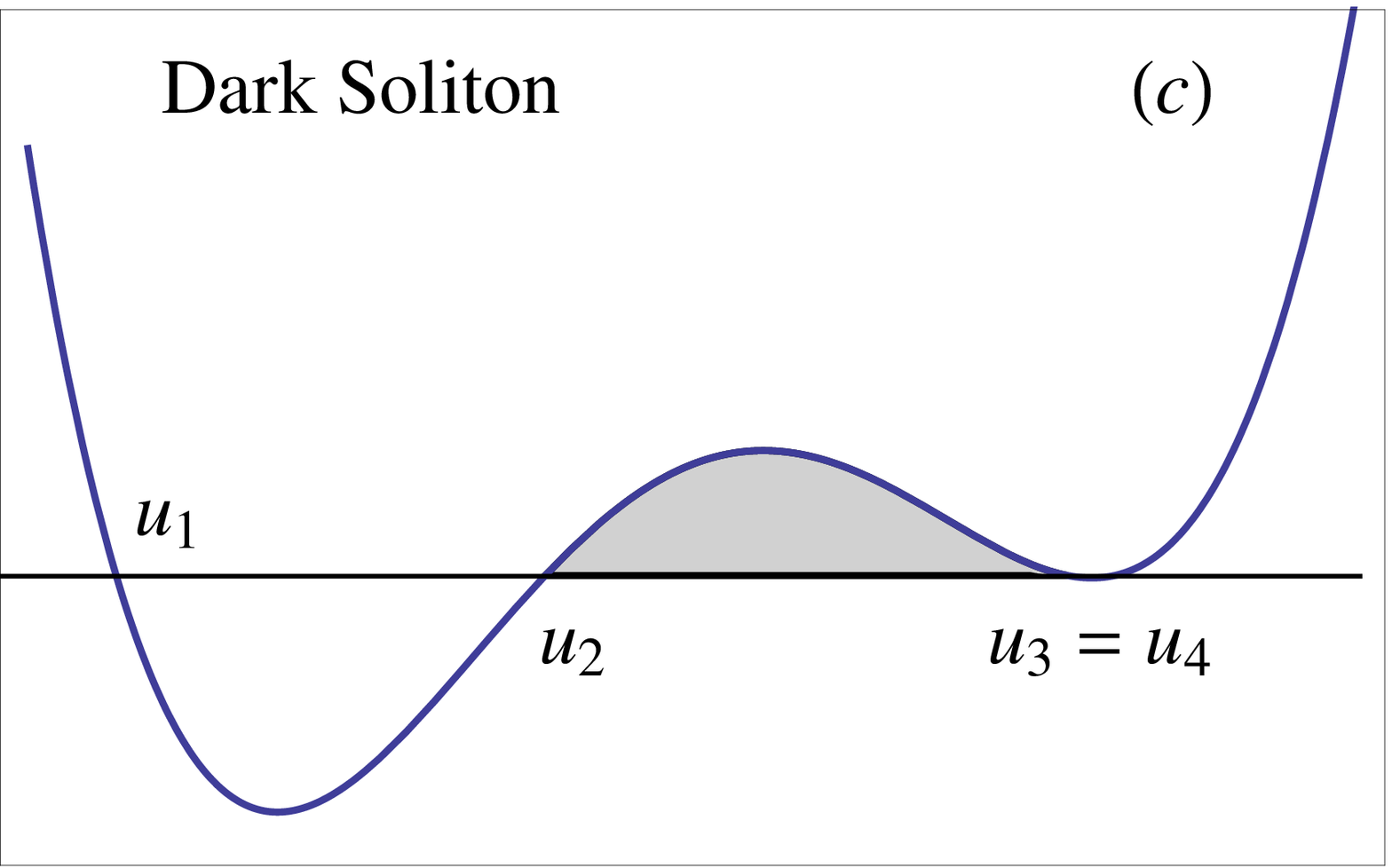}\\
\includegraphics[width=5.3cm, clip]{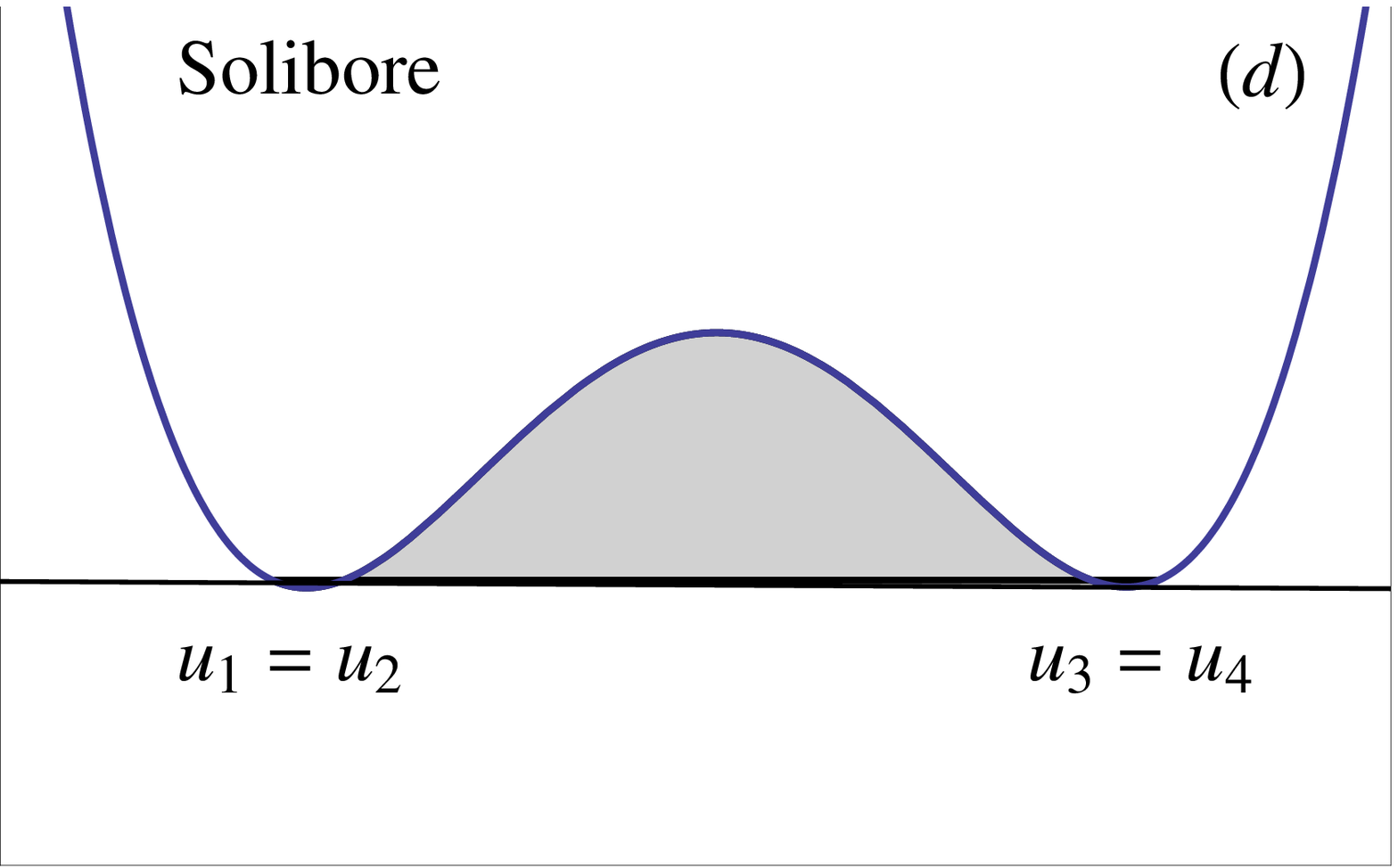}\\
\includegraphics[width=5.3cm, clip]{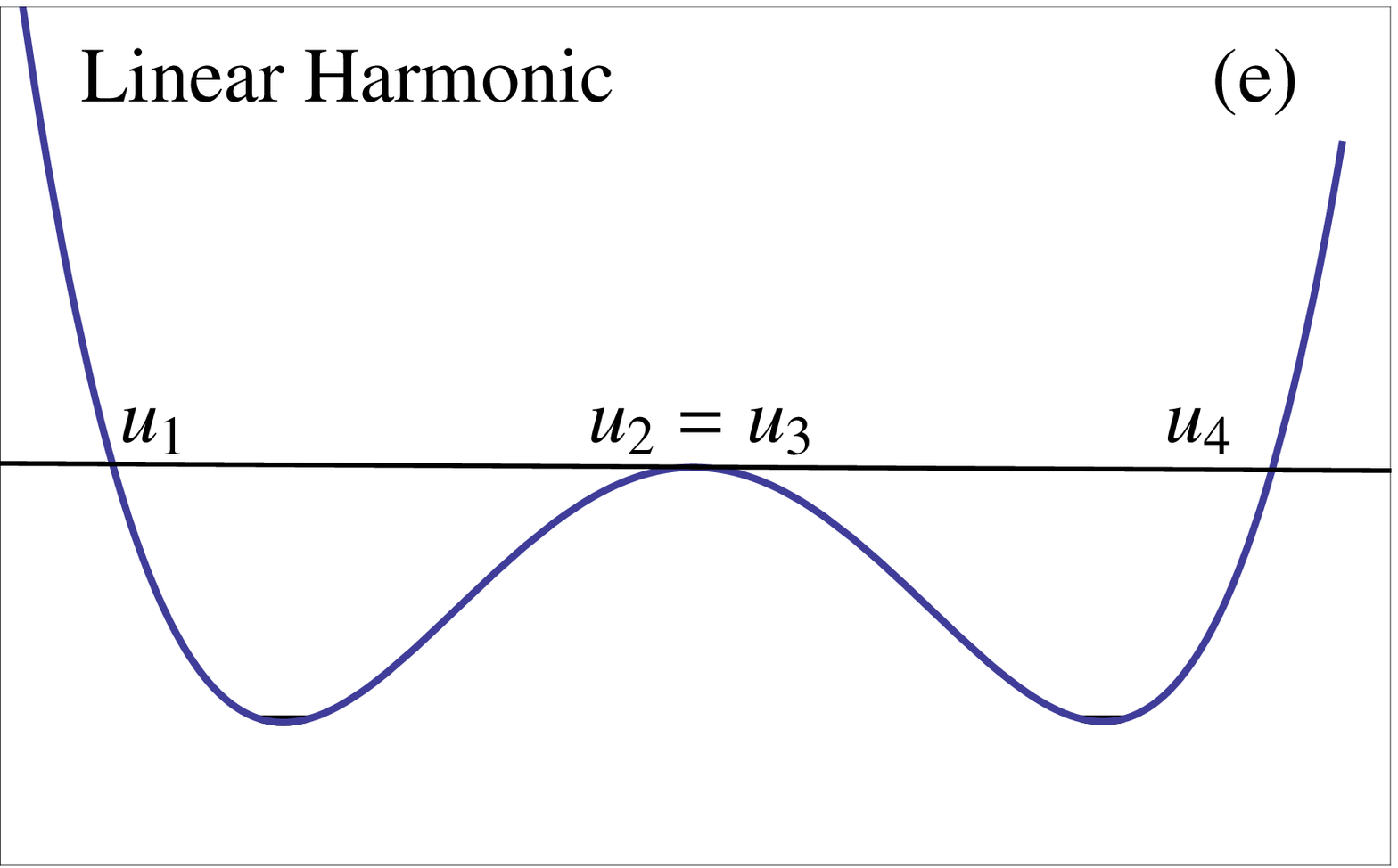}
\end{array}$
\end{center}
\caption{(Color online) Potential curve $Q(u)$ configurations for the travelling wave solutions of the Gardner equation with $\alpha>0$.
The oscillations occur between $u_2$ and $u_3$. (a) periodic (elliptic) solution; (b) bright soliton;
(c) dark soliton; (d) solibore; (e) linear wave.}
\label{pospotential}
\end{figure}

(a) Let $\al>0$.
Then the periodic solution corresponds to the oscillations in the interval
\begin{equation}\label{eq5}
    u_2\leq u\leq u_3 \, ,
\end{equation}
where the polynomial $Q(u)$ is positive and
\begin{equation}\label{eq6}
    \sqrt{\al}(\xi-\xi_0)=\int_u^{u_3}\frac{du}{\sqrt{(u-u_1)(u-u_2)(u_3-u)(u_4-u)}}.
\end{equation}
The possible configurations of the ``potential'' curve $Q(u)$ corresponding to qualitatively different travelling wave solutions are shown in Fig \ref{pospotential}

The integral in (\ref{eq6}) can be expressed in terms of the incomplete elliptic integral
of the first kind
and its inversion yields, after some algebra, the solution in terms of Jacobi elliptic functions:
\begin{equation}\label{el1}
    u=u_2+\frac{(u_3-u_2)\mathrm{cn}^2(\theta,m_1)}{1-
    \frac{u_3-u_2}{u_4-u_2}\mathrm{sn}^2(\theta,m_1)} \, ,
\end{equation}
where
\begin{equation}\label{el1a}
    \theta=\sqrt{\al(u_3-u_1)(u_4-u_2)}(x-Vt)/2 \, ,
\end{equation}
\begin{equation}\label{el2}
    m_1=\frac{(u_3-u_2)(u_4-u_1)}{(u_4-u_2)(u_3-u_1)},
\end{equation}
and $V$ is given by
\begin{equation}\label{el3}
    V=\al(u_1u_2+u_1u_3+u_1u_4+u_2u_3+u_2u_4+u_3u_4).
\end{equation}
The wavelength is given by the formula
\begin{equation}\label{el3a}
    L=\frac{4\K(m_1)}{\sqrt{\al(u_3-u_1)(u_4-u_2)}},
\end{equation}
where $\K(m_1)$ is the complete elliptic integral of the first kind.
The soliton limit $m_1 \to 1$ can be achieved in one of the two ways:
when $u_1 \to u_2$ or when $u_3 \to u_4$.

When $u_2 \to u_1$ we obtain the ``bright'' soliton of elevation propagating against a constant background $u=u_2$ (see Fig.~1b),
\begin{equation}\label{eq70}
    u(\xi)=u_1+\frac{u_3-u_1}{\cosh^2\theta-
    \frac{u_3-u_1}{u_4-u_1}\sinh^2\theta} \, .
\end{equation}
If, further, one has $u_4-u_3 \ll u_3-u_2,$ then the soliton (\ref{eq70}) becomes a wide,
``table-top" soliton.

Analogously, for $u_3 \to u_4$ we choose $\xi_0$ so that $u=u_2$ at $\xi=0$ and obtain
\begin{equation}\label{eq7a}
    u(\xi)=u_4-\frac{u_4-u_2}{\cosh^2\theta-
    \frac{u_4-u_2}{u_4-u_1}\sinh^2\theta}.
\end{equation}
This is a ``dark'' soliton  on the constant background $u=u_4$ (see Fig.~1c). If $u_2-u_1 \ll u_3-u_2$
it assumes the form of a  depression counterpart of the ``table-top'' soliton.

If both $u_2 \to u_1$ and $u_3 \to u_4$ then the polynomial $Q(u)$ in the right-hand side of (\ref{eq3})
has two double roots (see Fig.~1d), which implies that the solution assumes the form of a kink (a ``solibore'').
To study this limit, it is convenient to choose $\xi_0$ in such a way that $u=(u_1+u_4)/2$ at $\xi=0$.
As a result, an elementary integration of (\ref{eq3}) yields two possible solutions
\begin{equation}\label{solibore}
u(\xi)=u_4-\frac{u_4-u_1}{\exp[\pm\sqrt{\al}(u_4-u_1)\xi]+1} \, .
\end{equation}
The lower sign corresponds to the kink with $u \to u_4$ as $\xi\to-\infty$ and $u \to u_1$ at $\xi\to\infty$;
the upper sign yields the ``anti-kink'' with $u \to u_1$ as $\xi\to-\infty$ and $u \to u_4$ at $\xi\to\infty$.
As follows from (\ref{11-1}), the limiting constant states $u_1$ and $u_4$ are  related by the condition $u_1+u_4=1/\al$.
The speed of the kink (solibore) propagation in both cases is $c=\alpha^{-1}+2\alpha u_1u_4$, which agrees with
the shock speed obtained from the first conservation law $u_t+(3u^2 -2\alpha u^3)_x=0$ of the dispersionless
limit of the Gardner equation.

When $u_3\to u_2$ ($m_1\to 0$) (see Fig.~1e) the cnoidal wave (\ref{el1}) asymptotically transforms into a linear harmonic wave
\begin{equation}\label{eq7b}
\begin{split}
    &u\cong u_2+\tfrac12(u_3-u_2)\cos[k(x-Vt)],\\
    &k=\sqrt{\alpha(u_2-u_1)(u_4-u_2)},\\
    &V=4u_2+\alpha(u_1u_4-3u_2^2).
    \end{split}
\end{equation}

\begin{figure*}[ht]
\begin{center}$
\begin{array}{c}
\includegraphics[width=5.3cm, clip]{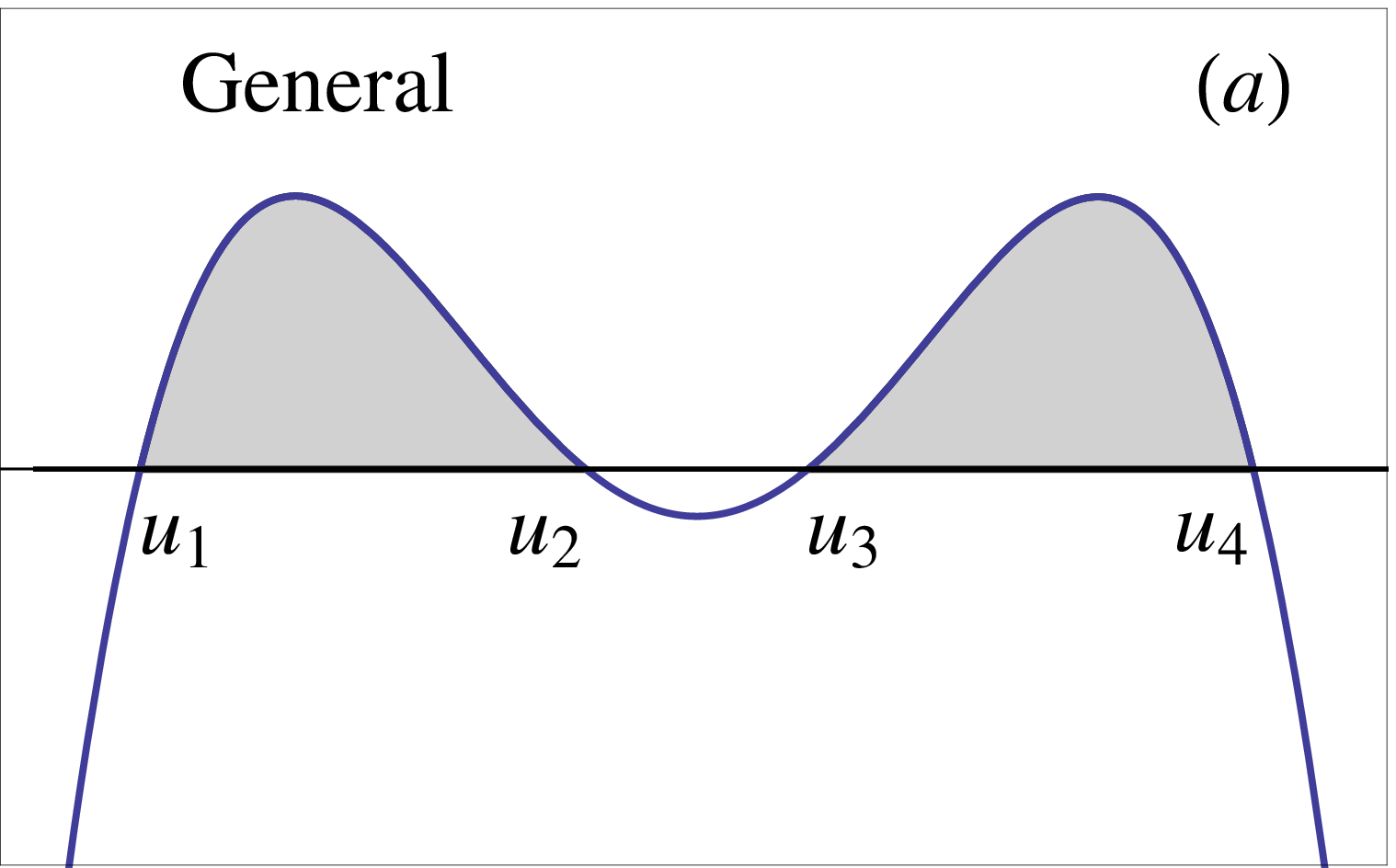} \
\includegraphics[width=5.3cm, clip]{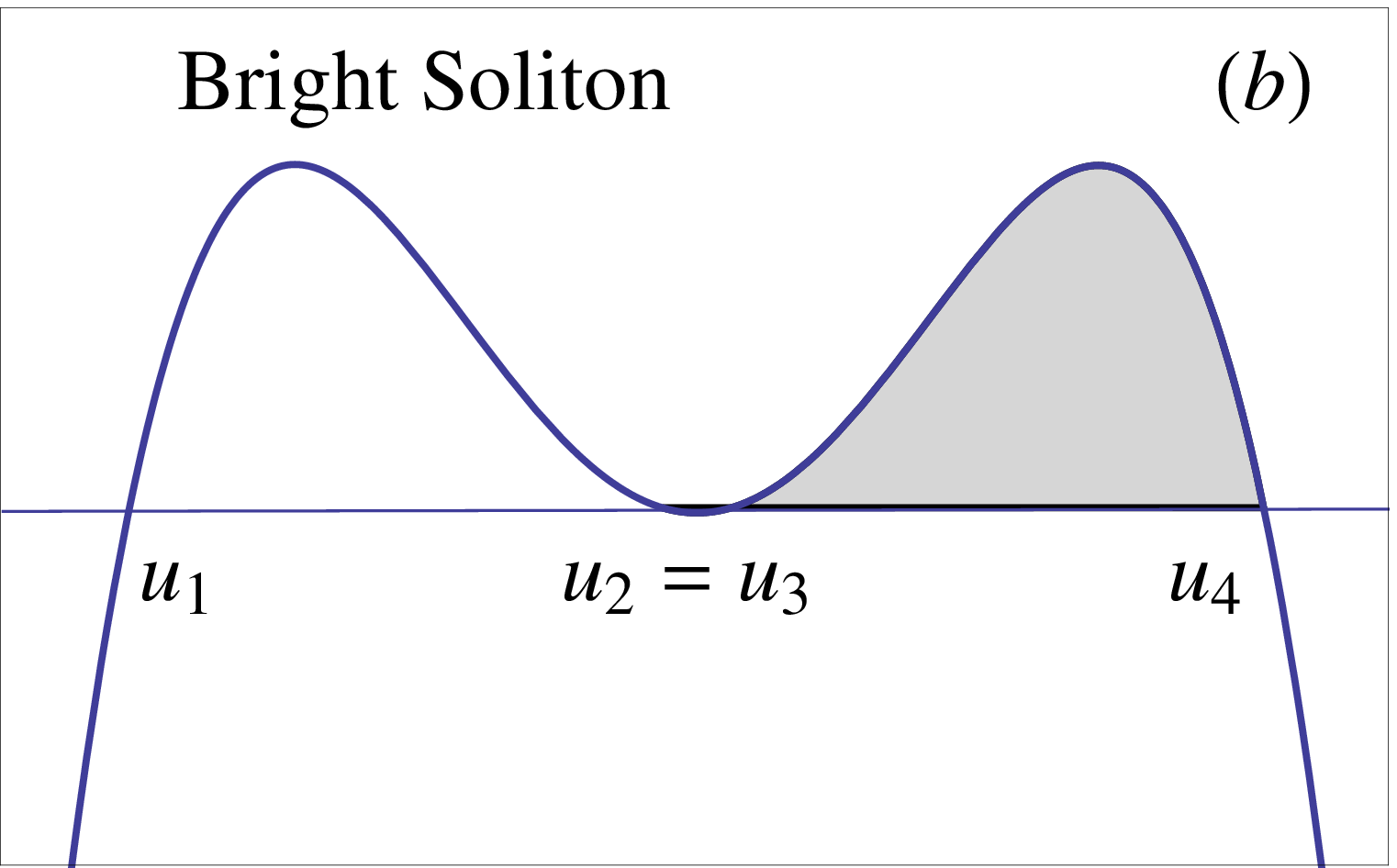} \
\includegraphics[width=5.3cm, clip]{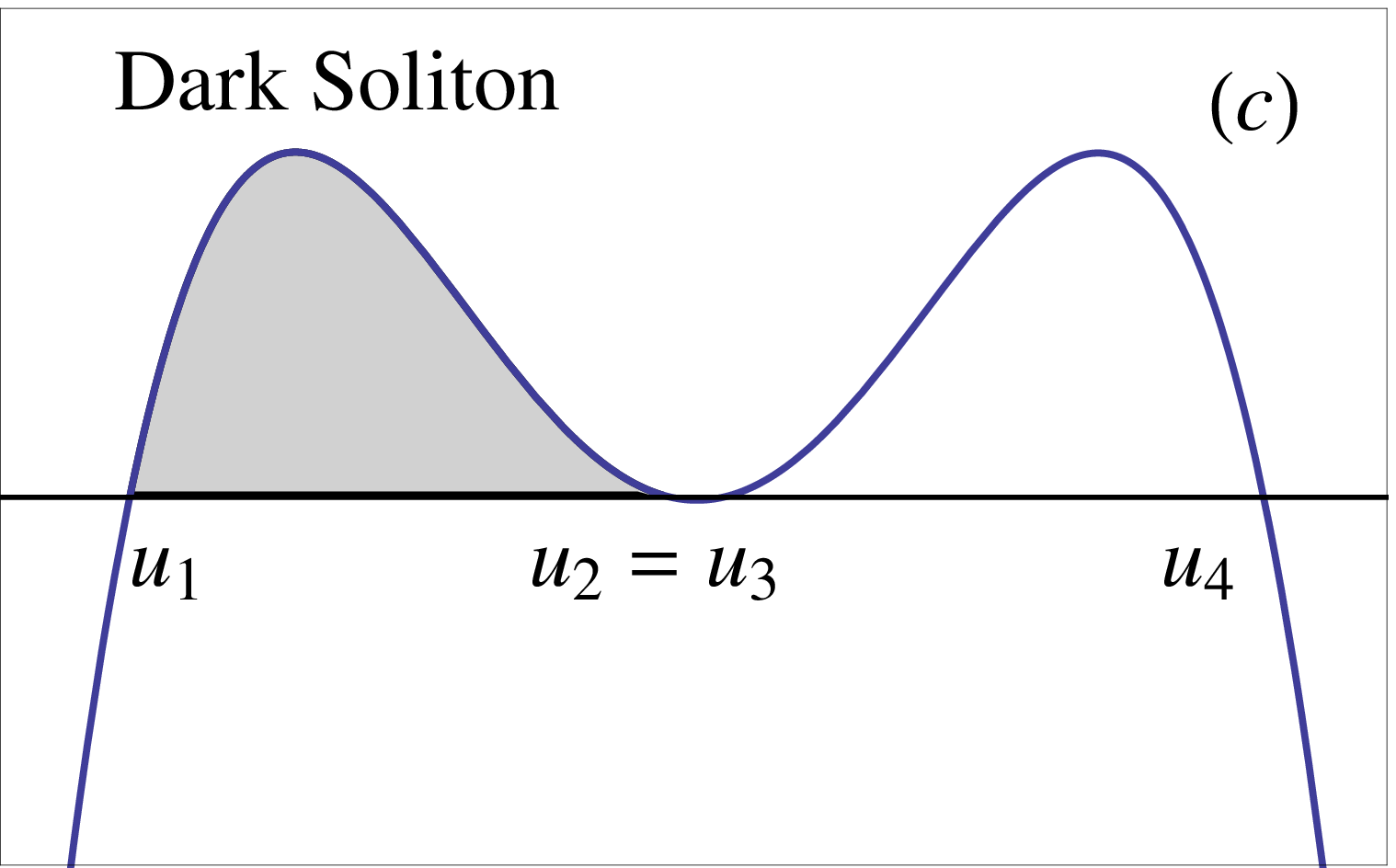}\\ \\
\includegraphics[width=5.3cm, clip]{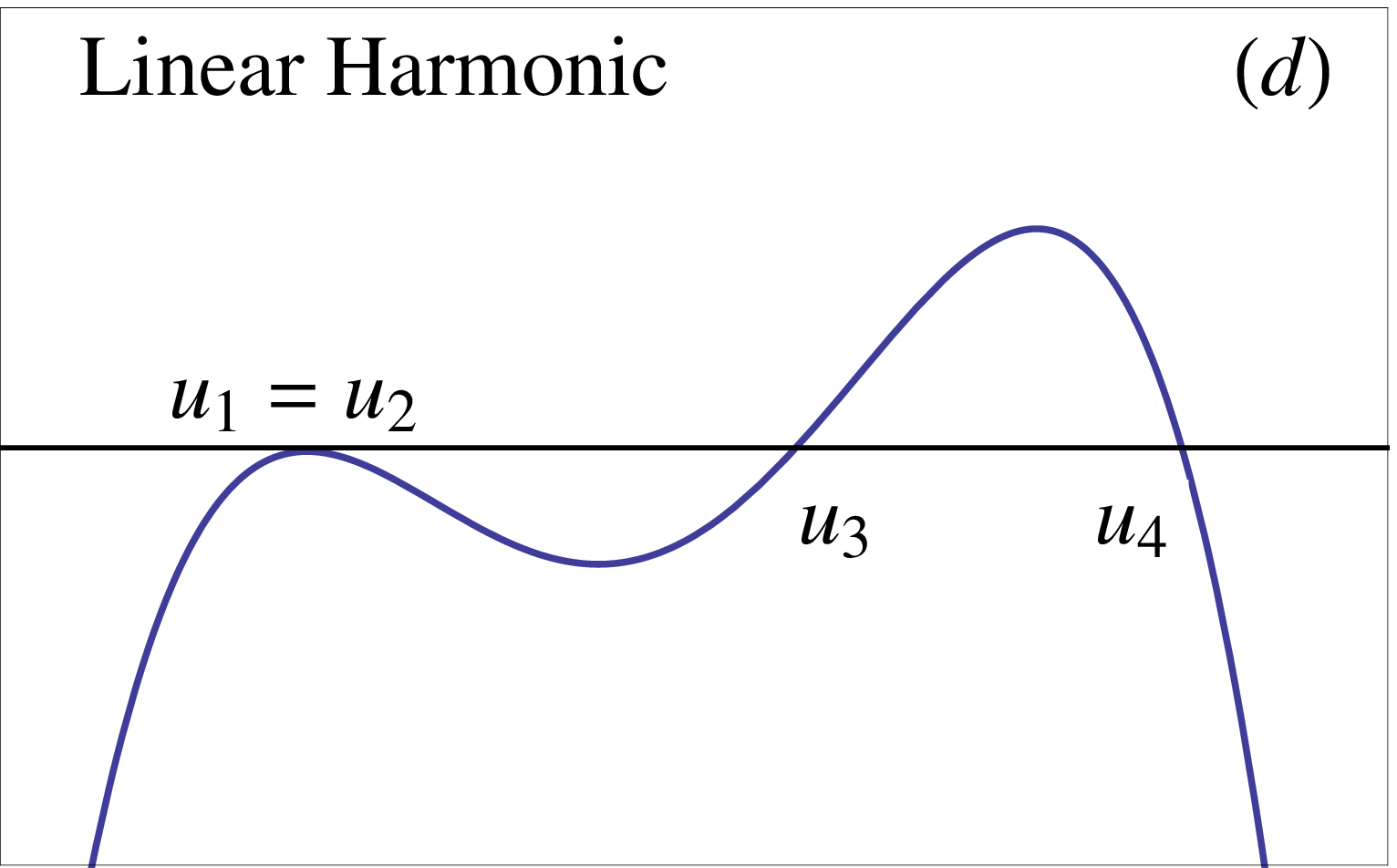} \quad
\includegraphics[width=5.3cm, clip]{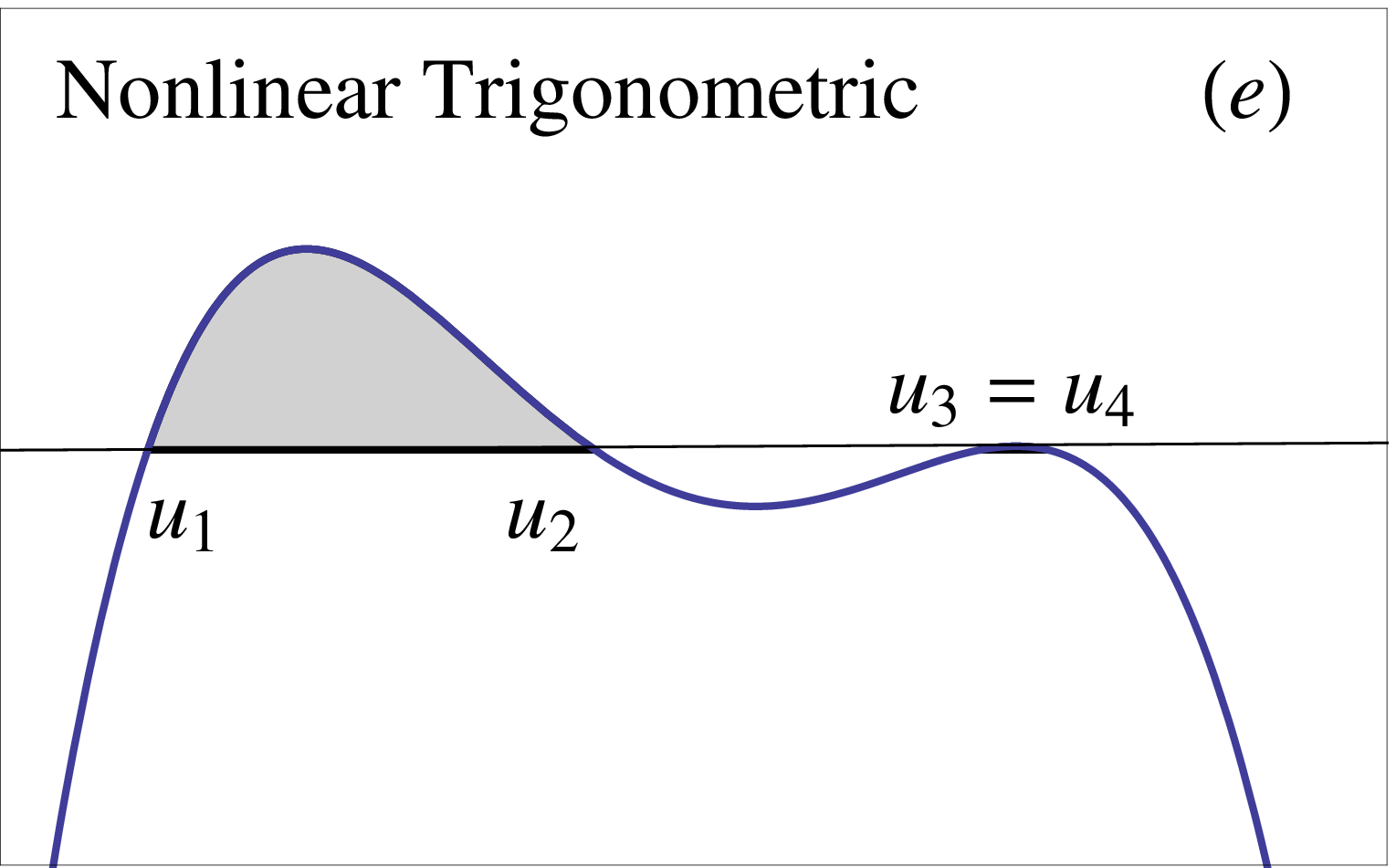}\\ \\
\includegraphics[width=5.3cm, clip]{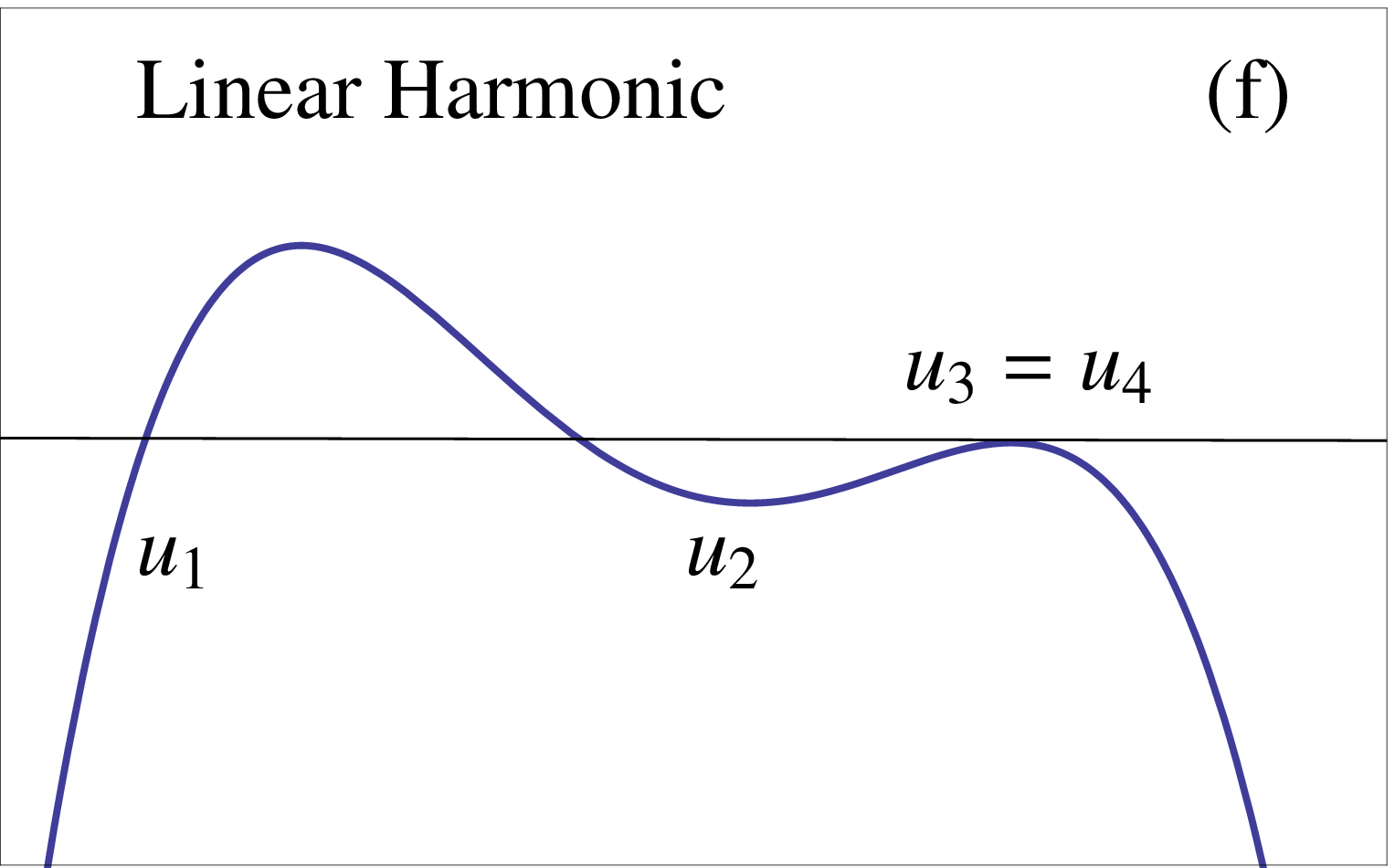} \quad
\includegraphics[width=5.3cm, clip]{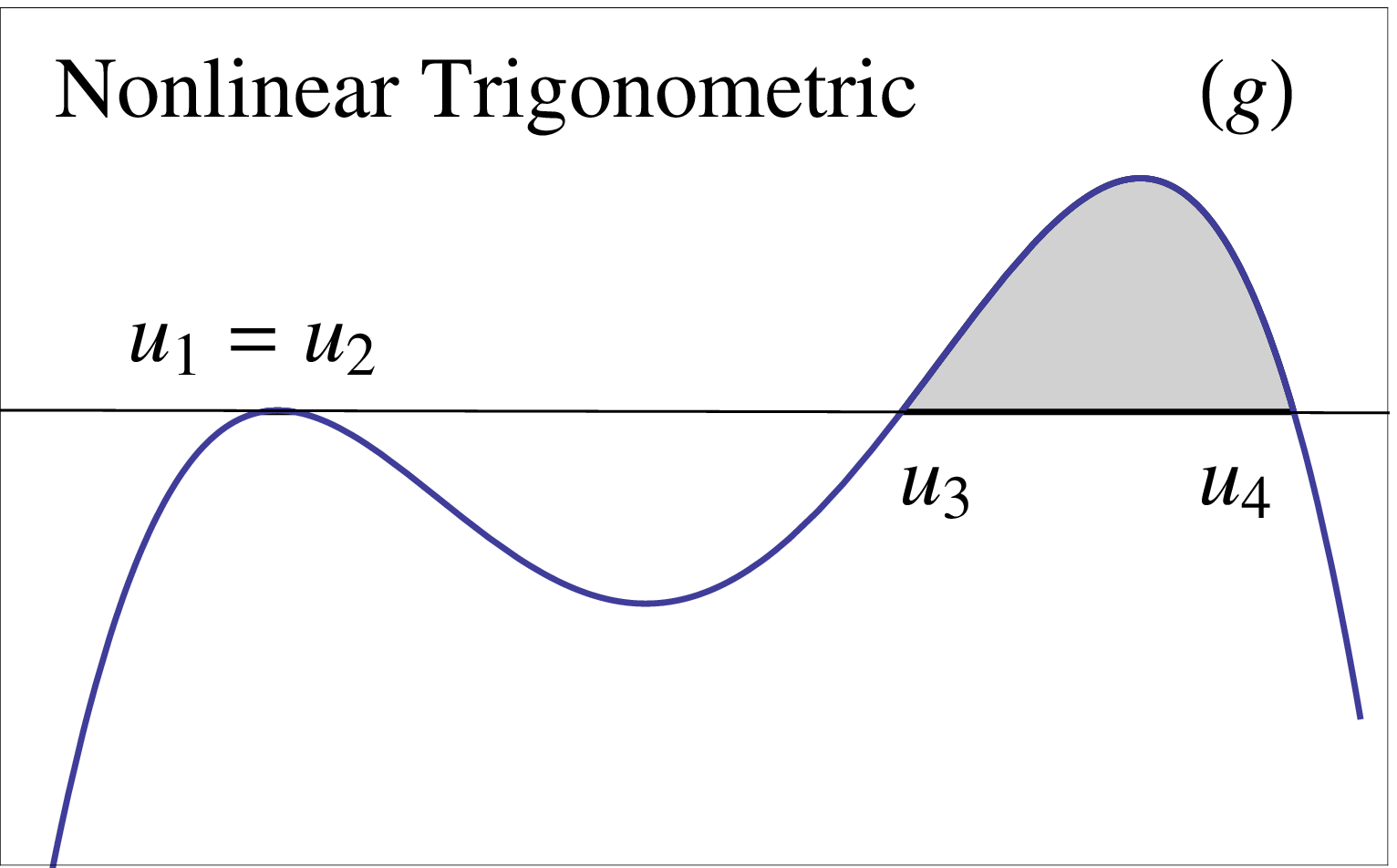}
\end{array}$
\caption{ (Color online) Potential curve $Q(u)$ configurations for the travelling wave solutions of the Gardner equation with $\alpha<0$.
(a) periodic (elliptic) solutions: $u_1\le u \le u_2$ or $u_3\le u \le u_4$;
(b) bright soliton; (c) dark soliton; (d) linear harmonic wave, $m_2=0$, $a=0$ propagating about the background $u=u_1=u_2$; (e) nonlinear trigonometric wave,
$m_2=0$, $a \ne 0$, $u_1 \le u \le u_2$; (f) linear harmonic wave, $m_2=0$, $a=0$ propagating about $u=u_3=u_4$; (g) nonlinear trigonometric wave,
$m_2=0$, $a \ne 0$, $u_3 \le u \le u_4$. }
\label{negpotentials1}
\end{center}
\end{figure*}

(b) Let now $\al<0$.
Then periodic solution corresponds to the oscillations in one of the two intervals,
\begin{equation}\label{eq9}
    u_1\leq u\leq u_2\quad\text{or}\quad u_3\leq u\leq u_4,
\end{equation}
where the polynomial $Q(u)$ assumes positive values. The possible configurations of the potential curve $Q(u)$ are shown in
Fig.~\ref{negpotentials1}.

First we consider the case
\begin{equation}\label{eq9a}
    u_1\leq u\leq u_2,
\end{equation}
so that
\begin{equation}\label{eq10}
    \sqrt{|\al|}(\xi-\xi_0)=\int_{u_1}^{u}\frac{du}{\sqrt{(u-u_1)(u_2-u)(u_3-u)(u_4-u)}}.
\end{equation}
A standard calculation yields
\begin{equation}\label{eq11}
    u=u_2-\frac{(u_2-u_1)\mathrm{cn}^2(\theta,m_2)}{1+
    \frac{u_2-u_1}{u_4-u_2}\mathrm{sn}^2(\theta,m_2)},
\end{equation}
where
\begin{equation}\label{el1b}
    \theta=\sqrt{|\al|(u_3-u_1)(u_4-u_2)}(x-Vt)/2,
\end{equation}
\begin{equation}\label{eq12}
    m_2=\frac{(u_4-u_3)(u_2-u_1)}{(u_4-u_2)(u_3-u_1)}.
\end{equation}
Now the wavelength is given by
\begin{equation}\label{eq3a}
    L=\frac{4\K(m_2)}{\sqrt{|\al|(u_3-u_1)(u_4-u_2)}}.
\end{equation}

In the soliton limit $u_3\to u_2$ ($m_2 \to 1$) we get
\begin{equation}\label{eq13}
\begin{split}
    &u=u_2-\frac{u_2-u_1}{\cosh^2\theta+
    \frac{u_2-u_1}{u_4-u_2}\sinh^2\theta},\\
    &V=2u_2+\alpha(u_1u_4-3u_2^2).
    \end{split}
\end{equation}
This is a ``dark'', depression soliton.

The limit $m_2\to 0$ can be reached by two ways.

{\it (1)} If $u_2\to u_1$  (see Fig.~2d) we get asymptotically
\begin{equation}\label{eq13b}
\begin{split}
    &u\cong u_2-\tfrac12(u_2-u_1)\cos[k(x-Vt)],\\
    &k=\sqrt{|\alpha|(u_3-u_1)(u_4-u_1)},\\
    &V=4u_1+\alpha(u_3u_4-3u_1^2).
    \end{split}
\end{equation}
This is a small-amplitude harmonic limit.

{\it (2)} If $u_4=u_3$, but $u_1 \ne u_2$  (see Fig.~2e) then we arrive at the nonlinear trigonometric solution
\begin{equation}\label{eq13c}
    u=u_2-\frac{(u_2-u_1)\cos^2\theta}{1+
    \frac{u_2-u_1}{u_3-u_2}\sin^2\theta}
\end{equation}
where
\begin{equation}\label{eq13d}
\begin{split}
    &\theta=\sqrt{|\al|(u_3-u_1)(u_3-u_2)}(x-Vt)/2,\\
    &V=4u_3+\alpha(u_1u_2-3u_3^2).
    \end{split}
\end{equation}
If we take the limit $u_2-u_1\ll u_3-u_1$ in this solution, then we return to
the particular case of the small amplitude limit (\ref{eq13b}) with $u_4=u_3$,
so that $k=\sqrt{|\alpha|}(u_3-u_1)$ and $V=4u_3+\alpha(u_1^2-3u_3^2)\equiv
4u_1+\alpha(u_3^2-3u_1^2)$.
On the other hand, if we take here the limit $u_2\to u_3=u_4$, then the argument
of trigonometric functions becomes small and we can approximate them by the first
terms of their series expansions to get the dark algebraic soliton
\begin{equation}\label{eq13e}
\begin{split}
    &u=u_2-\frac{u_2-u_1}{1+|\alpha|(u_2-u_1)^2(v-Vt)^2/4},\\
    &V=2u_2(1+\alpha u_1).
    \end{split}
\end{equation}

\smallskip

Now we consider the case
\begin{equation}\label{eq9b}
    u_3\leq u\leq u_4,
\end{equation}
so that
\begin{equation}\label{eq14}
    \sqrt{|\al|}(\xi-\xi_0)=\int_{u}^{u_4}\frac{du}{\sqrt{(u-u_1)(u-u_2)(u-u_3)(u_4-u)}}.
\end{equation}
Again, the standard calculation yields
\begin{equation}\label{eq15}
    u=u_3+\frac{(u_4-u_3)\mathrm{cn}^2(\theta,m_2)}{1+
    \frac{u_4-u_3}{u_3-u_1}\mathrm{sn}^2(\theta,m_2)}.
\end{equation}
In the soliton limit $u_3\to u_2$ ($m\to1$) we get
\begin{equation}\label{eq16}
    u=u_2+\frac{u_4-u_2}{\cosh^2\theta+
    \frac{u_4-u_2}{u_2-u_1}\sinh^2\theta},
\end{equation}
where
\begin{equation}\label{eq16a}
\begin{split}
    &\theta=\sqrt{|\al|(u_2-u_1)(u_4-u_2)}(x-Vt)/2,\\
    &V=4u_2+\alpha(u_1u_4-3u_2^2).
    \end{split}
\end{equation}
This is a ``bright'' elevation soliton.

Again, there are two ways for getting the limit $m_2\to0$.

{\it (1)} If $u_4\to u_3$ (see Fig.~2f), then we obtain a small-amplitude harmonic wave
\begin{equation}\label{eq16b}
    u=u_3+\tfrac12(u_4-u_3)\cos[k(x-Vt)],
\end{equation}
where
\begin{equation}\label{eq16c}
\begin{split}
    &k=\sqrt{|\al|(u_3-u_1)(u_3-u_2)}/2,\\
    &V=4u_3+\alpha(u_1u_2-3u_3^2).
    \end{split}
\end{equation}

{\it (2)} If $u_2\to u_1$ (see Fig.~2g), then we get another {\it nonlinear} trigonometric solution
\begin{equation}\label{eq16d}
    u=u_3+\frac{(u_4-u_3)\cos^2\theta}{1+
    \frac{u_4-u_3}{u_3-u_1}\mathrm{sin}^2\theta},
\end{equation}
where
\begin{equation}\label{eq16e}
\begin{split}
    &\theta=\sqrt{|\al|(u_3-u_1)(u_4-u_1)}(x-Vt)/2,\\
    &V=4u_1+\alpha(u_3u_4-3u_1^2).
    \end{split}
\end{equation}
If we assume here $u_4-u_3\ll u_4-u_1$, then we reproduce the small amplitude
asymptotics  (\ref{eq16b}) with $u_2 \cong u_1$, so that $k=\sqrt{|\alpha|}(u_3-u_1)$,
$V=4u_3+\alpha(u_1^2-3u_3^2)$. On the other hand, Eq.~(\ref{eq16d}) in the
limit $u_3\to u_2=u_1$ reduces to the algebraic bright soliton solution
\begin{equation}\label{eq13f}
\begin{split}
    &u=u_1+\frac{u_4-u_1}{1+|\alpha|(u_4-u_1)^2(v-Vt)^2/4},\\
    &V=2u_1(1+\alpha u_4).
    \end{split}
\end{equation}

This completes the classification of stable periodic solutions and their limiting cases
of the Gardner equation.

\section{Spectral parametrization of the periodic solution}

\subsection{Motivation}

The periodic solution derived in the previous section is parameterized by four ``integrals of motion''
$u_1\leq u_2\leq u_3\leq u_4$, which are related by the condition (\ref{11-1}).
In a strictly periodic solution these parameters $u_j$ are constants, but in a modulated wave,
which we are interested in, they become slow functions of space coordinate $x$ and time $t$.
Their evolution is then governed by the  Whitham modulation equations (see \cite{whitham1,whitham2,kamch2000})
which can be obtained by averaging  the conservation
laws of the Gardner equation over the periodic solution family (\ref{eq3}), and which, generally speaking,
have the form of a quasilinear (hydrodynamic type) system
\begin{equation}\label{11-3}
    \frac{\prt u_i}{\prt t}+\sum_{j} v_{ij}\frac{\prt u_j}{\prt x}=0,
    \quad i,j= \hbox{any three of}  \ \{1,2,3,4\}.
\end{equation}
Here the matrix elements $v_{ij}$ are functions of ${\bf u}=(u_1, u_2, u_3, u_4)$
(note that one of the variables $u_i$
can be eliminated with the help of Eq.~(\ref{11-1}) but then the symmetry of the above
expressions for the periodic solution will be lost). The modulation system in the form (\ref{11-3})
would be, however, completely impractical due to the highly complicated
structure of the matrix elements $v_{ij}$ --- this is already the case even for the KdV equation
(see \cite{whitham1,whitham2,kamch2000}). Fortunately, for the Gardner equation
this system  can be transformed to the
Riemann diagonal form
\begin{equation}\label{11-4}
    \frac{\prt r_k}{\prt t}+v_k({\bf r})\frac{\prt r_k}{\prt x}=0,\quad k=1,2,3,
\end{equation}
where  $r_k,\,k=1,2,3,$ are the Riemann invariants. This is possible due to the fact that the Gardner
equation is a completely integrable equation. Moreover, one can expect that, at least
for $\alpha>0$, the Whitham system for the Gardner equation will be closely related (or even equivalent)
to the Whitham system for the KdV equation. Indeed, for $\alpha>0$ the Gardner equation could be reduced,
by a simple change of variables (\ref{transmkdv1}), to the defocusing mKdV equation, which, in its turn, is connected with the KdV
equation by the Miura transform. As a result, for $\alpha>0$ the mKdV-Whitham system in the Riemann
form is equivalent to that of the KdV equation (the result first obtained in \cite{dron76}) and the
same is true for the Whitham-Gardner system \cite{pav95}. The Whitham equations (\ref{11-4}) can
be readily solved analytically, providing the necessary modulation solutions. The problem, however,
is that one still needs to know the dependence of $u_i$'s on the Riemann
invariants $r_1,r_2,r_3$ for the Gardner equation to be
able to find the modulations of the periodic traveling wave solutions obtained in the ``natural''
$u_j$-parametrization.
This dependence for a particular case of the travelling wave solution (\ref{el1}) was found in
\cite{pav95} but the description in \cite{pav95}, being merely an illustration of a more general theory,
is too brief and somewhat incomplete for our purposes, so below we present a detailed calculation, which also will not
be restricted to the case $\alpha>0$.

The transformation $u_i=u_i(\bf{r})$ is most conveniently found using  the spectral theory of the
Gardner equation (\ref{eq1}).
The method of obtaining periodic (generally, quasiperiodic) solutions via the linear spectral
problem associated with an integrable nonlinear dispersive equation is usually referred to as
the finite-gap integration method. It is based on the highly nontrivial properties of quasiperiodic
solutions of soliton equations which have only a finite number of bands (gaps) in their spectrum
when considered as potentials in the associated spectral problem (see e.g. \cite{ZMNP}). In the
context of the Whitham modulation theory the advantage of the finite-gap integration over the
direct procedure of finding periodic solutions is that, if the endpoints of the spectral bands
of the potential (quasiperiodic solution) are allowed to slowly vary with $x$ and $t$, they
become the {\it Riemann invariants} of the modulation equations. The full finite-gap integration
theory and the associated modulation theory, however, are quite technical and involve rather
complicated algebraic-geometrical constructions
on hyperelliptic Riemann surfaces (see e.g. seminal paper \cite{ffm-1980} where this theory was
developed for the first time for the KdV equation). However, in the single-phase periodic case
of our interest a more simple, reduced version of the finite-gap integration method is available
\cite{kamch2000} enabling one to derive the required  Riemann invariant parametrization for
periodic solutions of integrable equations associated with $(2 \times 2)$ linear spectral problems.

\subsection{Spectral theory}

The finite-gap integration method is based on the possibility to represent the Gardner
equation (\ref{eq1})
as a compatibility condition of two linear systems (see, e.g. \cite{ZMNP,kamch2000})
\begin{equation}\label{11-6}
    \Psi_x=\mathbb{U}\Psi,\quad \Psi_t=\mathbb{V}\Psi,
\end{equation}
where
\begin{equation}\label{11-7}
    \Psi=\left(
      \begin{array}{c}
        \psi_1 \\
        \psi_2 \\
      \end{array}
    \right), \quad
    \mathbb{U}=\left(
                 \begin{array}{cc}
                   F & G \\
                   H & -F \\
                 \end{array}
               \right), \quad
    \mathbb{V}=\left(
      \begin{array}{cc}
        A & B \\
        C & -A \\
      \end{array}
    \right),
\end{equation}
\begin{equation}\label{11-8}
    F=\la,\quad G=-(1-\al u),\quad H=u,
\end{equation}
\begin{equation}\label{12-1}
    \begin{split}
    &A= -4\la^3-2\la(1-\al u)u+u_x,\\
    &B= 4\la^2(1-\al u)-2\la\al u_x-\al u_{xx}+2(1-\al u)^2u,\\
    &C= -4\la^2u+2\la u_x-u_{xx}-2(1-\al u)u^2.
    \end{split}
\end{equation}
This means that the condition $\Psi_{xt}=\Psi_{tx}$ reduces to Eq.~(\ref{eq1}).

Calculations become somewhat simpler if we transform this matrix form of equations to their
scalar counterparts (see \cite{kk-02})
\begin{equation}\label{12-2}
    \psi_{xx}=\mathcal{A}\psi,\quad \psi_t=-\tfrac12\mathcal{B}_x\psi+\mathcal{B}\psi_x,
\end{equation}
where
\begin{equation}\label{12-3}
    \mathcal{A}=\left(\la+\frac{u_x}{2u}\right)^2-(1-\al u)u-\left(\frac{u_x}{2u}\right)_x,
\end{equation}
\begin{equation}\label{12-4}
    \mathcal{B}=-4\la^2+2\la\frac{u_x}u-\frac{u_{xx}}u-2(1-\al u)u.
\end{equation}
Then the second-order spectral equation in (\ref{12-2}) has two basis solutions $\psi_+$ and $\psi_-$
and the ``squared basis function''
\begin{equation}\label{12-5}
    g=\psi_+\psi_-
\end{equation}
satisfies the third-order equation with a well-known integral
\begin{equation}\label{12-6}
    \tfrac12gg_{xx}-\tfrac14g_x^2-\mathcal{A}g^2=P(\la).
\end{equation}
In the finite-gap integration method the periodic solutions are distinguished by the condition
that $P(\la)$ is a polynomial in $\la$. In our case we find that the one-phase periodic solution
corresponds to the polynomial
\begin{equation}\label{12-7}
    P(\la)=-\prod_{i=1}^3(\la^2-\la_i^2)=-(\la^6-s_1\la^4+s_2\la^2-s_3)
\end{equation}
and
\begin{equation}\label{12-8}
    g=\la^2-g_1\la+g_2,
\end{equation}
where the coefficients $g_1$ and $g_2$ are functions of $x$ and $t$. Substitution of
Eqs.~(\ref{12-7}) and (\ref{12-8}) into Eq.~(\ref{12-6}) with $\mathcal{A}$ given by
Eq.~(\ref{12-3}) and equating of the coefficients of equal degrees of $\la$ at both
sides of the resulting equation yields a set of equations for $g_1$, $g_2$,
as well as for $u$ and its
$x$-derivatives. Elimination of $g_1$, $g_2$ and of higher $x$-derivatives of $u$
from these equations gives after a somewhat tedious calculations the equation
\begin{equation}\label{12-9}
   \begin{split}
    &u_x^2=\al u^4-2u^3+2s_1u^2\\
    &+\left\{\frac{1}{\al^2}-\frac{2s_1}{\al} \mp
    \sqrt{(1-4\al\la_1^2)(1-4\al\la_2^2)(1-4\al\la_3^2)}\right\}u\\
    &-\frac1{2\al^3}\Bigg[1-2\al s_1-2\al^2(s_1^2-4s_2)\\
    &\mp     \sqrt{(1-4\al\la_1^2)(1-4\al\la_2^2)(1-4\al\la_3^2)}\Big].
    \end{split}
\end{equation}

The travelling periodic solution of Eq.~(\ref{eq1}) is then obtained by the
replacement $x\mapsto x-Vt$, where $V=-2s_1$, that is
\begin{equation}\label{13-1}
    V=2(\la_1^2+\la_2^2+\la_3^2)=2(r_1+r_2+r_3),
\end{equation}
where we have introduced $r_i=\la_i^2,\,i=1,2,3$. The parameters $r_j$ are expected to become the Riemann invariants
of the Whitham modulation equations.

We note that the representation (\ref{12-9}) of the ordinary differential equation for a periodic solution, unlike that given by its equivalent
(\ref{eq3}), contains two possible signs.  As a result, one needs to use different sets
of relationships between $\{u_i\}$ and $\{ r_i \}$ for different types of solutions, i.e. the mapping $\{u_i\} \mapsto \{ r_i\}$ is not one-to-one.
To express the original parameters $u_j$ (the zeroes of the polynomial in the right-hand side of equation (\ref{eq3})) in terms of the spectral parameters $r_i=\la_i^2$ we compare the two forms of the same ordinary differential equation defining the the periodic solution, namely, Eq.~(\ref{eq3}) and  Eq.~(\ref{12-9}).

First let us consider the case $\al>0$. The example of the mKdV equation
(see \cite{ksk-04}) suggests the following
expressions which can be verified by direct calculations:
\begin{equation}\label{13-2}
    \begin{split}
    &u_1=\frac1{2\al}\left(1-\sqrt{1-4\al r_1}-\sqrt{1-4\al r_2}+\sqrt{1-4\al r_3}\right),\\
    &u_2=\frac1{2\al}\left(1-\sqrt{1-4\al r_1}+\sqrt{1-4\al r_2}-\sqrt{1-4\al r_3}\right),\\
    &u_3=\frac1{2\al}\left(1+\sqrt{1-4\al r_1}-\sqrt{1-4\al r_2}-\sqrt{1-4\al r_3}\right),\\
    &u_4=\frac1{2\al}\left(1+\sqrt{1-4\al r_1}+\sqrt{1-4\al r_2}+\sqrt{1-4\al r_3}\right),
    \end{split}
\end{equation}
in the case of the upper sign in Eq.~(\ref{12-9}) and
\begin{equation}\label{13-3}
    \begin{split}
    &u_1=\frac1{2\al}\left(1-\sqrt{1-4\al r_1}-\sqrt{1-4\al r_2}-\sqrt{1-4\al r_3}\right),\\
    &u_2=\frac1{2\al}\left(1-\sqrt{1-4\al r_1}+\sqrt{1-4\al r_2}+\sqrt{1-4\al r_3}\right),\\
    &u_3=\frac1{2\al}\left(1+\sqrt{1-4\al r_1}-\sqrt{1-4\al r_2}+\sqrt{1-4\al r_3}\right),\\
    &u_4=\frac1{2\al}\left(1+\sqrt{1-4\al r_1}+\sqrt{1-4\al r_2}-\sqrt{1-4\al r_3}\right)
    \end{split}
\end{equation}
in the case of the lower sign. In both cases the zeroes $u_i$ are ordered according to (\ref{eq4})
provided
\begin{equation}\label{13-4}
    \la_1^2\leq \la_2^2\leq \la_3^2\quad \text{or, equivalently,}\quad r_1\leq r_2\leq r_3.
\end{equation}
For both cases (\ref{13-2}) and (\ref{13-3}) the inverse formulae are simply
\begin{equation}\label{ru}
\begin{split}
 r_1=\frac{\alpha}4(u_1+u_2)(u_3+u_4), \\
 r_2=\frac{\alpha}4(u_1+u_3)(u_2+u_4), \\
 r_3=\frac{\alpha}4(u_2+u_3)(u_1+u_4).
\end{split}
\end{equation}
The existence of two sets (\ref{13-2}), (\ref{13-3}) of the travelling wave parameters corresponding to the same set of the spectral parameters $r_j$ is due to the invariance  of the Gardner equation with respect to the transformation (\ref{inv}). Indeed, the set of relationships (\ref{13-3}) can be obtained from (\ref{13-2}) by applying the transformation $u_j \to 1/\alpha - u_j$ and then reordering the resulting set.

In the case of $\al<0$ the expressions remain the same but their order corresponding to (\ref{eq4})
is different. For the upper sign in (\ref{12-9}) we obtain
\begin{equation}\label{13-2a}
    \begin{split}
    &u_1=\frac1{2\al}\left(1+\sqrt{1-4\al r_1}+\sqrt{1-4\al r_2}+\sqrt{1-4\al r_3}\right),\\
    &u_2=\frac1{2\al}\left(1-\sqrt{1-4\al r_1}-\sqrt{1-4\al r_2}+\sqrt{1-4\al r_3}\right),\\
    &u_3=\frac1{2\al}\left(1-\sqrt{1-4\al r_1}+\sqrt{1-4\al r_2}-\sqrt{1-4\al r_3}\right),\\
    &u_4=\frac1{2\al}\left(1+\sqrt{1-4\al r_1}-\sqrt{1-4\al r_2}-\sqrt{1-4\al r_3}\right),\\
    \end{split}
\end{equation}
and for the lower sign
\begin{equation}\label{13-3a}
    \begin{split}
    &u_1=\frac1{2\al}\left(1-\sqrt{1-4\al r_1}+\sqrt{1-4\al r_2}+\sqrt{1-4\al r_3}\right),\\
    &u_2=\frac1{2\al}\left(1+\sqrt{1-4\al r_1}-\sqrt{1-4\al r_2}+\sqrt{1-4\al r_3}\right),\\
    &u_3=\frac1{2\al}\left(1+\sqrt{1-4\al r_1}+\sqrt{1-4\al r_2}-\sqrt{1-4\al r_3}\right),\\
    &u_4=\frac1{2\al}\left(1-\sqrt{1-4\al r_1}-\sqrt{1-4\al r_2}-\sqrt{1-4\al r_3}\right),\\
    \end{split}
\end{equation}
Now the inverse formulae are
\begin{equation}\label{ru-a}
\begin{split}
 r_1=\frac{\alpha}4(u_1+u_4)(u_2+u_3), \\
 r_2=\frac{\alpha}4(u_1+u_3)(u_2+u_4), \\
 r_3=\frac{\alpha}4(u_1+u_2)(u_3+u_4).
\end{split}
\end{equation}

It is essential that the expressions (\ref{el2}) and (\ref{eq12}) for the modulus of the elliptic function
for the cases $\alpha>0$ and $\alpha<0$ respectively, reduce to the same formula in terms of
$r_k$'s:
\begin{equation}\label{mod-a}
    m_1=m_2=m=\frac{r_2-r_1}{r_3-r_1} \, ,
\end{equation}
and in both cases the wavelength is given by the formula
\begin{equation}\label{14-1}
    L=\frac{2\K(m)}{\sqrt{r_3-r_1}}.
\end{equation}

The periodic solutions obtained in the previous
subsection can now be written down directly in terms of $r_j$'s.
This would lead, however, to rather cumbersome expressions so it is better to
keep the original $u_j$-parametrization in the periodic solutions and use the relationships
(\ref{13-2}), (\ref{13-3}) or (\ref{13-2a}), (\ref{13-3a}) for imposing slow
modulation more conveniently represented in terms of $r_j$'s.

\subsection{The Whitham modulation equations}

The Whitham modulation equations in the Riemann form (\ref{11-4}) can be derived using the well-established procedure
of averaging the generating equation for conservation laws (see e.g. \cite{kamch2000}). This procedure for
the Gardner equation, however, is not as straightforward as it is for the KdV or mKdV equations, so, to avoid lengthy
calculations, we make a plausible assumption that the roots $r_j$ of the ``spectral polynomial'' (\ref{12-7}) are
the Riemann invariants of the associated modulation system (\ref{11-4}) as it is the case for the related KdV and
mKdV equations and other integrable systems. Then we observe that expressions (\ref{13-2})--(\ref{13-3a}) do agree
with the corresponding KdV \cite{whitham1} and mKdV (\cite{dron76}) expressions  in the limits as $\alpha \to 0$
and $\alpha \to \pm \infty$ respectively. We also note that the particular set (\ref{13-2}) of the relations
between $u_i$'s and $r_i$'s was actually obtained (up to some obvious misprints) in \cite{pav95} using the traditional
finite-gap method; the remaining expressions (\ref{13-3}), (\ref{13-2a}), (\ref{13-3a}) have the same structure and
can be simply derived as extensions of that result.

Having established the modulation Riemann invariants, the expressions for the corresponding characteristic speeds
$v_i({\bf r})$ can be derived directly from the wavenumber conservation law, by-passing thus the detailed averaging
procedure. To this end we consider the wavenumber conservation law, which is a generic modulation equation
(see, e.g., \cite{whitham1,whitham2,kamch2000}),
\begin{equation}\label{wcl}
k_t+(kV)_x=0\, ,
\end{equation}
where $k({\bf r})=2\pi/L$ and $V({\bf r})$ are the wavenumber and the phase velocity
expressed in terms of the Riemann invariants.
Since Eq. (\ref{wcl}) must be consistent with the diagonal system (\ref{11-4}),
one readily obtains the ``potential'' representation
\begin{equation}\label{13-6}
    v_i=\left(1-\frac{L}{\prt_iL}\prt_i\right)V,\quad \prt_i\equiv\frac{\prt}{\prt r_i}\, .
\end{equation}
The function $V({\bf r})$ is given by (\ref{13-1}) and the dependence $L({\bf r})$
by (\ref{14-1}).
As a result, Eqs.~(\ref{13-6}) yield the Whitham characteristic velocities
\begin{equation}\label{14-5}
    \begin{split}
    &v_1=2(r_1+r_2+r_3)+\frac{4(r_2-r_1)\K(m)}{\E(m)-\K(m)},\\
    &v_2=2(r_1+r_2+r_3)-\frac{4(r_2-r_1)(1-m)\K(m)}{\E(m)-(1-m)\K(m)},\\
    &v_3=2(r_1+r_2+r_3)+\frac{4(r_3-r_2)\K(m)}{\E(m)}.
    \end{split}
\end{equation}
One can see that the characteristic velocities (\ref{14-5}) coincide with those for the KdV
modulation system (see e.g. \cite{ZMNP,kamch2000}). This, however, does not imply that
the dynamics of the  modulated periodic waves in the Gardner equation will necessarily
be the same or even qualitatively similar to the KdV case. Indeed, to obtain the
modulated solution for the Gardner equation, one needs first to convert the solution for $r_i(x,t)$ into the dynamics
of the original modulation parameters $u_i(x,t)$ and then substitute $u_j$'s into the relevant
periodic solution from Section II (which could be drastically different from the KdV cnoidal wave).
Moreover, since the mapping $\{r_k \} \mapsto \{ u_i\}$ is not one-to-one,
the same modulation solution $\{ r_{k}(x,t)\}$
can give rise to two completely different modulations $\{ u_i(x,t)\}$ of the periodic
solutions to the Gardner equation.
This becomes evident already on the level of the dispersionless limit of the Gardner equation,
\begin{equation}\label{dG}
u_t+6u(1 -\alpha u)u_x=0 \, ,
\end{equation}
which is related to the dispersionless limit of the KdV equation (the Hopf equation)
\begin{equation}\label{dK}
w_t+6ww_x=0
\end{equation}
via the quadratic mapping $u \mapsto w$ specified by the function $w=u(1-\alpha u)$. Indeed, a constant solution $w=a$
of equation(\ref{dK}) gives rise to two different constant solutions
of (\ref{dG}) found as roots $u^{\pm}$ of the quadratic equation $w(u)=a$ (obviously $u^++u^-=1/\alpha)$.
These two constant solutions can also be combined into a weak solution in the form of a propagating step:
$u=u_-$ for $x<6at$ and $u=u_+$ for $x>6at$ (we note that this step solution transforms into a smooth kink
(or solibore)  solution (\ref{solibore}) of the full Gardner equation  if dispersion is taken into account).

It is clear that a
one-to-one correspondence between the solutions of (\ref{dG}) and (\ref{dK}) is only possible in one of the
restricted domains of the function $w(u)$:  $u<1/2\alpha$ or $u>1/2\alpha$, where
$w(u)$ is monotone. Thus, one can expect significantly different, compared to the KdV case, dynamics if
the range of the initial function $u(x,0)$ would include an open interval containing the turning point
$u=1/2\alpha$ of the characteristic velocity of the dispersionless Gardner equation (\ref{dG}).

The above non-uniqueness in the correspondence between the modulation solutions
of the KdV and Gardner equations
is due to the invariance of the Gardner equation with respect to the transformation (\ref{inv}).
As a result, the Whitham-Gardner system  in ``natural'' modulation variables, unlike the KdV-Whitham system, is neither genuinely
nonlinear nor strictly hyperbolic (see \cite{kod2008} for the detailed analysis  of a similar issue in the context
of the closely related complex mKdV system). This results in the occurrence of  much richer modulation dynamics
for the Gardner equation than those for the KdV equation.
A very similar issue was also recently discussed in \cite{ep-11} where the dam-break and lock-exchange flows were
studied for the  Miyatta-Choi-Camassa (MMC) system \cite{cc99} describing fully nonlinear long dispersive interfacial
waves in a two-layer fluid. This is not surprising, of course, as the Gardner equation with $\alpha>0$ represents
a weakly nonlinear approximation of the MMC system obtained under an additional requirement that the layers depth
ratio is close to the critical value defined by the square root of the ratio of the respective fluid densities.

In conclusion we note that the Whitham modulation system associated with the Gardner equation with $\alpha<0$
can be elliptic (this is possible since the spectral eigenvalues $\la_i$ can be pure imaginary so the  the squared
Riemann invariants $r_i$ are negative and the mapping  $\{ r_i\} \mapsto \{u_i \}$ is generally complex).
It is indeed the case for
the related focusing mKdV equation $u_t+6u^2u_x+u_{xxx}=0$ (see e.g. \cite{dron76}) so modulational instability
is generally to be expected in this case.
However, it was shown in \cite{ercolani} that for the focusing mKdV equation with real initial data, the modulation
equations arising in the zero-dispersion
limit of the initial-value problem, are hyperbolic which guarantees modulational stability for such problems.
This property was recently used in \cite{march2008} to construct undular bore solutions to the focusing mKdV equation.
Since the focusing mKdV equation and the Gardner equation with $\alpha <0$
are related by the simple transformation (\ref{transmkdv1}), we shall be assuming hyperbolicity of the Gardner modulations
for the resolution of a step problem considered in the next section.

\section{Classification of the solutions for the step problem}

We now consider the Gardner equation (\ref{eq1}) with the initial conditions in the form of a step,
\begin{equation}\label{5-3}
    u(x,0)=\left\{
    \begin{array}{cc}
    {u}^-, & \quad x<0,\\
    {u}^+, & \quad x>0.
    \end{array}
    \right.
\end{equation}
It is clear that, due to the form of the nonlinear term in the Gardner equation, the structure of the solutions
to the initial value problem (\ref{eq1}),
 (\ref{5-3}) will strongly depend on the positions of the initial step parameters $u^+$, $u^-$
relative to the  turning point $u=1/2\alpha$ of the dispersionless characteristic velocity $6u(1-\alpha u)$.

\subsection{Key ingredients}

We first describe several particular solutions of the Gardner equation playing the role of  ``building blocks''
in the full solutions to the dispersive Riemann problem (\ref{eq3}), (\ref{5-3}) for different values of $u^{\pm}$.
These solutions are: cnoidal undular bores, rarefaction waves, solibores (for $\alpha >0$) and trigonometric undular
bores (for $\alpha <0$).

\medskip
a) {\it Cnoidal undular bores: Gurevich-Pitaevskii solution}.

The local structure of the simple undular bore  is described by one of the periodic solutions obtained in Section II:
solution (\ref{el1}) for $\alpha>0$ and solution
(\ref{eq11}) for $\alpha<0$. The corresponding modulations are expressed in terms of the parameters $r_1,r_2,r_3$
satisfying the Whitham equations
(\ref{11-4}), (\ref{14-5}). It is clear that, in the problem of dispersive resolution of an initial discontinuity  we are interested
in the similarity modulation solutions
where the modulation variables depend on $s=x/t$ alone (both initial data and the modulation equations are invariant
with respect to the scaling transformation $x \to Cx$, $t \to C t$).

The classical Gurevich-Pitaevskii similarity solution of the modulation system (\ref{11-4}), (\ref{14-5}) has the form (see \cite{GP1,ZMNP,kamch2000})
\begin{equation}\label{GPconst}
r_1=r^+, \qquad r_3=r^- \, ,
\end{equation}
where $r^+,r^-$ are some constants, while the dependence $r_2(x,t)$ is given implicitly by
\begin{equation}\label{14-6}
    v_2(r^+,r_2,r^-)=\frac{x}{t} \, .
\end{equation}
It is essential that, since $r_3>r_1$, one must have
\begin{equation}\label{condr}
r^->r^+\, .
\end{equation}
A typical modulation solution is presented in Fig.~6.

\begin{figure}[h]
\begin{center}
\includegraphics[width=8cm, clip]{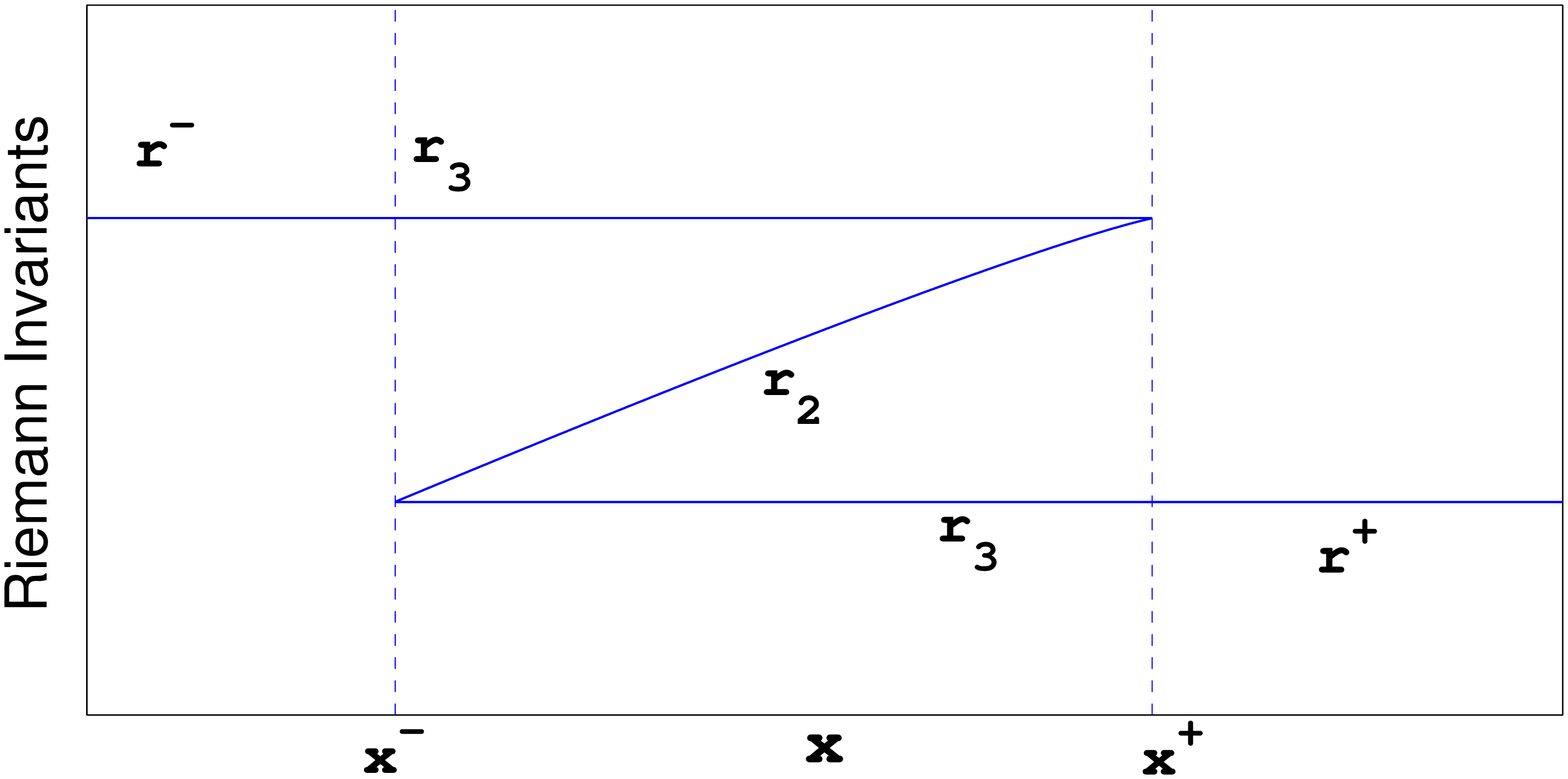} \
\caption{(Color online) Typical behaviour of the Riemann invariants $r_j$ in the modulation solution for a simple cnoidal undular bore.}
\end{center}
\label{fig3}
\end{figure}

The undular bore described by the Gurevich-Pitaevskii modulation solution (\ref{GPconst}), (\ref{14-6}) occupies an expanding
region $x^-<x<x^+$ whose edges $x^{\pm}=s^{\pm}t$ propagate with constant velocities $s^{\pm}$. The trailing (harmonic) edge
is defined by the condition $m=0$ (i.e. $r_2=r_1=r^+$ see (\ref{14-1}) and the leading (soliton) edge---by the condition $m=1$
(i.e. $r_2=r_3=r^-$).  Then the velocities $s^{\pm}$ are found from (\ref{14-6}), (\ref{GPconst}), (\ref{14-5}) as
\begin{equation}\label{gp-1}
   s^-= \left.v_2\right|_{r_2=r_1}=12r_1-6r_3 = 12r^+-6r^-\, ,
\end{equation}
\begin{equation}\label{gp-2}
 s^+=   \left.v_2\right|_{r_2=r_3}=2r_1+4r_3 = 2r^+ + 4r^-.
\end{equation}
 The above solution (\ref{GPconst}), (\ref{14-6}) coincides with the modulations in the  undular bore arising as a result
 of the resolution of an initial discontinuity: $r(x>0,0)=r^+$ and $r(x<0,0)=r^-$ for the KdV equation $r_t+6rr_x+r_{xxx}=0$
 \cite{GP1}. However, in the KdV context  the modulation solution (\ref{GPconst}), (\ref{14-6}) uniquely characterises
 the asymptotic solution for KdV undular bore due to the one-to-one correspondence between the Riemann invariants
 $\{r_j\}$ and the physical parameters $\{u_j \}$ of the travelling wave solution \cite{whitham1}.
In the case of the Gardner equation (\ref{eq1}) there are two possibilities for each signs of $\alpha$
due to different possible relationships between the $\{r_k\}$
and $\{u_i\}$ for different types of travelling wave solutions described in Section IIIB---see relationships
(\ref{13-2}), (\ref{13-3}) for $\alpha>0$ and  (\ref{13-2a}), (\ref{13-3a}) for $\alpha <0$.
The actual choice depends on the positions of the initial step parameters $u^+$ and $u^-$ relative to the turning point
$u= {1}/(2\alpha)$ of the function $w(u)=u(1-\alpha u)$ (see Section III C). We shall consider all possible cases
in the next two sections.

\bigskip
b) {\it Rarefaction waves.}

The rarefaction waves  are asymptotically described by the similarity solution
\begin{equation}\label{a4}
    \frac{x}t=6u(1-\al u)
\end{equation}
 of the dispersionless limit (\ref{dG}) of the Gardner equation.
The two possible roots of (\ref{a4}) describe two types of rarefaction waves: ``normal'' and ``reverse''. We first consider the case $\alpha>0$.

The normal simple rarefaction wave connects two constant states $u=u^l$ (left) and $u=u^r$ (right) satisfying the condition
$u_l<u_r<1/2\alpha$ and has a structure similar to the  rarefaction solution of the ``rightward-propagating'' KdV equation, in which $\partial u/\partial x >0$. It is described by the equation
\begin{eqnarray}\label{normrare}
u &=& u^l \quad \hbox{for} \quad x<s^l t \,, \nonumber \\
u &=& \frac1{2\al}\left(1-\sqrt{1-\frac{ 2\alpha x} {3t}} \right)  \quad \hbox{for} \quad  s^l t < x <  s^r t \,, \nonumber \\
u &=& u^r,  \quad \hbox{for}  \quad   x > s^r  t    \,. \label{rref}
\end{eqnarray}
We shall be using a symbolic diagram $\{u^l$ {\bf RW}  $\rightarrow u^r\}$ for the normal rarefaction wave.

The reversed simple rarefaction wave connects two constant states $u=u^l$ and $u=u^r$ satisfying the condition
$u_r<u_l<1/2\alpha$ and is described by the equation
\begin{eqnarray}\label{revrare}
u &=& u^l \quad \hbox{for} \quad x<s_l t \,, \nonumber \\
u &=& \frac1{2\al}\left(1+\sqrt{1-\frac{ 2\alpha x} {3t}}\right)  \quad \hbox{for} \quad  s^l t < x <  s^r t \,, \nonumber \\
u &=& u^r,  \quad \hbox{for}  \quad   x > s^r  t    \,. \label{lref}
\end{eqnarray}
The reversed rarefaction wave  is similar to the rarefaction wave in the ``leftwards-propagating'' KdV equation, in which  $\partial u/\partial x <0$. The symbolic diagram for this wave is $\{u_l  \leftarrow$ {\bf RW} $  u^r\}$.

The speeds $s^{l,r}$ of the left and right boundaries for both normal and reversed rarefaction wave are given by
\begin{equation}\label{slr}
s^{l,r}=6u^{l,r}(1-\alpha u^{l,r}), \quad s^l<s^r \, .
\end{equation}
Solutions (\ref{lref}) and (\ref{rref}) have weak discontinuities at the corners $x=s^{l,r}t$. These are smoothed out by
small-amplitude oscillatory wavetrains which are generated if the dispersive term of the Gardner equation is taken
into account (cf. \cite{ZMNP} for the KdV case).

For $\alpha< 0$, the rarefaction waves are described by the same solutions, however, formula (\ref{rref}) would then
describe the reversed wave with $u^r<u^l<-1/(2 |\alpha|)$ and (\ref{lref})---the normal one with $u^r>u^l>-1/(2|\alpha|)$.

\bigskip
c) {\it Solibores ($\alpha>0$)}.

The solibore (kink) solutions for the Gardner equation with $\alpha >0$ are given by formulae (\ref{solibore}).
Solibores provide the smooth transition between two constant states $u_l$ and $u_r$  satisfying the condition
$u_l+u_r=1/\alpha$. Using the terminology introduced above for rarefaction waves, we shall refer to the solibore
as  ``normal'' when we have $\partial u/\partial x<0$ (``-'' sign in solution (\ref{solibore})) and ``reverse'',
when $\partial u/\partial x >0$ (``+'' sign in (\ref{solibore})). The corresponding diagrams are:
$\{u_l \ {\bf SB}  \rightarrow u_r\}$ for the normal solibore and $\{u_l \leftarrow {\bf SB} \  u_r\}$ for the the reversed one.

\bigskip

d) {\it Trigonometric undular bores} ($\alpha <0$).

This type of undular bores, not encountered in the KdV theory, was first reported in \cite{march2008} where the
evolution of a step problem was studied for the focusing mKdV equation  (see also a similar solution for the complex
modified mKdV equation in \cite{kod2008}). The trigonometric undular bores of the Gardner equation with $\alpha <0$
are described by the modulated finite-amplitude nonlinear periodic solutions (\ref{eq13c}) or (\ref{eq16d}) so that
$m=0$ throughout the wavetrain. At one of the edges of the trigonometric bore the amplitude vanishes and at the
opposite edge it assumes some finite value.  Generically, as will be explained later trigonometric undular bores are realised  as parts of
composite solutions (either a combination of cnoidal and trigonometric bores or a combination of a trigonometric
bore and a rarefaction wave). As with other wave patterns arising for the Gardner equation, one can have two types
of trigonometric bores: normal, \{$u^l  \ \hbox{\bf TB} \rightarrow \ u^r$\}  and reversed,
\{$u^l \ \leftarrow \hbox{\bf TB} \ u^r$\}. The normal trigonometric bore is locally described by solution
(\ref{eq16d}) while for the reversed one solution  (\ref{eq13c}) should be used.

\begin{figure}[ht]
\begin{center}
\includegraphics[width=8cm]{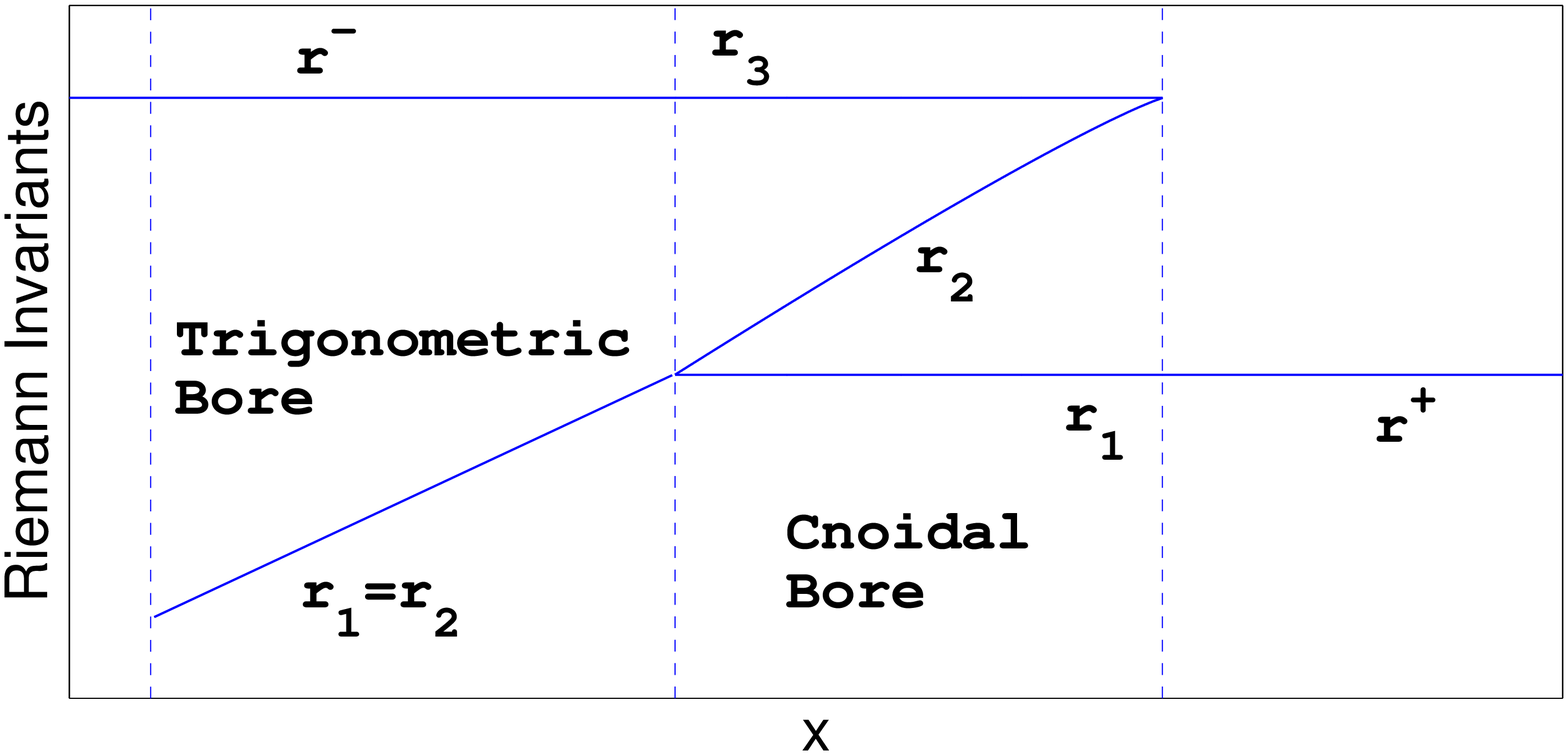}
\caption{(Color online) The sketch of the ``square'' Riemann invariants $r_1,\,r_2,\,r_3$ behaviour in the composite
cnoidal-trigonometric undular bore}
\label{fig4}
\end{center}
\end{figure}

A typical configuration of the Riemann invariants $r_j$ in the composite cnoidal-trigonometric bore is shown in Fig.~4.
Since in the region of the trigonometric bore one has $r_2=r_1=r$, the corresponding similarity modulation solution
$v_2(r,r,r^-)=x/t$ is degenerate in the sense that it does not allow one to reconstruct uniquely the modulations
for $u_j(x,t)$ in the trigonometric bore and to provide the necessary matching between $u^l$ and $u^r$ across the bore.
Therefore, the ``square'' Riemann invariants $r_j$ (\ref{ru}) used so far, are not suitable for the description of
trigonometric bores.  Instead, motivated by the results in \cite{march2008} we introduce the classical Whitham combinations
\begin{equation}\label{eq5a}
    \begin{split}
    &R_1=\tfrac12(u_2+u_3)=\frac1{2\al}\left(1+\sqrt{1-4\al r_1}\right),\\
    &R_2=\tfrac12(u_2+u_4)=\frac1{2\al}\left(1-\sqrt{1-4\al r_2}\right),\\
    &R_3=\tfrac12(u_3+u_4)=\frac1{2\al}\left(1-\sqrt{1-4\al r_3}\right),\\
    \end{split}
\end{equation}
($R_3 \ge R_2 \ge R_1$)  for the normal trigonometric undular bore (see the relationships (\ref{13-3a})) and
\begin{equation}\label{eq5b}
    \begin{split}
    &R_3=\tfrac12(u_1+u_2)=\frac1{2\al}\left(1+\sqrt{1+4\al r_3}\right),\\
    &R_2=\tfrac12(u_1+u_3)=\frac1{2\al}\left(1+\sqrt{1+4\al r_2}\right),\\
    &R_1=\tfrac12(u_2+u_3)=\frac1{2\al}\left(1-\sqrt{1-4\al r_1}\right),\\
    \end{split}
\end{equation}
($R_1 \ge R_2 \ge R_3$) for the reversed trigonometric undular bore (see the relationships (\ref{13-3})).

Obviously, the quantities $R_j(r_j)$ are the Riemann invariants of the modulation system (any function of the Riemann
invariant alone is also a Riemann invariant).

\medskip
(i) \underline{Normal trigonometric bores}.

We now construct the modulation solution for the normal trigonometric bore, where the oscillations occur between the roots
$u_3$ and $u_4$ of the travelling wave solution polynomial $Q(u)$ (\ref{eq3}).
The amplitude in such a bore is
\begin{equation}\label{ampt}
a=u_4 - u_3 = 2(R_1 - R_2)
\end{equation}
and it gradually increases from $a=0$ at the left (harmonic) edge, say $x=x^l$, to some nonzero value $a=a^r$ at the right edge $x=x^r$.

When $a=0$ we have from (\ref{eq16d}) $u=u_3=u_4$. Hence, since the trigonometric bore must match with $u=u^l$ at $x=x^l$
we obtain from (\ref{eq5a})
\begin{equation}\label{r3l}
R_3=u_4=u^l \qquad \hbox{at} \quad x=x^l,
\end{equation}
and, therefore, $R_3=u^l$ everywhere within the trigonometric bore. Hence, from (\ref{eq5a}) we obtain
\begin{equation}\label{r3}
r_3=u^l(1-\al u^l) \, .
\end{equation}
Now, within the trigonometric undular bore we have $r_1=r_2=r(x/t)$, which is determined by the degenerate similarity
solution of the Whitham system (\ref{11-4}),
\begin{equation}\label{eq5s}
    \left.v_2\right|_{m=0}=12r-6r_3=\frac{x}t \ .
\end{equation}
Substituting the value of $r_3$ (\ref{r3}) we obtain
\begin{equation}\label{eq5c}
    r=\frac1{12}\left[\frac{x}t+6u^l(1-\al u^l)\right]
\end{equation}
and so from (\ref{eq5a}) we get
\begin{equation}\label{eq5d}
    \begin{split}
    &R_1=\frac1{2\al}\left\{1+\sqrt{2\al^2\left(u^l-\frac1{2\al}\right)^2+\frac12\left(1-\frac{2\al x}{3t}\right)}\right\},\\
    &R_2=\frac1{2\al}\left\{1-\sqrt{2\al^2\left(u^l-\frac1{2\al}\right)^2+\frac12\left(1-\frac{2\al x}{3t}\right)}\right\}.
    \end{split}
\end{equation}
One can see that  $R_1+R_2=1/\alpha$ for all $x$ in the trigonometric bore.

If a trigonometric undular bore is fully realised (i.e. is not part of the composite cnoidal-trigonometric bore) then
at the leading edge $x=x^r$ it must assume the limiting waveform of a {\it bright algebraic soliton} (\ref{eq13f})
(otherwise the matching with constant or smooth external solution would not be possible). This implies that we have
$u_3=u_2=u_1$ and so $R_2=R_3=u^l$ at the leading edge $x=x^r$. The algebraic soliton rides on the background $u_1$
so the relevant matching condition becomes
\begin{equation}\label{}
u_1=u^r \qquad \hbox{at} \quad x=x^r \, ,
\end{equation}
which, by $u_1=u_2=u_3$ and the first formula in (\ref{eq5a}) implies $R_1=u^r$ at $x=x^r$. Therefore, the trigonometric bore can only connect the states $u^l$ and $u^r$ satisfying the condition
\begin{equation}\label{cond00}
u^l+u^r= \frac{1}{\alpha},
\end{equation}
i.e. a single isolated trigonometric bore can be realised as a result of the step evolution only in the special cases when
the parameters of the initial step satisfy the condition (\ref{cond00})
(note that for normal trigonometric bore one must have $u^l>u^r$ which follows from the ordering $R_3>R_1$). Thus,
for the Gardner equation with $\alpha<0$ the trigonometric undular bores play the role similar to that played by
solibores in the step problem for the the Gardner equation with $\alpha>0$.

The speed $s^l$ of the trailing edge of the trigonometric bore  is found from the condition that at the trailing edge $R_2=R_1=\frac{1}{2\alpha}$ which implies by (\ref{eq5d})
\begin{equation}\label{trt}
s^l=\frac{3}{\alpha}-6\alpha u^l(1-\alpha u^l)\, .
\end{equation}
At the leading edge $x=s^r t$ of the trigonometric undular bore we have $R_2=R_3=u^l$. Then, substituting $R_2=u^l$,
$x/t=s^r$ into the second equation (\ref{eq5d}) we obtain the speed of the leading edge
\begin{equation}\label{srt}
s^r=6u^l(1-\alpha u^l) = 6u^r(1-\alpha u^r)\, ,
\end{equation}
which coincides with the characteristic speed of the dispersionless Gardner equation at $u=u^r$. This implies,
in particular, that the trigonometric undular bore can be joined at the leading edge to a simple rarefaction wave solution.

The amplitude of the algebraic soliton (\ref{eq13f}) at the leading edge follows from (\ref{ampt}) where we set $R_2=R_3$:
\begin{equation}\label{ampalg}
a^r=2(R_3 - R_1)|_{x=x^r} = 2(u^l-u^r)= 2(2u^l-1/\alpha)\, .
\end{equation}

(ii) \underline{Reversed trigonometric bores}.

In the reversed trigonometric bore the oscillations occur between the roots $u_1$ and $u_2$ of the polynomial $Q(u)$ in (\ref{eq3}).
The amplitude is given by
\begin{equation}\label{ampt1}
a=u_2 - u_1 = 2(R_1 - R_2)
\end{equation}
In terms of the Riemann invariants $r_j$ the modulation solution for the reversed bore is given by the same formula (\ref{eq5c}),
which is then translated to $R_j$'s (\ref{eq5b}) as
(cf.(\ref{eq5d}))
\begin{equation}\label{eq500}
\begin{split}
&R_3=u^l \\
&R_2=\frac1{2\al}\left\{1+\sqrt{2\al^2\left(u^l-\frac1{2\al}\right)^2+\frac12\left(1-\frac{2\al x}{3t}\right)}\right\}, \\
 &R_1=\frac1{2\al}\left\{1-\sqrt{2\al^2\left(u^l-\frac1{2\al}\right)^2+\frac12\left(1-\frac{2\al x}{3t}\right)}\right\}.
\end{split}
\end{equation}
Similar to the normal trigonometric bore, the reversed trigonometric bore has a restriction (\ref{cond00}) for the
admissible boundary values $u^l$ and $u^b$.
At the leading edge $x=x^r$ the reversed bore assumes the limiting form of a {\it dark algebraic soliton} (\ref{eq13e}),
which has the amplitude $a^r=2(2u^l-1/\alpha)$.
The trailing and the leading edge speeds are given by the same expressions (\ref{trt}) and (\ref{srt}) respectively.

\medskip
As was already mentioned, the trigonometric undular bore (normal or reversed) can occur as part of the composite
cnoidal-trigonometric bore. In that case it is realised only partially and
does not contain an algebraic soliton at the leading edge.
The two bores match at the trailing edge of the cnoidal bore, which is defined by (\ref{gp-1}) with $r^-=u^l(1-\alpha u^l)$
and $r^+=u^r(1-\alpha u^r)$
\begin{equation}\label{s*}
s^*=12u^r(1-\al u^r)-6u^l(1-\al u^l)
\end{equation}

The amplitude at the matching point for the normal composite bore is (see (\ref{ampt}), (\ref{eq5d}))
\begin{equation}\label{}
a^*=2(R_2(s^*) - R_1(s^*)) =4\left| \frac{1}{2\alpha} - u^r \right| \, .
\end{equation}
The same result obviously holds for the reversed composite bore, for which we use (\ref{ampt1}) and (\ref{eq500}).
The trailing edge speeds for both types of composite bores is given by
(\ref{trt}).

\bigskip

\subsection{Classification for $\al>0$}
We can now proceed with the full classification of the solutions to the step problem.

\begin{figure}[h]
\centerline{\includegraphics[width=8cm, clip]{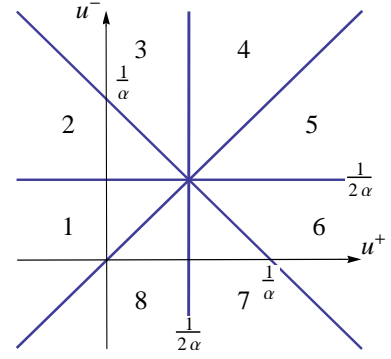}}
\caption{(Color online) Parametric map of solutions of the step problem for the Gardner equation with $\alpha>0$.
The resolution diagrams corresponding to each of the cases on the plane of the initial step parameters $u^-$ and $u^+$ are the following:
Region 1: $\{u^-  \ {\bf UB} \rightarrow u^+\}$; Region 2: $\{u^-  \leftarrow {\bf UB} \ (u^*) \ {\bf SB} \rightarrow u^+\}$;
Region 3: $\{u^- \leftarrow {\bf RW} \ (u^*) \ {\bf SB} \rightarrow u^+\}$;   Region 4: $\{u^-$ {\bf $\leftarrow$ RW } $u^+ \}$;
Region 5: $\{u^-$ {\bf $\leftarrow$ UB } $u^+ \}$; Region 6: $\{u^-$ {\bf UB} $\rightarrow \ (u^*)$ $\leftarrow$ {\bf SB} $u^+\}$;
Region 7: $\{u^-$ {\bf RW} $\rightarrow \ (u^*)$ $\leftarrow$ {\bf SB} $u^+\}$; Region 8:  $\{u^-$ {\bf RW $\rightarrow$ } $u^+\}$.
In all relevant cases the intermediate state $u^*=1/\alpha -u^+$.}
\label{chart1}
\end{figure}
We first present a detailed classification of solutions to the evolution of a step problem for the Gardner equation with $\alpha>0$.
The parametric map of solutions is constructed on the $(u^+, u^-)$ plane  of the initial step parameters (see Fig.~5).
The whole $(u^-, u^+)$-plane is split into 8 regions, each corresponding either to one of the basic patterns listed in
Section IV A (``pure'' solutions) or to the combination of two patterns (``composite'' solutions). To represent the result
of the evolution of an initial step for each region we shall be using symbolical diagrams introduced in the previous Section.
Say, the resolution diagram for Region 7, $\{u^-$ {\bf UB} $\rightarrow \ (u^*)$ $\leftarrow$ {\bf SB} $u^+\}$, denotes a normal
undular bore connecting the left state $u^-$ with an intermediate state $u^*$ which is further connected to the right state
$u^+$ via the reversed solibore.

The lines separating the regions are:
\begin{itemize}
\item $u^-=u^+$ --- separates the regions of pure undular bores and pure rarefaction waves;
\item $u^-=1/\alpha - u^+$  --- corresponds to the steps resolving into single solibores and separates the regions of
composite solutions of different types: undular bore + solibore and rarefaction wave + solibore;
\item $u^-=1/2\alpha$  --- separates regions of pure (undular bore) and composite (undular bore + solibore) solutions.
\end{itemize}

 We note that the classification for $\alpha>0$ is qualitatively similar to that presented in \cite{ep-11} for the
 Myatta-Choi-Camassa (MMC) system describing fully nonlinear interfacial dispersive waves in a two-layer fluid.
 In \cite{ep-11}, the analytic method of \cite{el-2005} was used to obtain the locations of undular bore boundaries
 and the leading solitary wave amplitude. However, the full modulation solutions are not available for the MMC system
 due to complexity of the corresponding Whitham equations.  The theory presented in this paper has an obvious advantage
 of greater simplicity and universality due to the integrable nature of the problem and availability of exact analytic solutions.
 At the same time, in the context of internal water waves, the Gardner equation with $\alpha>0$, being a weakly nonlinear
 approximation of the MCC system, is quantitatively valid only for the waves of sufficiently small amplitude.

\bigskip
The classification is most conveniently performed using the function $w(u)=u(1-\alpha u)$ defining the mapping from the
dispersionless Gardner equation (\ref{dG}) to the dispersionless limit of the KdV equation (the Hopf equation (\ref{dK})
(see Section IIIC). We shall illustrate each wave pattern in the classification by presenting the analytical (modulation theory) solutions along with respective direct numerical solutions of the Gardner equation.
In our numerics, equation (\ref{eq1}) was solved using the method of lines (see e.g. \cite{schiesser91}) where the spatial derivatives are  discretised using second order accurate finite difference approximation to reduce the governing partial differential equation to a system of ordinary differential equations. This system is then solved using the fourth order Runge-Kutta method.

\bigskip

{\bf Region 1}, $u^+<u^- \le \frac{1}{2\alpha}$, \ $\{u^-  \ {\bf UB} \rightarrow u^+\}$

\medskip
Both values $u^-$ and $u^+$ lie in the domain where the function $w(u)=u(1-\alpha u)$ is monotonically increasing so
there is one-to one correspondence between the  dispersionless limits of the Gardner and the KdV equations.
This suggests that in the Region 1 initial discontinuity can be resolved by a single normal ``shallow-water'' undular bore
of the KdV type with the bright soliton at the leading edge and the linear wavepacket at the trailing edge.

We shall use the travelling wave solution (\ref{el1}) and the relationships (\ref{13-2}) between $u_j$'s and $r_i$'s
to construct the desired modulated travelling wave solution for the undular bore and show that it indeed provides
the required matching between $u^-$ and $u^+$.
The parameters $r^{\pm}$ entering the modulation solution (\ref{14-5}), (\ref{14-6}) can be expressed in terms of
the initial step parameters $u^{\pm}$ using the relationships (\ref{13-2}).

It follows for the small-amplitude limit (\ref{eq7b}) of the travelling wave solution (\ref{el1}) that the trailing
edge of the undular bore ($m=0$) propagates against the background
$u=u_2=u_3$ (this can be inferred directly from the ordinary differential equation (\ref{eq3}), where we set $u_2 \to u_3$ (see Fig.~2b). Similarly,
for the soliton edge $m \to 1$ we have that the leading bright soliton propagates on the background $u=u_1=u_2$
(see (\ref{eq7}) and Fig.~1b).
Thus, if the step is resolved by a single undular bore we must require
\begin{equation}\label{tru}
u_2=u_3=u^-
\end{equation}
at the trailing edge and
\begin{equation}\label{leadu}
u_2=u_1=u^+
\end{equation}
at the leading edge.

Considering the same limits in the relationships Eqs.~(\ref{13-2}) we have
\begin{equation}\label{gp-3i-a}
      \left.u_2\right|_{m=0}=\left.u_3\right|_{m=0}=\frac1{2\al}\left(1-\sqrt{1-4\al r_3}\right)
\end{equation}
and
\begin{equation}\label{gp-3i-b}
 \left.u_1\right|_{m=1}=\left.u_2\right|_{m=1}=\frac1{2\al}\left(1-\sqrt{1-4\al r_1}\right)\, .
\end{equation}
According to (\ref{GPconst}) $r_3=r^-$ and $r_1=r^+$ in the Gurevich-Pitaevskii solution. Then, from
(\ref{tru})--(\ref{gp-3i-b})  we have
\begin{equation}\label{wupm}
\begin{split}
r^-=u^-(1-\alpha u^-) = w(u^-), \\
r^+=u^+(1-\alpha u^+) =w(u^+)\, .
\end{split}
\end{equation}
Since for the considered Region 1 we have $w(u^-)>w(u^+)$, the condition (\ref{condr}) of the Riemann invariant
ordering is satisfied, therefore our construction is consistent throughout.

\begin{figure}[h]
\begin{center}
\includegraphics[width=8cm,height=5cm,clip]{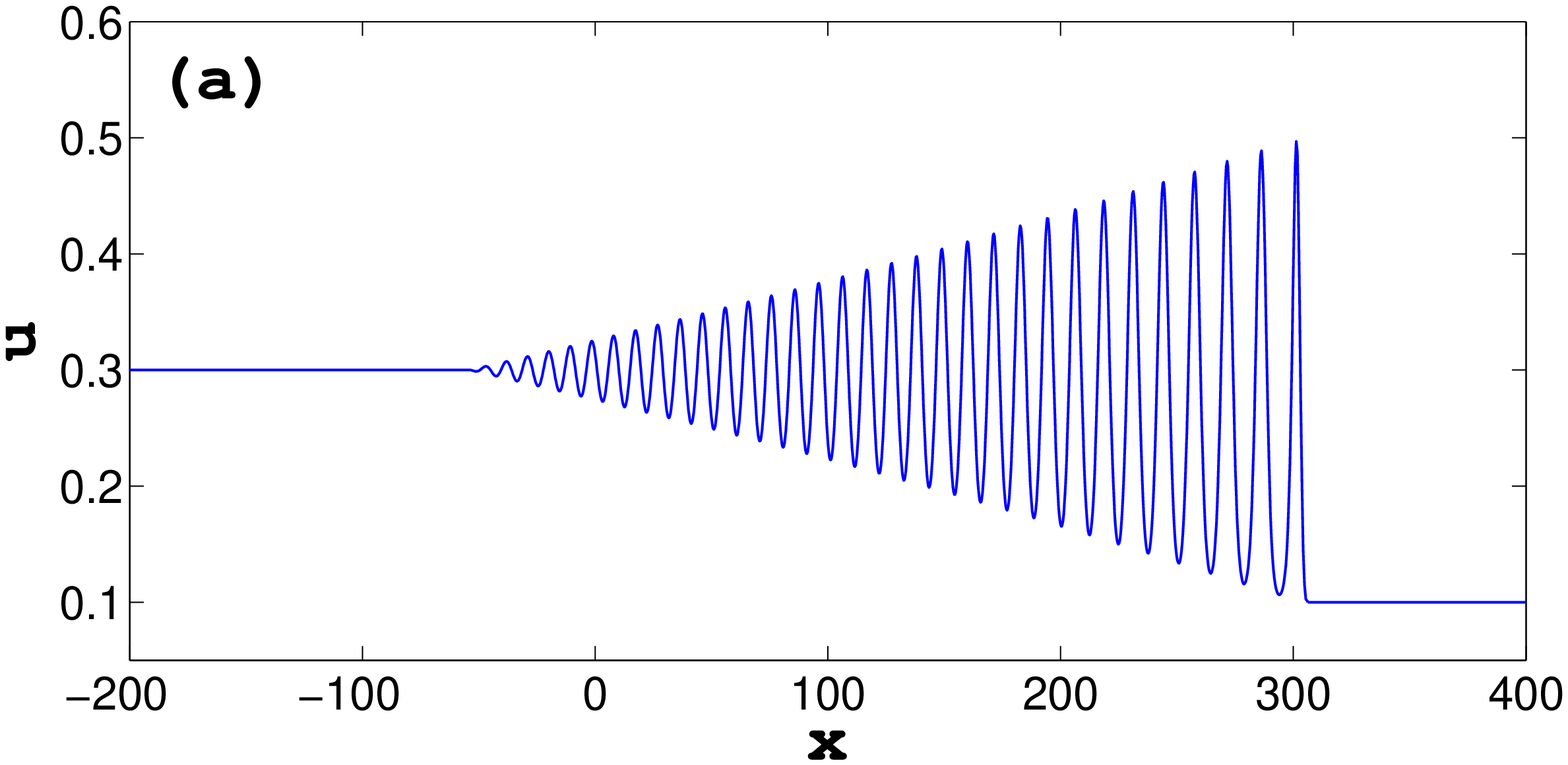}
\includegraphics[width=8cm,height=5cm,clip]{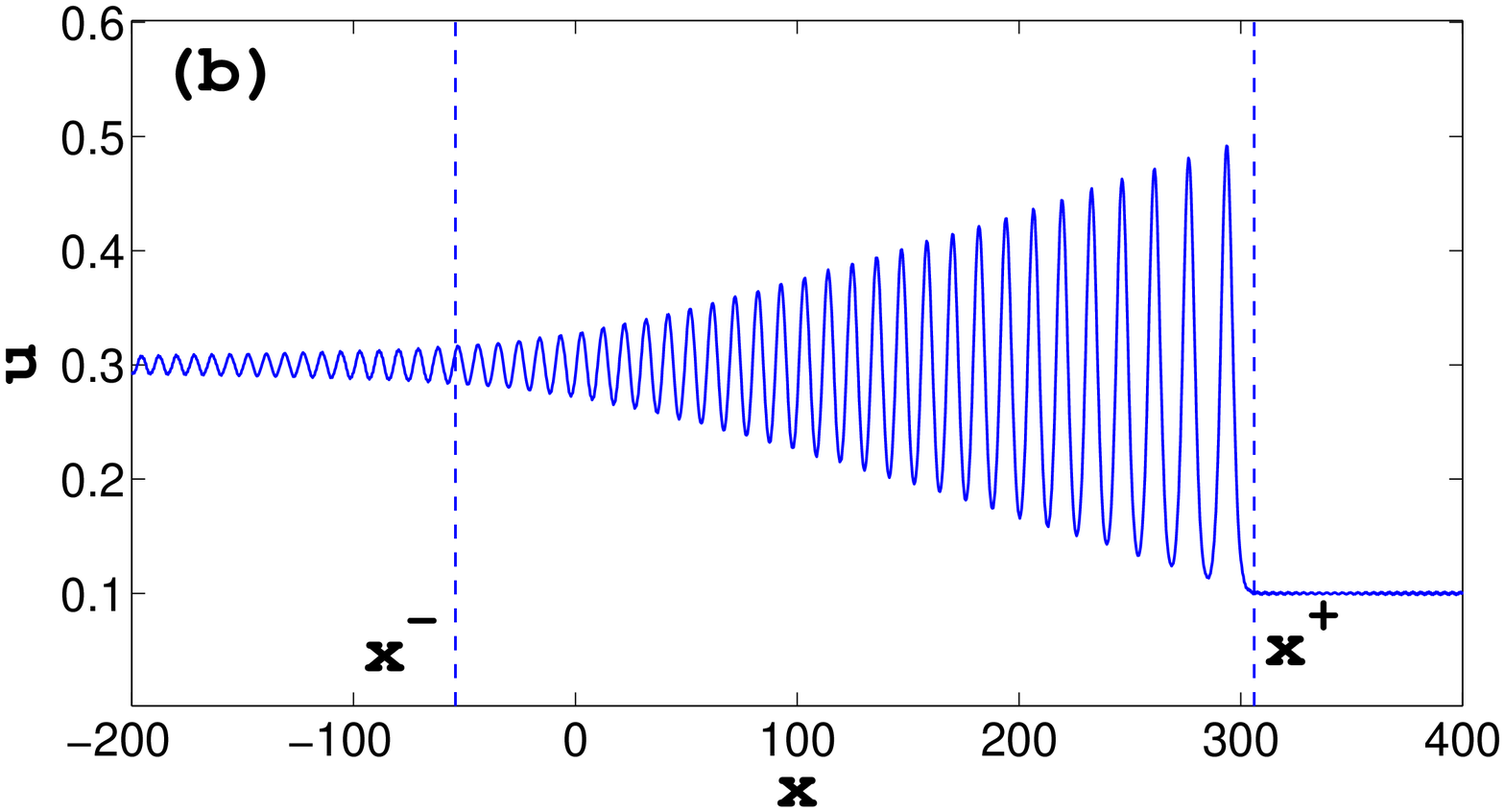}
\caption{(Color online) Evolution of an initial discontinuity for the Gardner equation with $\alpha=1$.
Region 1: $\{ u^- {\bf UB \rightarrow } u^+ \}$). The step parameters are $u^-=0.3$, $u^+=0.1$. (a) analytical solution in the form of a modulated periodic wave;
(b) numerical solution; the analytically found edges $x^{\pm}=s^{pm} t$ of the undular bore are shown by dashed lines.
Both plots correspond to $t=300$. }
\end{center}
\label{fig6}
\end{figure}

The undular bore occupies the region $s^-t<x<s^+t$ where the edge speeds $s^{\pm}$ are obtained from
(\ref{gp-1}) and (\ref{gp-2})
\begin{equation}\label{ss12}
    \begin{split}
    &s^-=12u^+(1-\al u^+)-6u^-(1-\al u^-),\\
    &s^+=2u^+(1-\al u^+)+4u^-(1-\al u^-).
    \end{split}
\end{equation}
The width of the undular bore is then
\begin{equation}\label{}
\begin{split}
\Delta&=(s^+-s^-)t=10(r^- -  r^+)t \\
&= 10 (u^--u^+)(1+ \alpha(u^-+u^+))t\, .
\end{split}
\end{equation}
One can see that the Gardner undular bore is wider than its KdV counterpart, for which one has $\Delta_{KdV}=10(u_- - u^+)t$
for the same initial conditions. As a matter of fact the KdV result is reproduced when $\alpha=0$.
The amplitude of the lead soliton in the undular bore is (see (\ref{eq70}), (\ref{13-2}) and (\ref{wupm}))
\begin{equation}\label{ls1}
a^+= \left. (u_3-u_1) \right|_{m=1}= 2(u^--u^+) \, .
\end{equation}
This result coincides with the classical KdV formula for the lead soliton amplitude.

The constructed solution is illustrated in Fig.~6 where the plot (a) of the analytical  (modulation theory) solution is presented along
with the direct numerical solution of the Gardner equation (plot (b)).
One can see that agreement is very good. The presence in the numerical plot of an extended small-amplitude oscillatory tail stretching
behind the trailing edge as defined by the modulation theory is a well-known feature of undular bore solutions
observed in the early comparisons of the KdV modulation solutions with numerics (see e.g. \cite{fw78}) and recently
studied in detail in \cite{gk07}.

\bigskip

{\bf Region 2}, $u^+<\frac{1}{2\alpha}<u^- <1/\alpha- u^+$. \  $\{u^- \leftarrow$ {\bf UB} $(u^*)$ {\bf SB} $\rightarrow u^+\}$.

\medskip
 Since $u^+$ and $u^-$ now lie in different regions of monotonicity of the function $w(u)=u(1-\alpha u)$, the Region 1 solution in the form of a single normal undular bore is not able to provide the necessary continuous matching between the given states. Instead, a reversed undular bore is generated between  $u^-$ and the intermediate state $u^*=(1/\alpha - u^+)>u^-$. This intermediate state  is found from the condition $w(u^*) = w(u^+)$ and corresponds to the required boundary value  for the Riemann invariant, $r^+=u^+(1-\alpha u^+)$ but lies in the same as $u^+$  region of monotonicity of the mapping function $w(u)$ (see Fig.~7a).  The further connection between $u^*$ and $u^+$ is provided  by a normal solibore. The corresponding wave pattern is presented in Fig.~8.
\begin{figure}[ht]
\begin{center}\label{wuplane}
\includegraphics[width=5cm,clip]{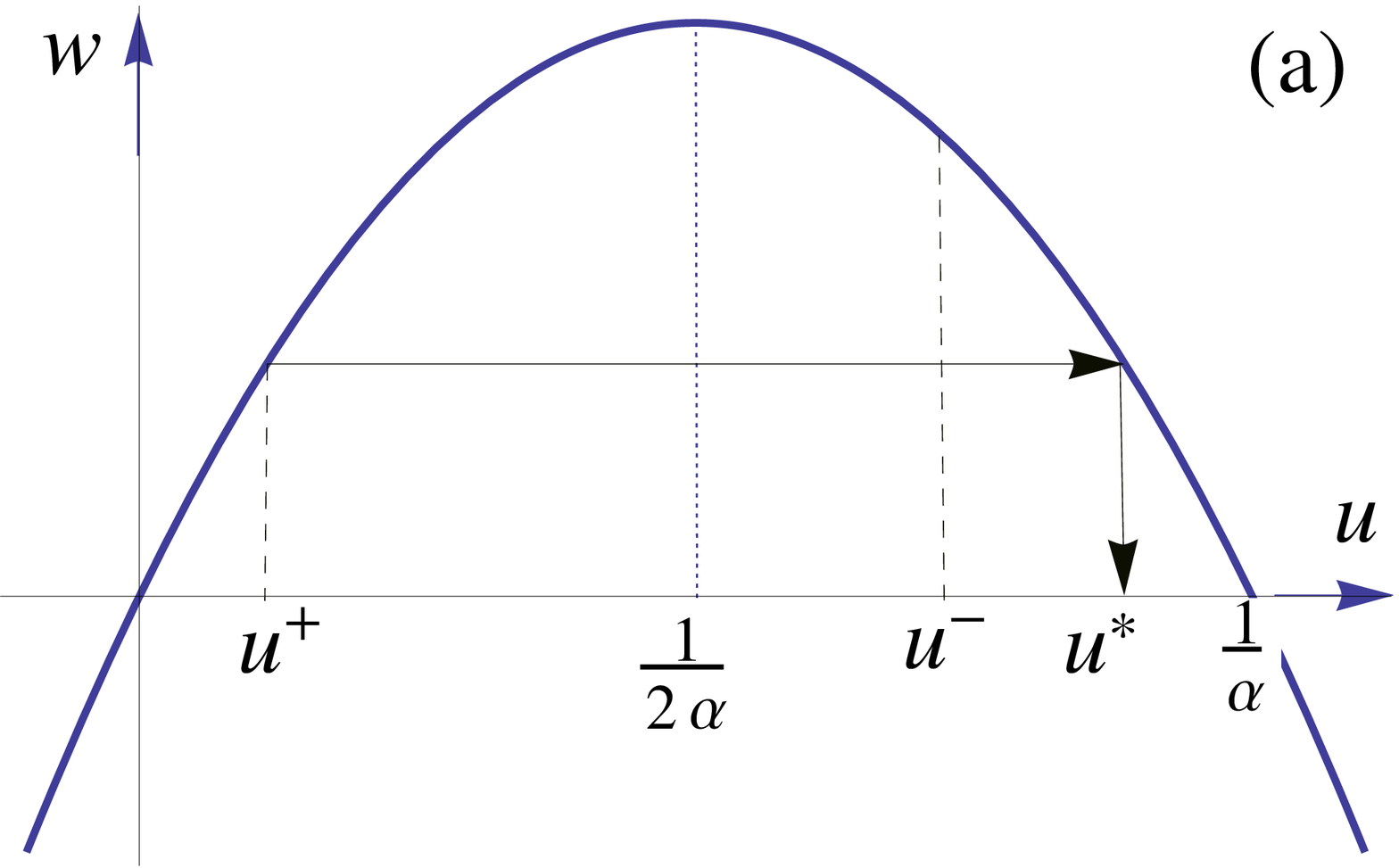}  \vspace{0.5cm}
\includegraphics[width=5cm,clip]{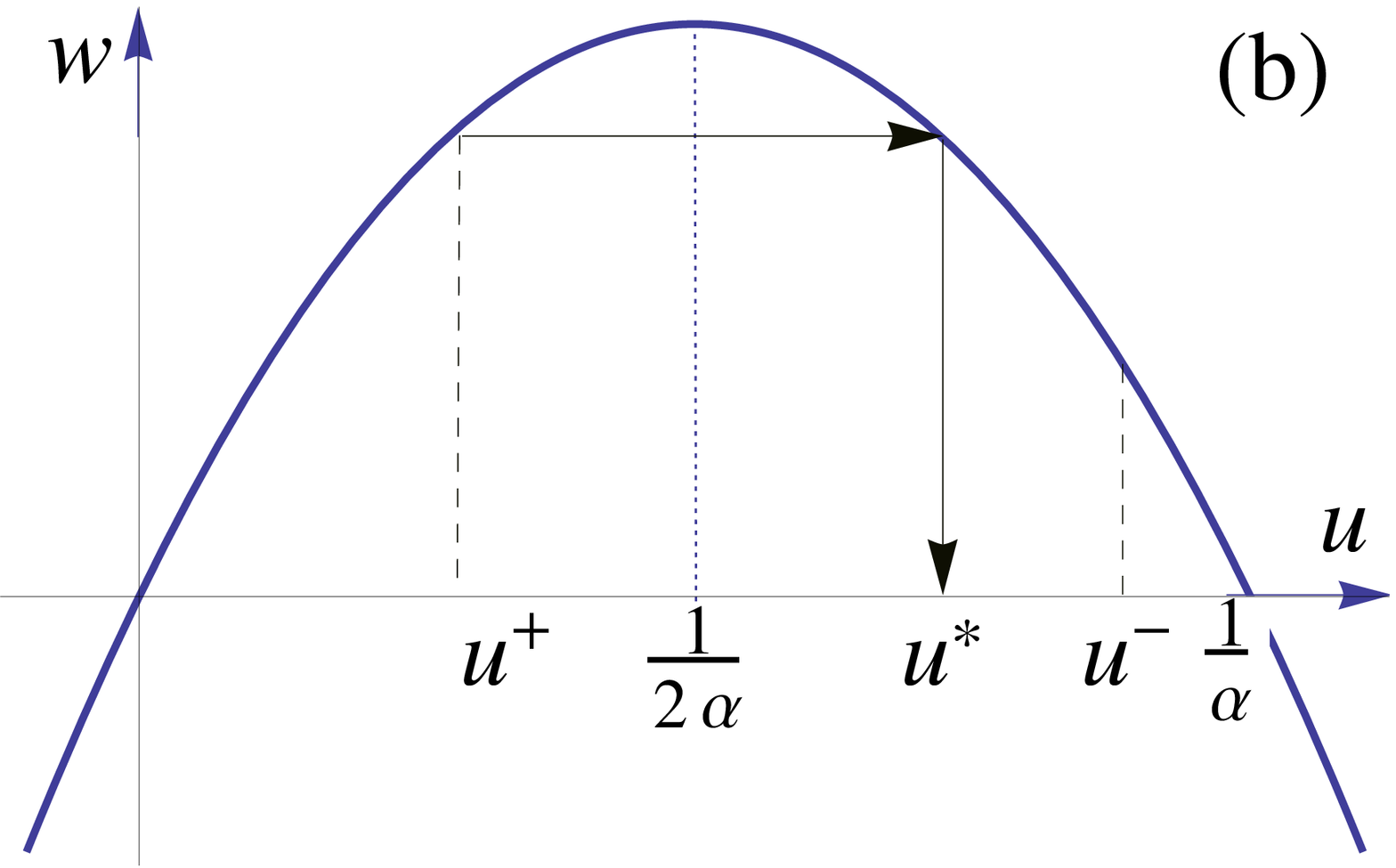}
\caption{(Color online) Finding the intermediate state $u=u^*$. (a) Region 2 diagram, $u^*>u^-$;
(b) Region 3 diagram, $u^*<u^-$. }
\end{center}
\end{figure}

\begin{figure}[h]
\begin{center}
\includegraphics[width=8cm,clip]{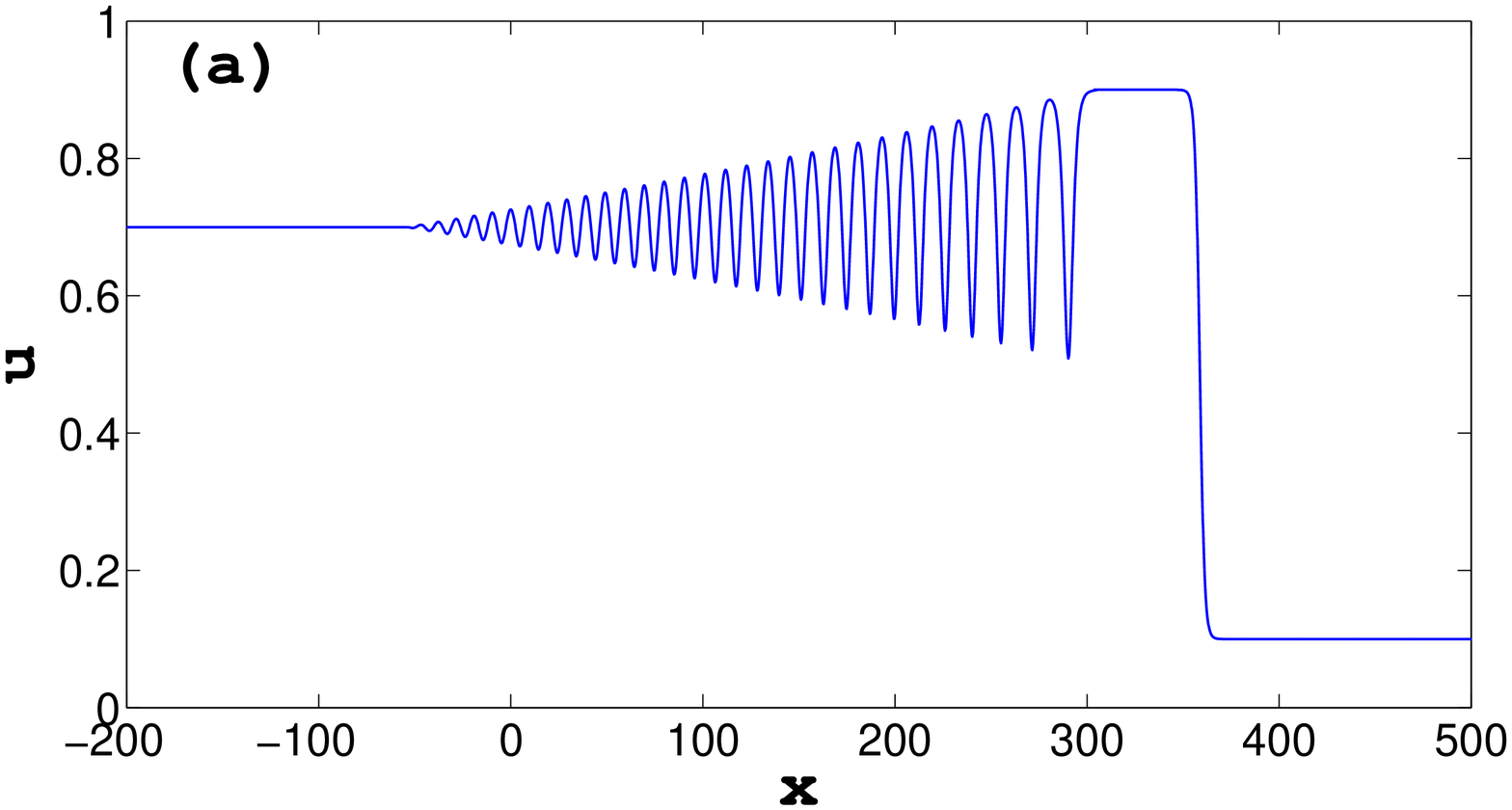}
\includegraphics[width=8cm,clip]{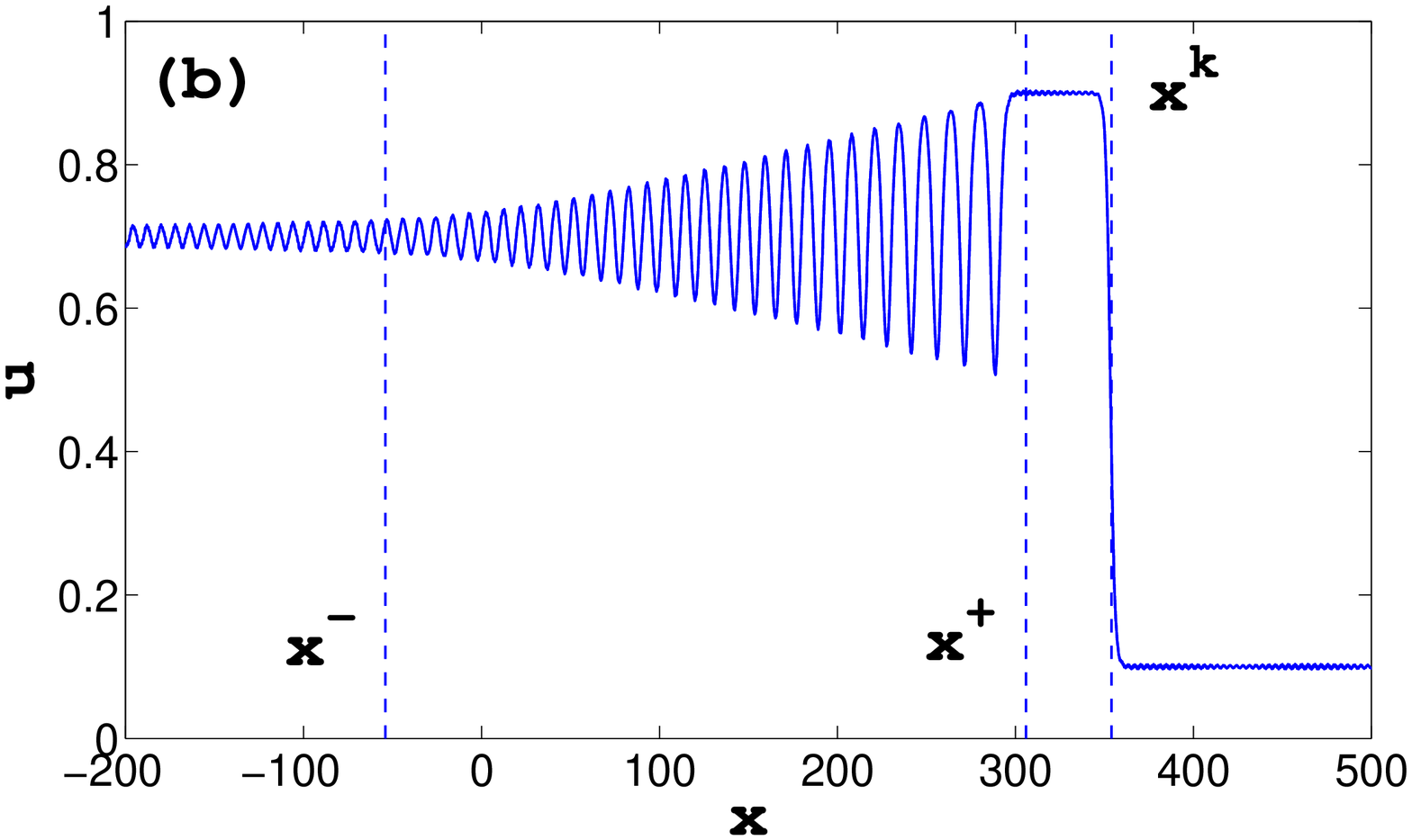}
\caption{(Color online) Evolution of an initial discontinuity for the Gardner equation with $\alpha=1$. Region 2, $\{u^-  \leftarrow {\bf UB} \ (u^*) \ {\bf SB} \rightarrow u^+\}$.
The initial step parameters are $u^-=0.7$, $u^+=0.1$.
(a) analytical solution in the form of a modulated periodic wave connected to a solibore; (b) numerical solution of the Gardner equation; the analytically found edges $x^{\pm}=s^{\pm} t$ and $x^k=s^k t$ of the undular bore and solibore respectively are shown by dashed lines. Both plots correspond to $t=300$.}
\end{center}\label{}
\end{figure}
The relationship between the Riemann invariants $\{r_i\}$ and the undular bore parameters $\{ u_i\}$ is now described by formulae (\ref{13-3}).
The trailing edge ($m=0$) of the bore propagates on a background $u=u_3=u_2=u^-$. At the leading edge  ($m=1$) we have a dark soliton (\ref{eq7a}) propagating against the background $u=u_3=u_4$.
Thus, for the undular bore we have
\begin{equation}\label{85}
 u_3=u_2=u^-
\end{equation}
and
\begin{equation}\label{86}
 u_3=u_4=u^*=\frac{1}{\alpha} - u^+ \, .
\end{equation}
Then from (\ref{13-3}) we obtain
\begin{equation}\label{87}
   \left.u_2\right|_{m=0}=\left.u_3\right|_{m=0}=\frac{1}{2\al}\left( 1+\sqrt{ 1-4\al r_3 }    \right)
\end{equation}
and
\begin{equation}\label{88}
   \left.u_3\right|_{m=0}=\left.u_4\right|_{m=0}=\frac{1}{2\al}\left( 1+\sqrt{ 1-4\al r_1 }    \right)\, .
\end{equation}
Again, since $r_1=r_+$ and $r_3=r_-$ in the modulation solution for the undular bore, we have from (\ref{85})--(\ref{88}):
\begin{equation}\label{}
\begin{split}
r_3=r^-=u^-(1-\alpha u^-) = w(u^-), \\
r_1=r^+=u^*(1-\alpha u^*) =u^+(1-\alpha u^+)\, .
\end{split}
\end{equation}
Thus, the expressions for the undular bore speeds remain the same (cf. (\ref{ss12})):
\begin{equation}\label{ss120}
\begin{split}
s^-=12u^+(1-\al u^+)-6u^-(1-\al u^-)\, , \\
s^+=2u^+(1-\al u^+)+4u^-(1-\al u^-)\ \, .
\end{split}
\end{equation}
The front solibore connecting the states $u^*$ and $u^+$ is described by formula (\ref{solibore}). The solibore
speed is given by
\begin{equation}
  s^k=\frac{1}{\alpha}+2\alpha u^+ u^*=\frac{1}{\alpha} + 2u^+(1-\alpha u^+).
  \end{equation}
Now it is not difficult to see that
\begin{equation}\label{ds}
 s^k-s^+=4\al (u^- -\frac{1}{2\al})^2 \ge 0 \, ,
 \end{equation}
that is the solibore always propagates ahead of the undular bore as expected.
\bigskip
\begin{figure}[h]
\begin{center}
\includegraphics[width=7cm,height=4cm, clip]{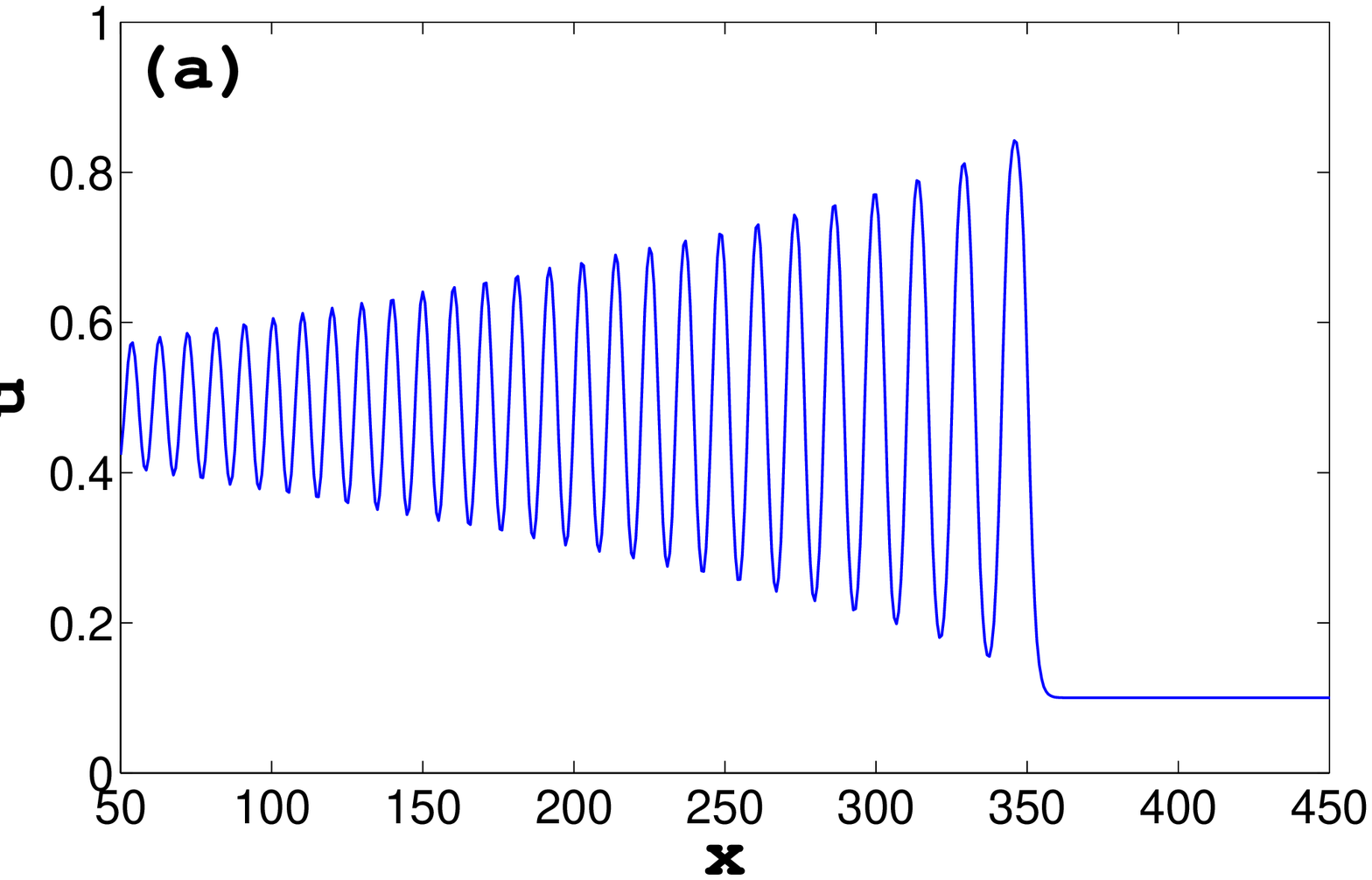}
\includegraphics[width=7cm,height=4cm, clip]{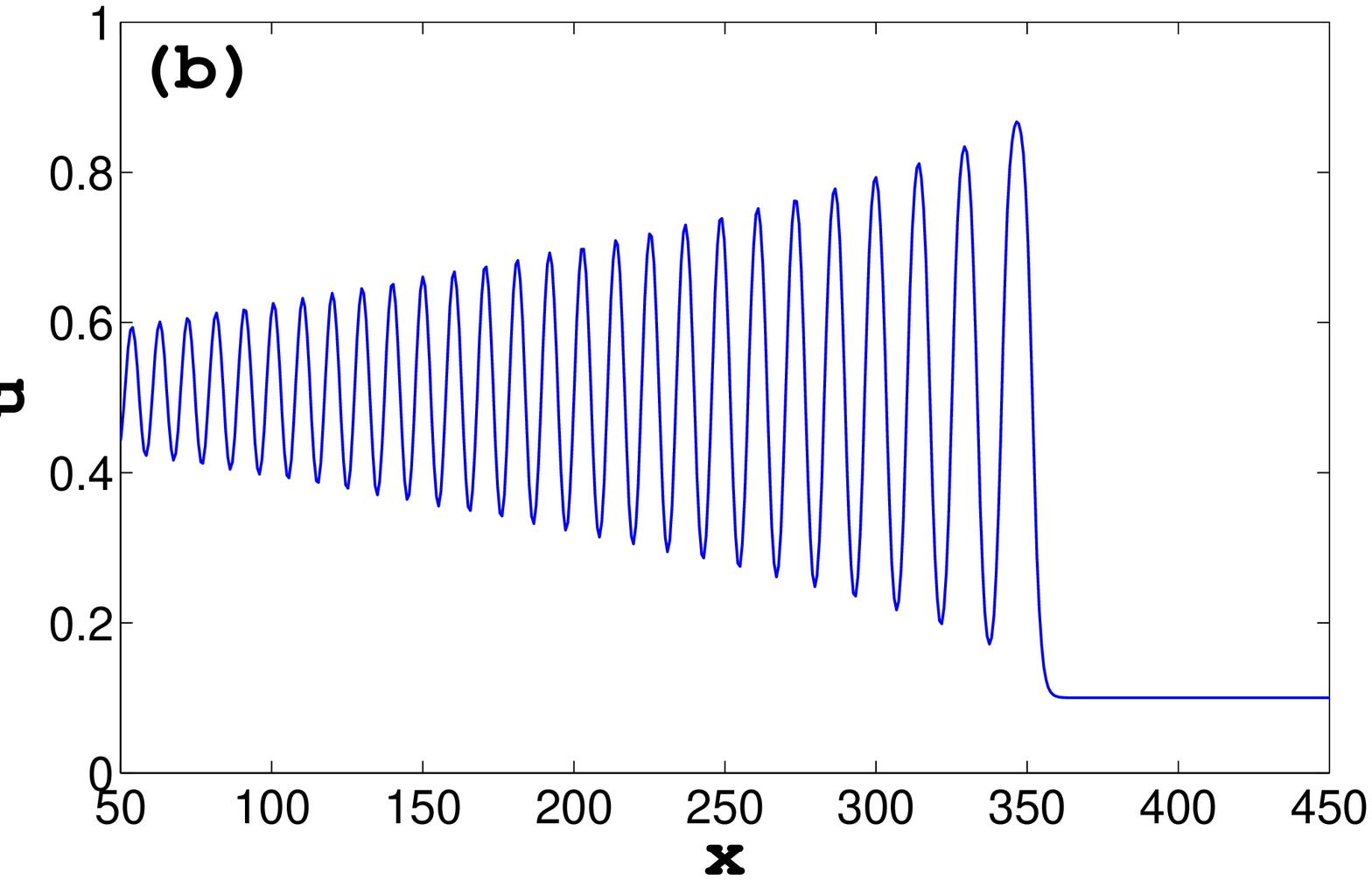}
\includegraphics[width=7cm,height=4cm, clip]{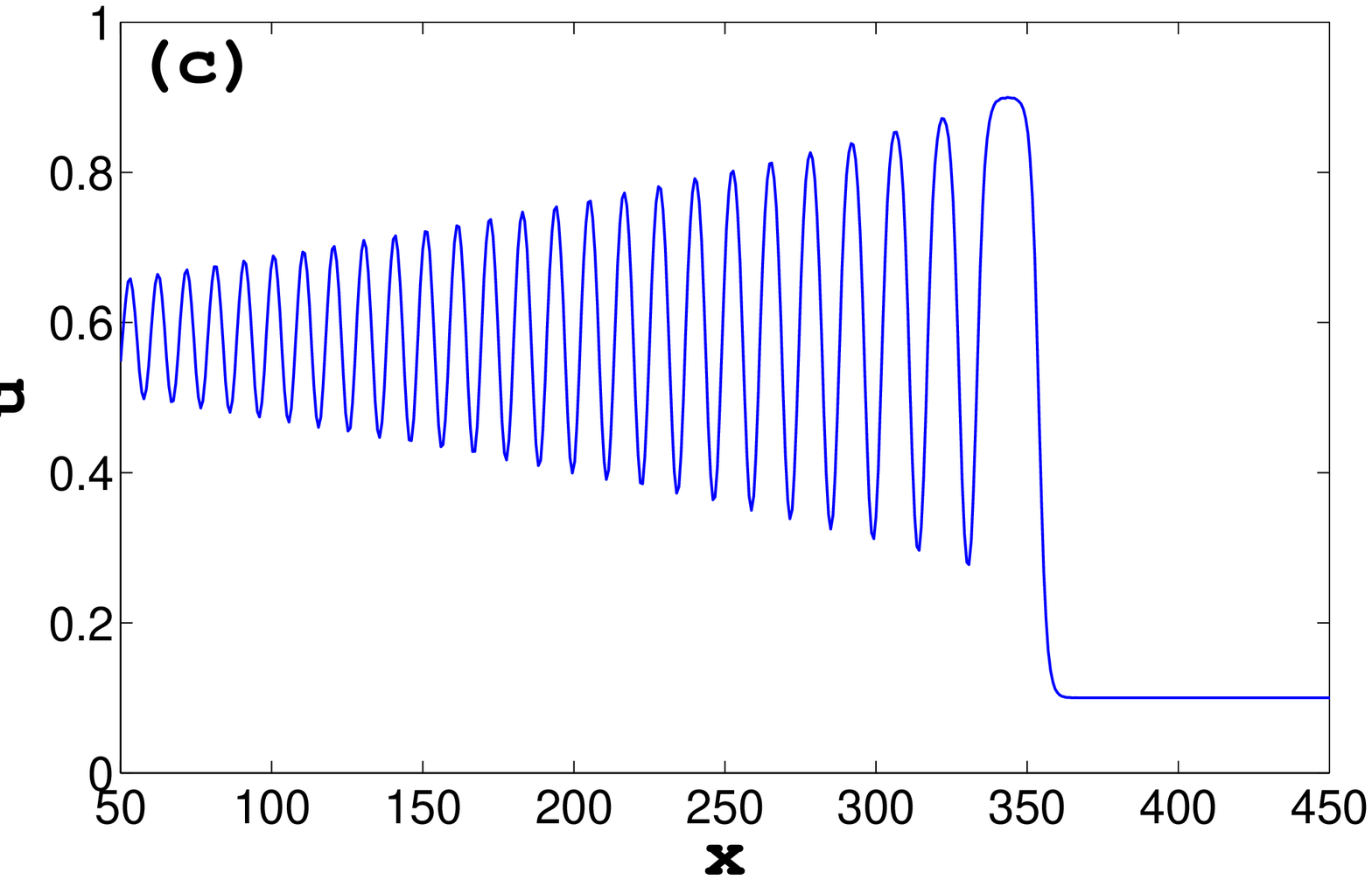}
\includegraphics[width=7cm,height=4cm, clip]{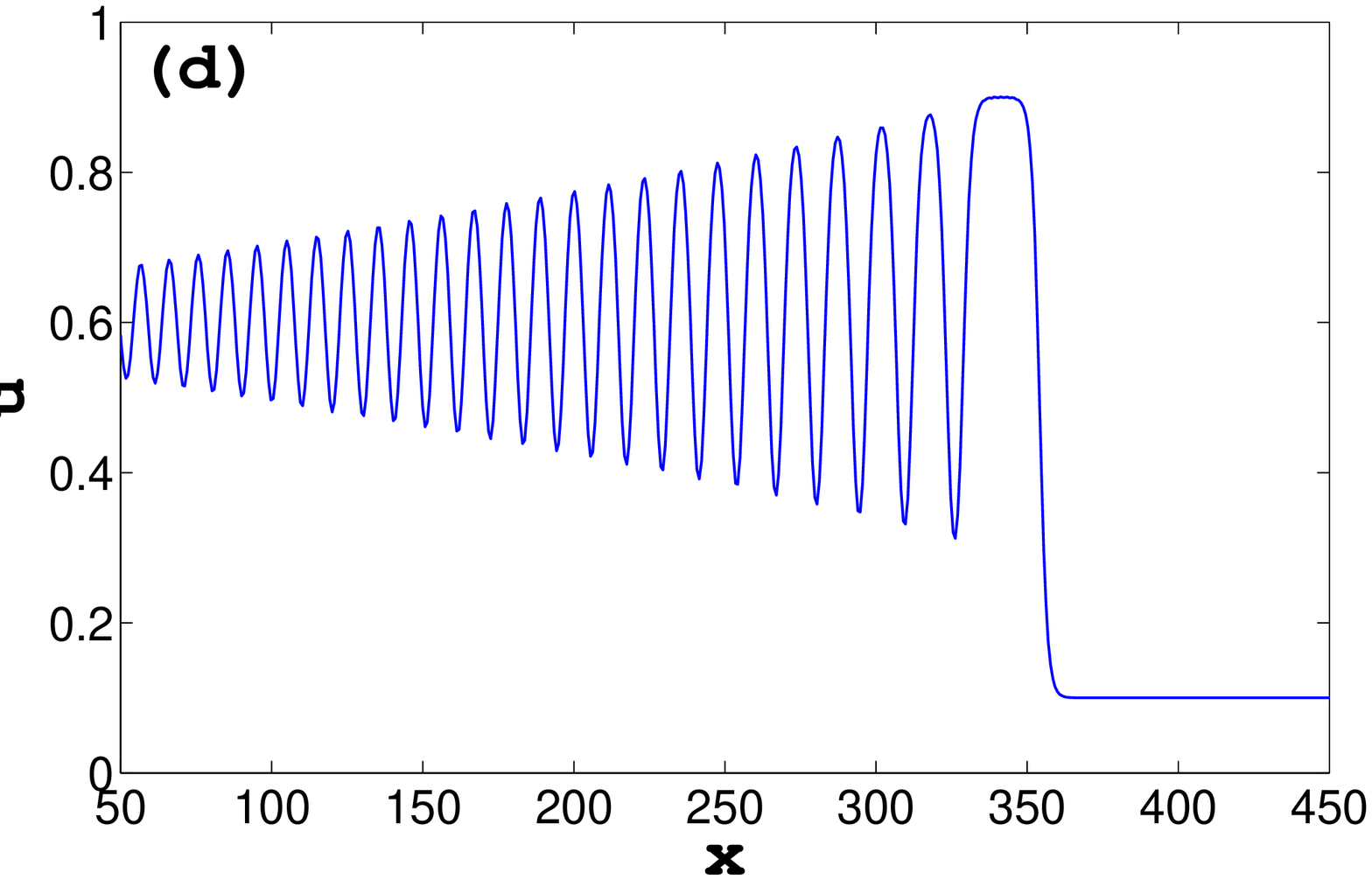}
\caption{(Color online) The transformation of the undular bore structure from the normal, ``bright soliton", pattern  in Region 1
(plot a) to the reversed, ``dark soliton'', pattern in Region 2 (plot d). Numerical simulations of the Gardner equation
with $\alpha =1$. The downstream state,  $u^+=0.1$, is the same for all cases, the upstream state $u^-$ is taken in
the range $u^-=0.49<1/2\alpha$ (Region 1) to $u^-=0.58>1/2\alpha$ (Region 2), t=300}
\end{center}\label{fig9}
\end{figure}
Since there is a qualitative change of the wave pattern in the transition from Region 1 to Region 2
it is necessary to look closer at what happens near the boundary between these two regions determined by the value of
the left state $u^-=1/{2\alpha}$. The change of the pattern is illustrated in Fig.~9 where several numerical solutions
of the Gardner equation with $\alpha = 1$ are presented
for the evolution of initial discontinuities with the same right state,  $u^+=0.1$, while the left state $u_-$ was taken
in the range $u_-=0.49<1/2\alpha$ (Region 1) to $u_-=0.58>1/2\alpha$ (Region 2).

One can see from (\ref{ds}) that for $u^-=1/{2\alpha}$ the speed of the solibore coincides with the speed of
the leading soliton in the undular bore so the ``borderline'' wave pattern in Fig.~9b can be interpreted in both ways:
as a normal, ``bright'', undular bore or, equivalently,  as a reversed, ``dark'', undular bore with an attached solibore.
When we increase $u^-$, the solibore separates from the undular bore, which in its turn acquires the reversed waveform
with a distinct dark soliton structure near the leading edge (see Fig.~9d).

\bigskip
 {\bf Region 3}, $1/\alpha - u^-<u^+<\frac{1}{2\alpha}<u^-$,  \ $u^-+u^+>1/\alpha$. $\{u^-  \leftarrow {\bf RW} \ (u^*) \ {\bf SB} \rightarrow u^+\}$

\medskip
\begin{figure}[h]
\begin{center}
\includegraphics[width=8cm,height=5cm,clip]{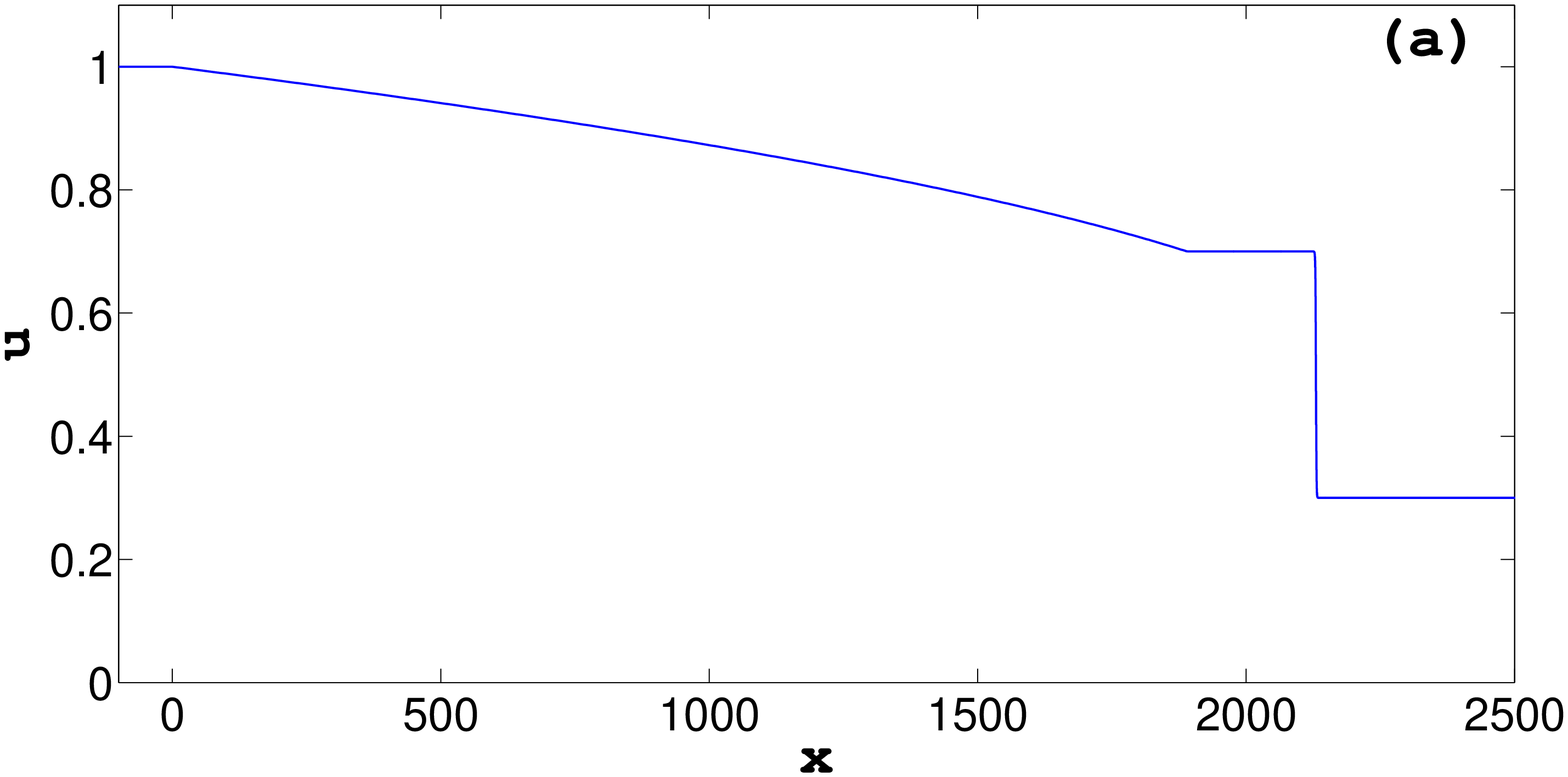}
\hspace{0.5cm}
\includegraphics[width=8cm,height=5cm,clip]{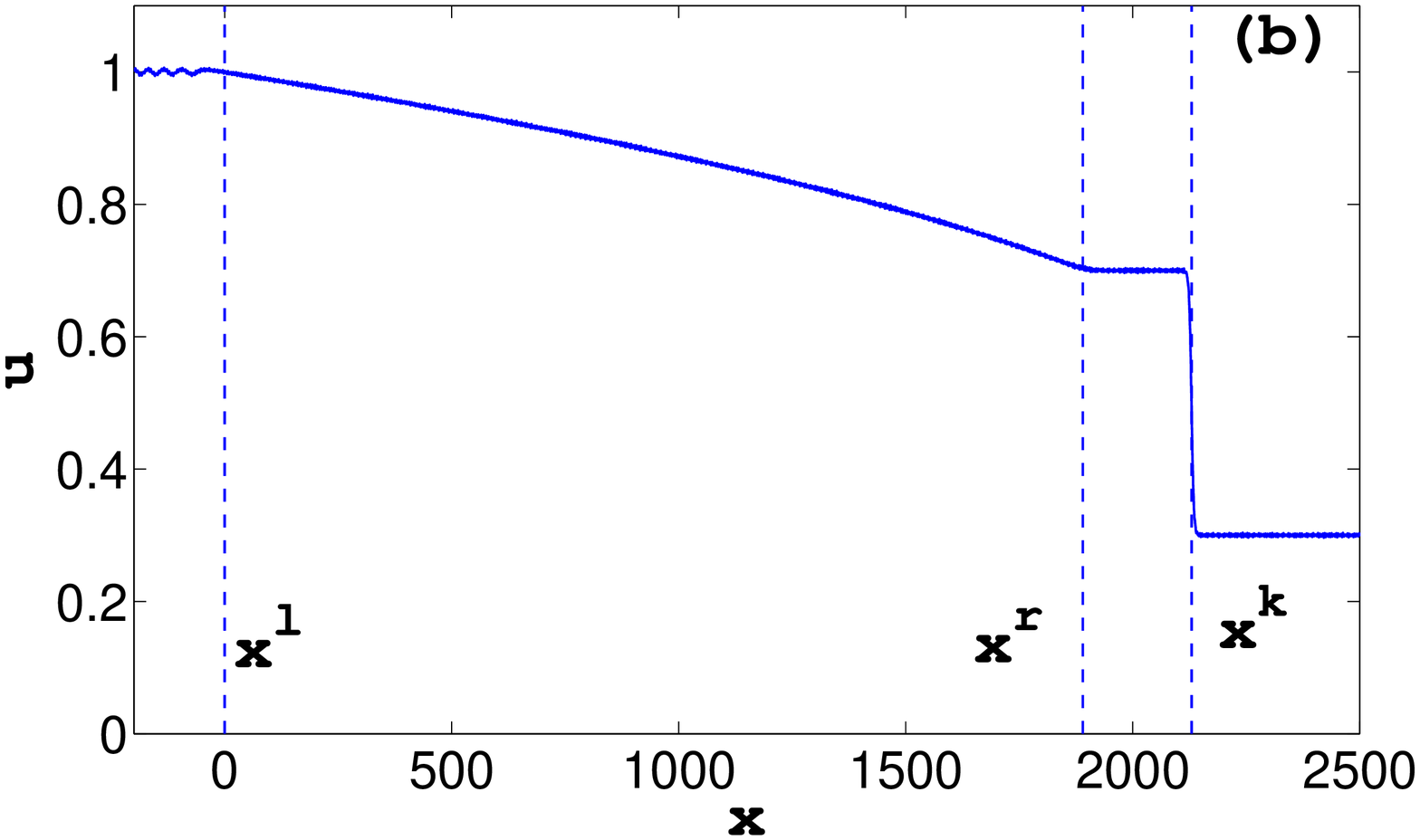}
\caption{(Color online) Evolution of an initial discontinuity for the Gardner equation with $\alpha=1$. Region 3: $\{u^- \leftarrow {\bf RW} \ (u^*) \ {\bf SB} \rightarrow u^+\}$. The initial step parameters are: $u^-=1.0$, $u^+=0.3$; (a) analytical (dispersionless limit) solution; (b) numerical solution of the Gardner equation. Dashed lines in (b) correspond to the analytically found locations of the rarefaction wave edges $x^{l,r}=s^{l,r}t$ and solibore position $x^k=s^kt$. Both plots are made for $t=1500$.
}
\end{center}
\end{figure}
This region is analogous to Region 2, since the values $u^-$ and $u^+$ again lie in different domains of monotonicity
of the function $w(u)$, thus a single-wave resolution is not possible.
However, now the intermediate state $u^*=1/\alpha -u^+<u^+$ (see Fig.~7b) so a reversed rarefaction wave is generated
instead of reversed undular bore. The solution for the rarefaction wave is given by formula (\ref{lref}), where
$u_l=u^-$ and $u^r=u^*$. The solibore solution connecting $u^*$ and $u^+$ is the same as in Region 2.
The analytical and numerical plots corresponding to Region 3 are  presented in Fig.~10.  We note that,
since $\max{u^+(1-\alpha u^+)} = \frac{1}{4\alpha}$, the speed of the solibore $s_k= \frac{1}{\alpha} + 2u^+(1-\alpha u^+)$
is always greater than  that of the right edge  of the rarefaction wave, $s^+=s_r=6u^+(1-\alpha u^+)$  (see (\ref{slr})).
At the boundary between Regions 3 and 4, when $u^+=1/2\alpha$, we have $s_k=s_r$ and the solibore gets ``attached''
to the right edge of the rarefaction wave.

\bigskip
{\bf Region 4}, $1/2\alpha \le u^+<u^-$.  $\{u^-$ {\bf $\leftarrow$ RW } $u^+ \}$.

\medskip
\begin{figure}[bt]
\begin{center}
\includegraphics[width=8cm,height=5cm,clip]{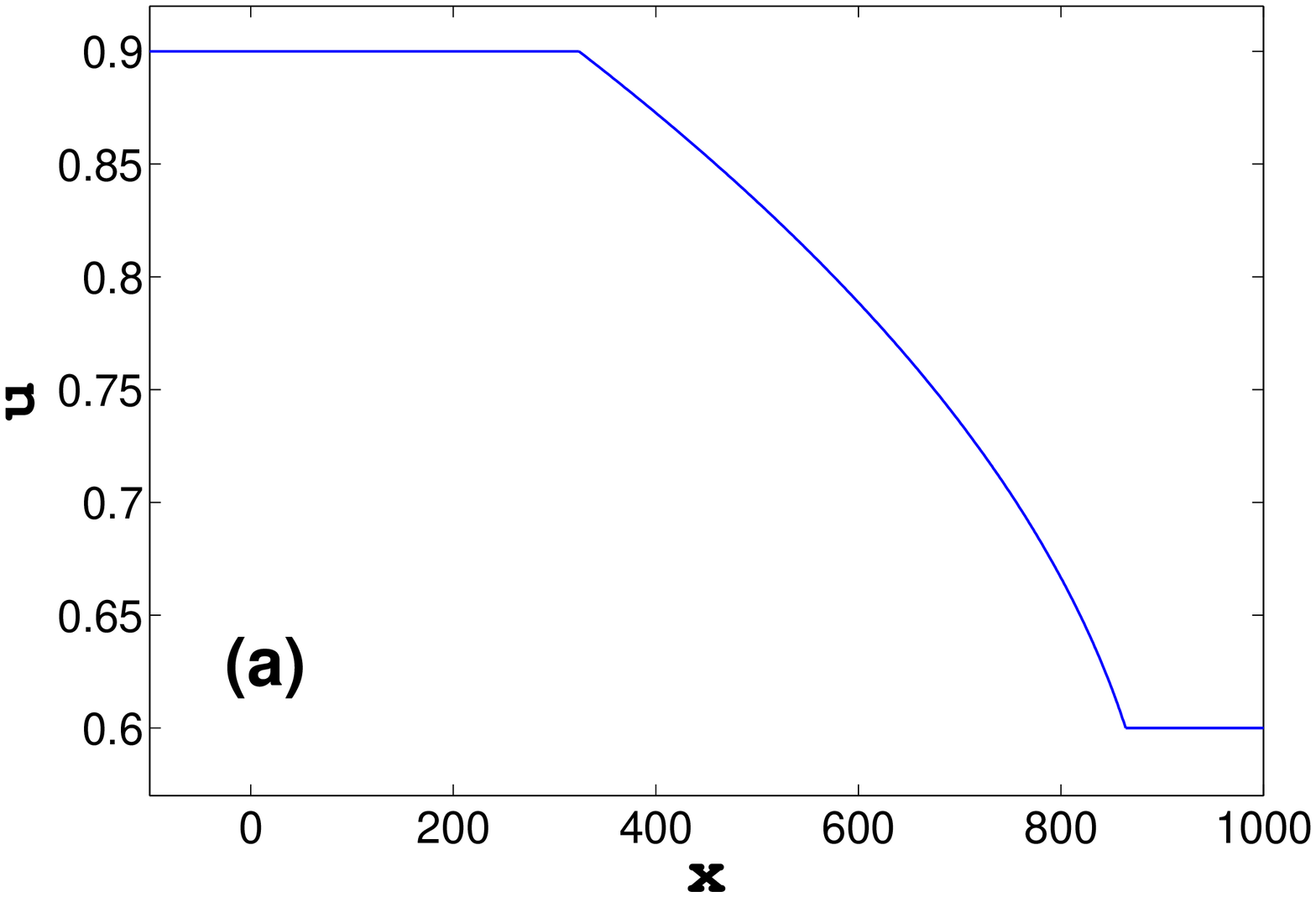}
\includegraphics[width=8cm,height=5cm,clip]{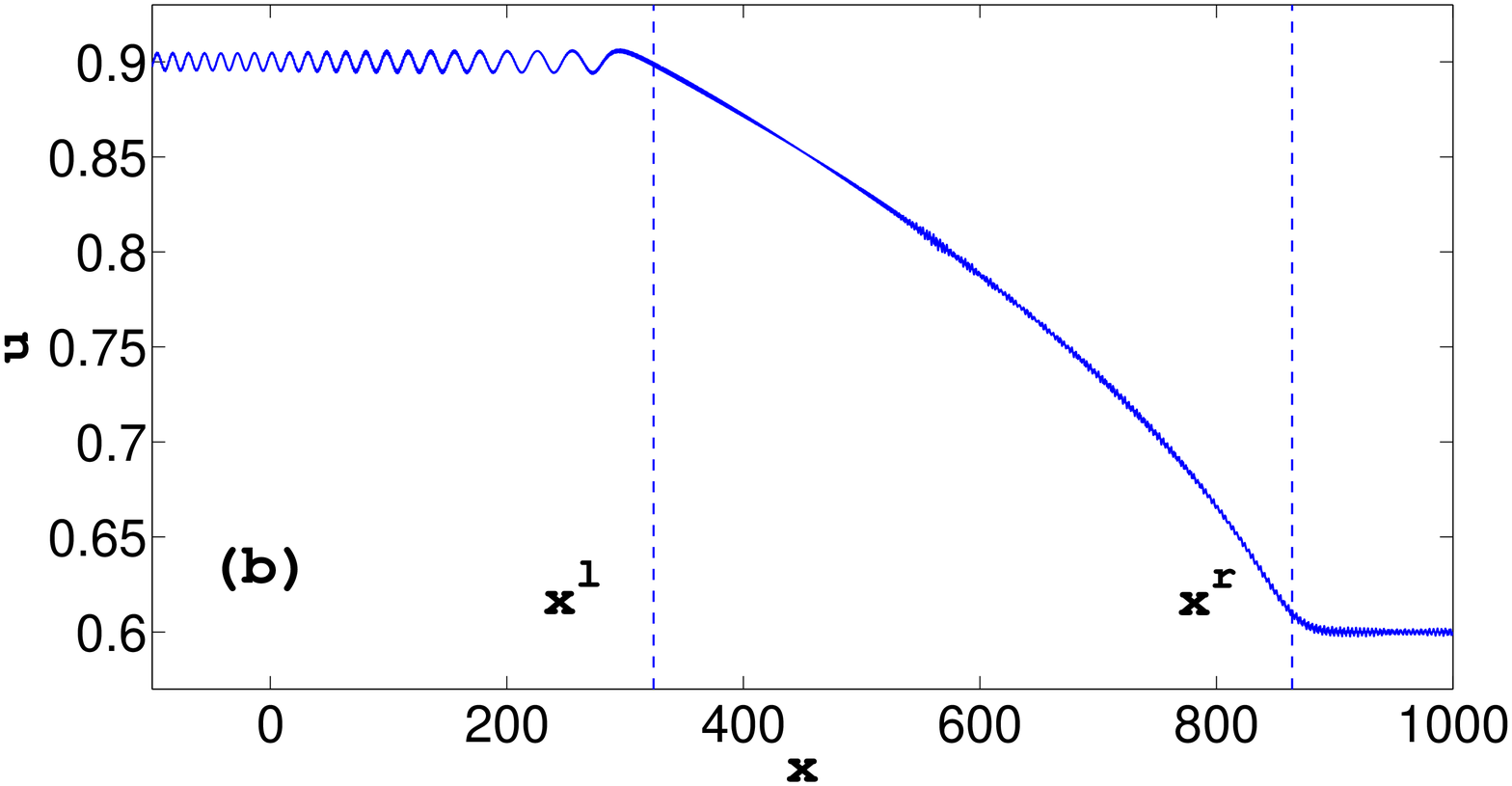}
\caption{(Color online) Evolution of an initial discontinuity for the Gardner equation with $\alpha=1$. Region 4: $\{u^-$ {\bf $\leftarrow$ RW } $u^+ \}$.   The initial step parameters are: $u^-=0.9$, $u^+=0.6$.  (a) analytical (dispersionless limit) solution; (b) numerical solution. Dashed lines in (b) correspond to the analytically found locations $x^{l,r}$ of the rarefaction wave edges.
Both plots correspond to $t=600$.
}
\end{center}
\end{figure}

A single reversed rarefaction  wave is produced. It is described by the solution (\ref{lref}) with $u_l=u^-$, $u_r=u^+$.
The corresponding typical analytical and numerical solution are presented in Fig.~11. Note the small-amplitude dispersive wave train
seen to the left of the rarefaction wave in the numerical plot in Fig.~11b. This wavetrain is necessary to resolve weak discontinuity at the left edge of the rarefaction wave,
while another weak discontinuity at the right edge is smoothed out (see \cite{GP1} for the detailed description of similar effects in the KdV theory).
\bigskip

{\bf Region 5 - 8}. \quad The wave patterns corresponding to Regions 5 -- 8 represent the ``reflections'' of the patterns arising in Regions 1 -- 4. More precisely, the counterpart solutions correspond to the ``opposite'' regions in the parametric map in Fig.~5 and are related to each other by the transformation (\ref{inv}). From this viewpoint, the solutions for Regions 5 - 8 are not ``new''. At the same time, the change of the solution polarity (e.g. from the ``bright undular bore'' to the ``dark undular bore'') due to the change of initial data is not trivial physically so it deserves separate description. For this reason and for the reader to be able to identify the arising wave patterns directly, without the need to invoke intermediate transformations, we shall proceed with the descriptions of the Regions 5 - 8 in the same format that was used for Regions 1 - 4.

\bigskip

{\bf Region 5}, \ \ $1/2\alpha < u^-<u^+ < \frac{1}{\alpha}$. $\{u^-$ {\bf $\leftarrow$ UB } $u^+ \}$

\medskip
In this region both values $u^-$ and $u^+$ are in the domain when the function $w(u)=u(1-\alpha u)$ is monotonically
decreasing so a single reversed undular bore is produced. The modulation description of this undular bore is identical
to that in Region 2 but to obtain the oscillatory structure one now needs to use the relations (\ref{13-3}) between
the Riemann invariants and parameters of the periodic solution (\ref{el1}). The amplitude of the lead dark soliton
$a^+=(u_4-u_2)|_{m=1}$ is given by by the same expression (\ref{ls1}). The analytical and numerical solution plots for Region 5
are presented in Fig.~12.

\begin{figure}[bt]
\begin{center}
\includegraphics[width=8cm,height=5cm,clip]{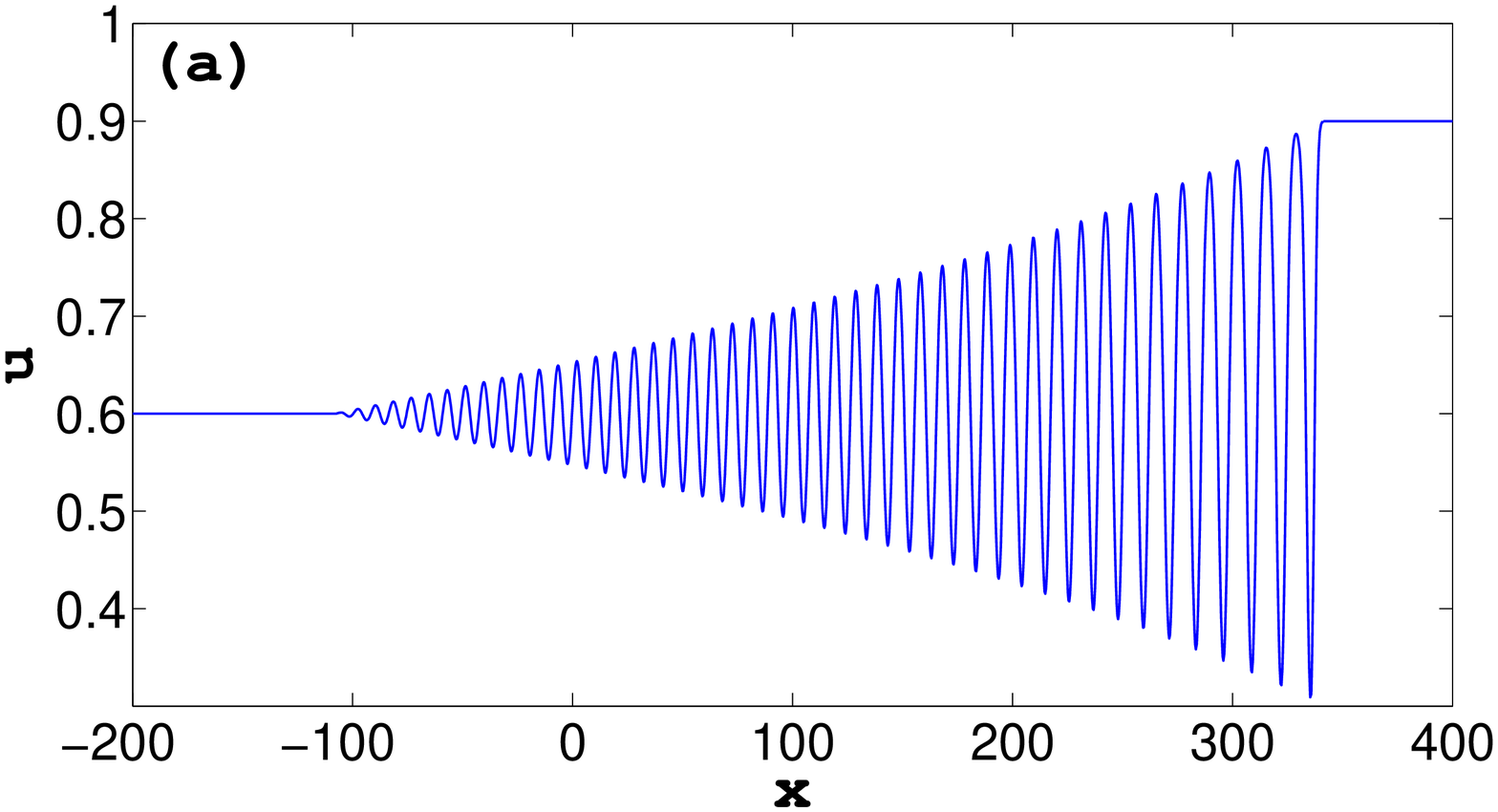}
\hspace{0.5cm}
\includegraphics[width=8cm,height=5cm,clip]{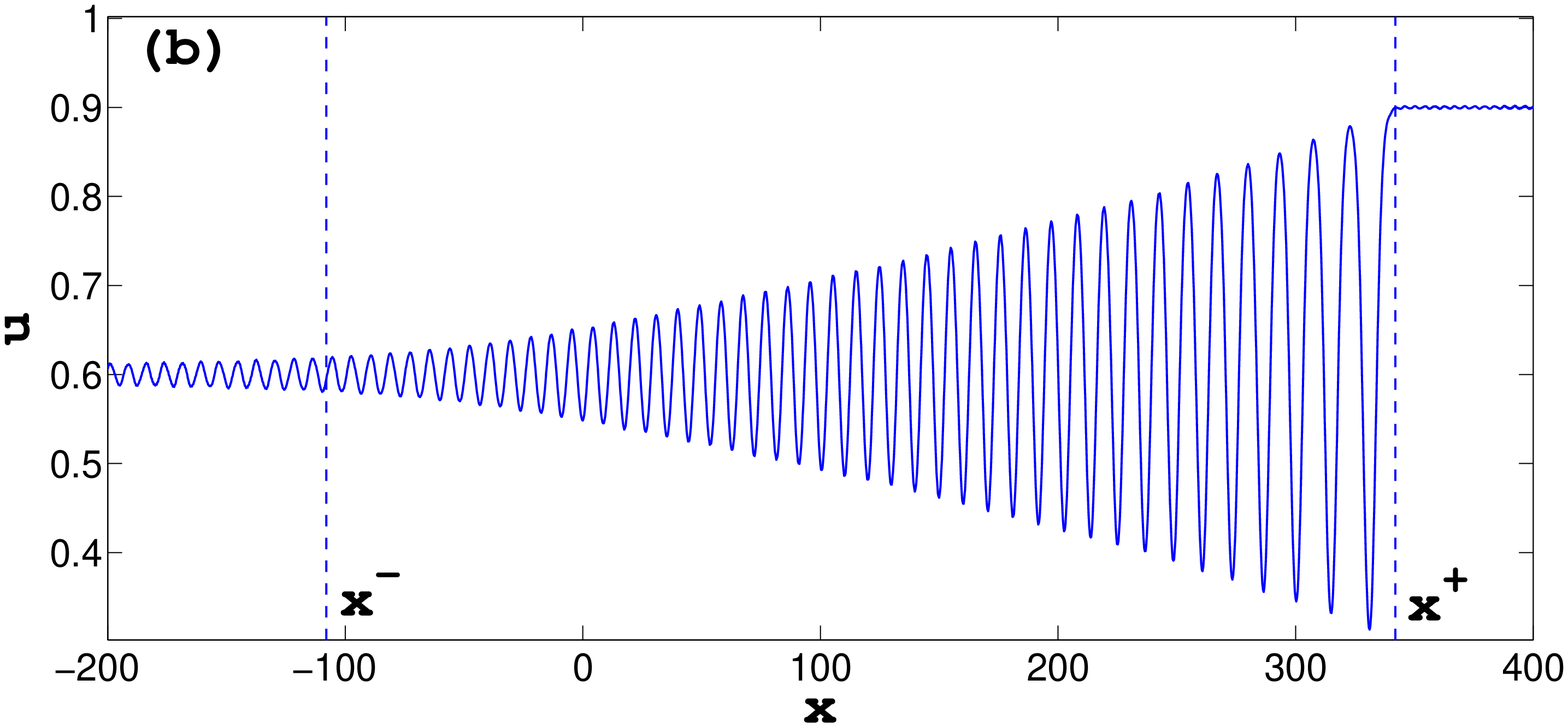}
\caption{(Color online) Evolution of an initial discontinuity for the Gardner equation with $\alpha=1$. Region 5: $\{u^-$ {\bf $\leftarrow$ UB } $u^+ \}$. The initial step parameters are:
$u^-=0.6$, $u^+=0.9$.
(a) analytical (modulation theory) solution;
(b) numerical solution of the Gardner equation. Dashed lines in (b) correspond to the analytically found locations $x^{\pm}$
of the undular bore edges. Both plots correspond to $t=300$.}
\end{center}\label{fig12}
\end{figure}

\bigskip
{\bf Region 6}, \ $ \frac1{\al}-u^+<u^-<\frac1{2\al}<u^+$. $\{u^-$ {\bf UB} $\rightarrow \ (u^*)$ $\leftarrow$ {\bf SB} $u^+\}$

\medskip
A combination of the normal undular bore and a reversed solibore is produced.

The undular bore connects the state $u^-$ and an intermediate state $u^*=1/\alpha - u^+$ and is described by the same
set of formulae as a single normal undular bore in Region 1. The reversed solibore further connects the intermediate
state $u^*$ with the downstream state $u^+$. It is described  by formula (\ref{solibore}) with ``+'' sign.
The plots of the analytical and numerical solutions are shown in Fig.~13.
\begin{figure}[bt]
\begin{center}
\includegraphics[width=8cm,height=5cm,clip]{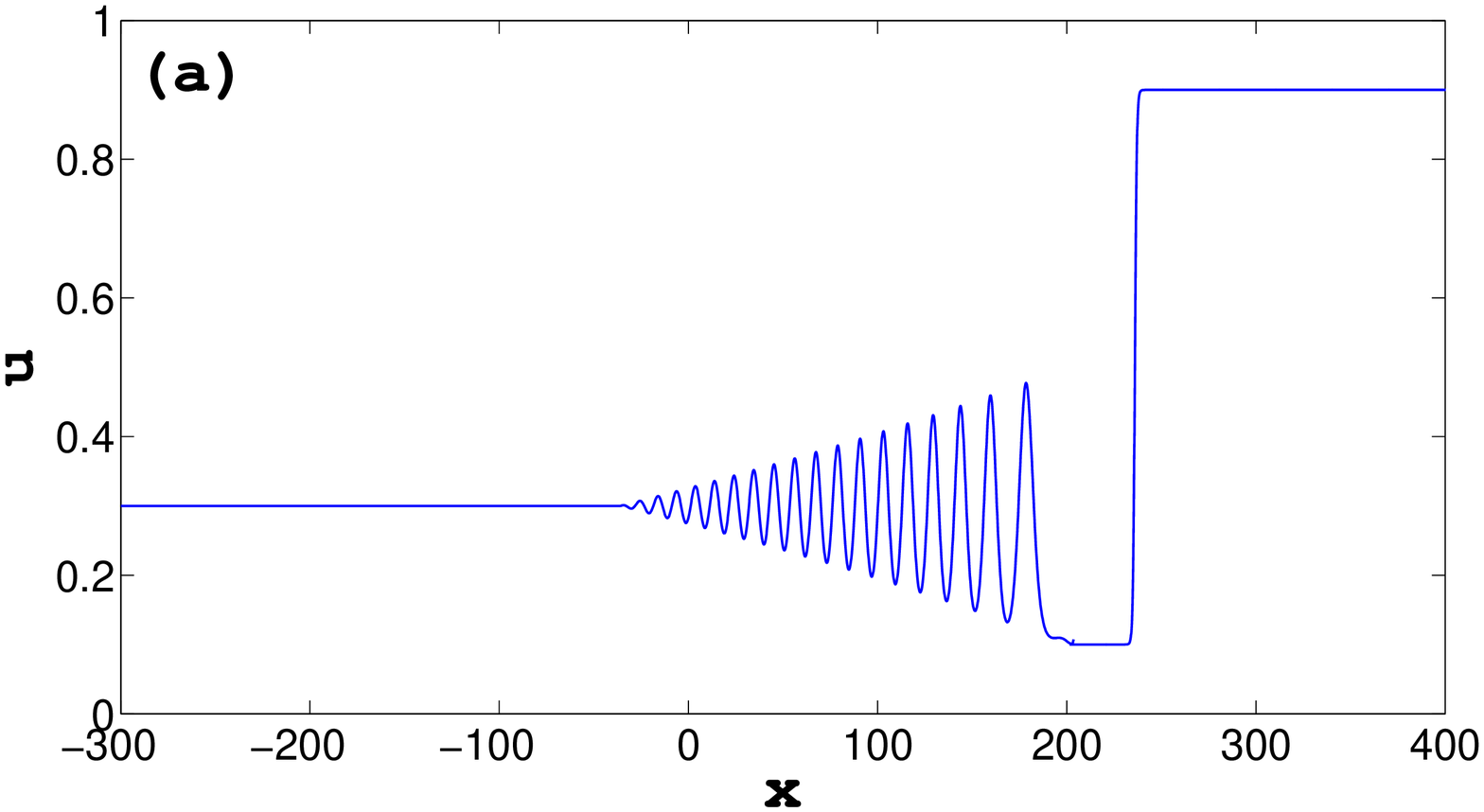}
\hspace{0.5cm}
\includegraphics[width=8cm,height=5cm,clip]{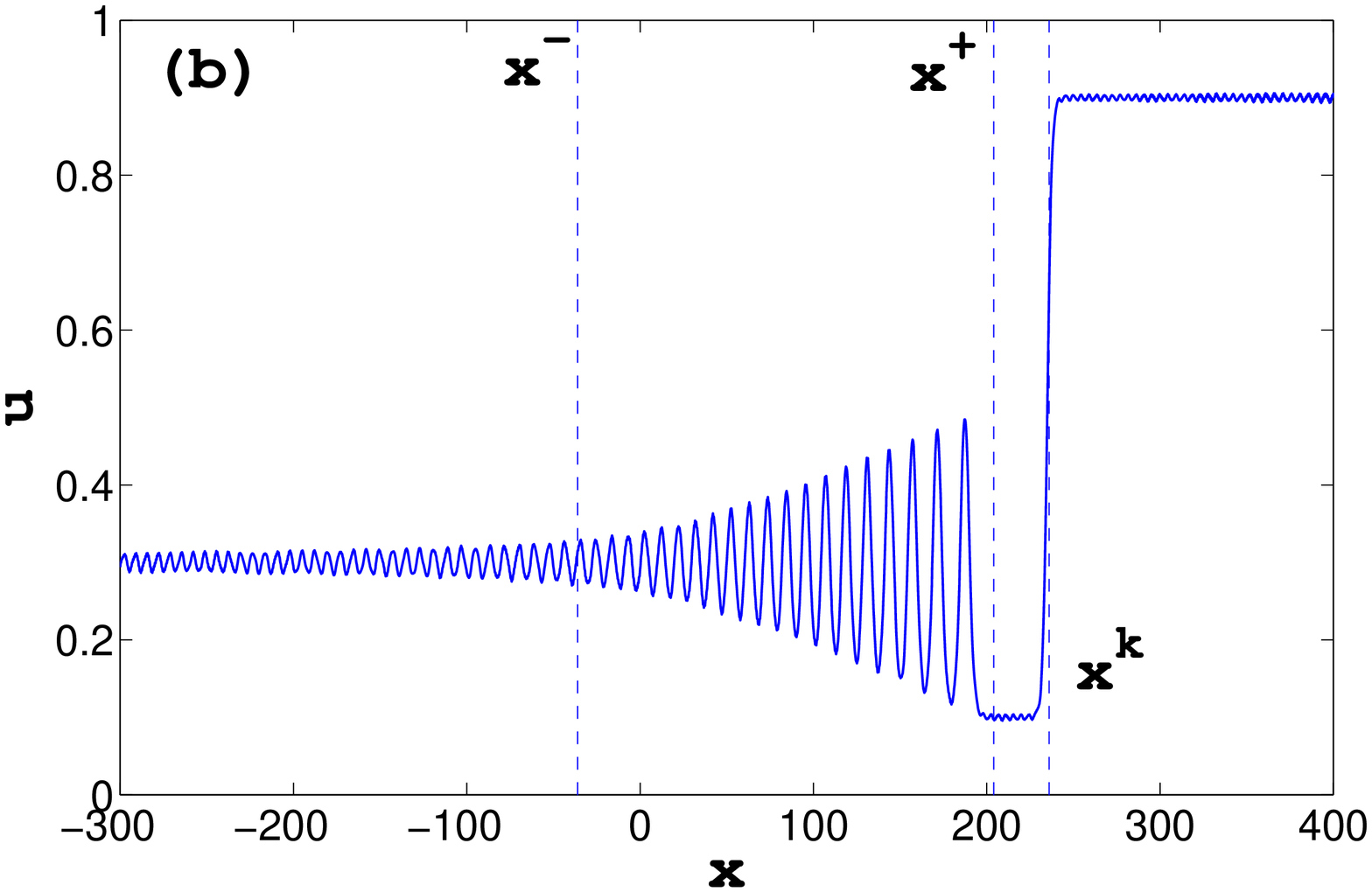}
\caption{(Color online) Evolution of an initial discontinuity for the Gardner equation with $\alpha=1$.
The initial step parameters are: $u^-=0.3$, $u^+=0.9$.  Region 6: $\{u^-$ {\bf UB} $\rightarrow \ (u^*)$
$\leftarrow$ {\bf SB} $u^+\}$).
(a) analytical (modulation theory) solution; (b) numerical solution of the Gardner equation.  Dashed lines in (b) correspond to the analytically found locations
of the undular bore edges $x^{\pm}$ and the solibore $x^k$. Both plots correspond to $t=200$.}
\end{center}
\label{fig13}
\end{figure}

\bigskip

{\bf Region 7}, $u^-<\frac{1}{2\alpha}<u^+<1/\alpha- u^-$.  $\{u^-$ {\bf RW} $\rightarrow \ (u^*)$ $\leftarrow$ {\bf SB} $u^+\}$

\medskip
The resolution pattern is similar to that in Region 6 but, instead of the normal undular bore, a normal rarefaction wave
described by formula (\ref{rref}) is generated. The boundary states are $u_l=u^-$, $u_r=u^*=1/\alpha - u^+$.
The corresponding analytical and numerical plots are presented in Fig.~14. Note that, similar to Region 3, the solibore always propagates ahead of the
rarefaction wave and gets attached to the right edge of the rarefaction wave when $u^+=1/2\alpha$.

\bigskip
{\bf Region 8}, $u^-<u^+\le \frac{1}{2\alpha}$.

\medskip
A single normal rarefaction wave is produced, $\{u^-$ {\bf RW $\rightarrow$ } $u^+\}$.  It is described by the solution (\ref{rref}) with $u^l=u^-$, $u^r=u^+$. The plots of the analytical and and numerical solutions for Region 8 are shown in Fig.~15.

\begin{figure}[bt]
\begin{center}$
\begin{array}{cc}
\includegraphics[width=8cm,height=5cm,clip]{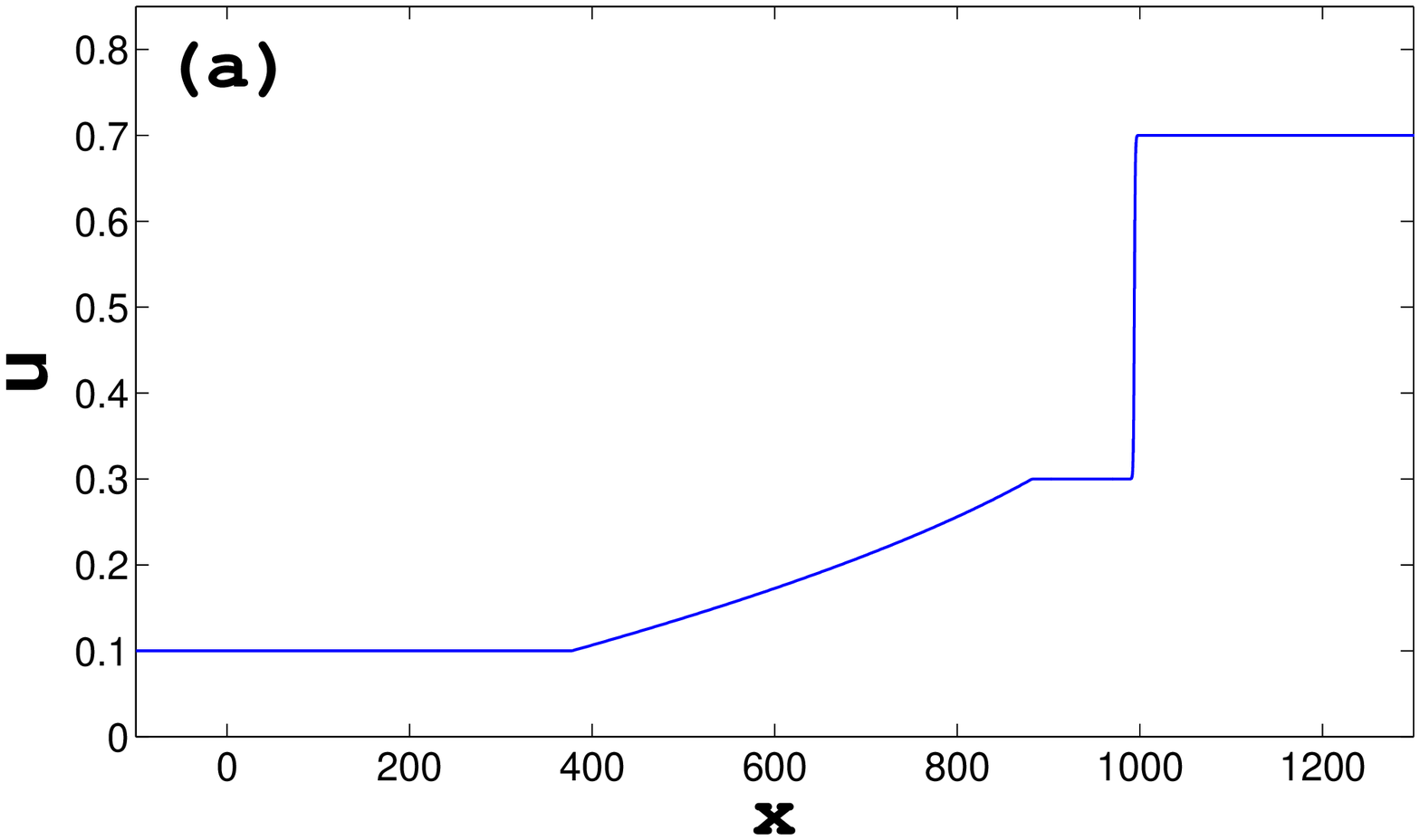}\\
\includegraphics[width=8cm,height=5cm,clip]{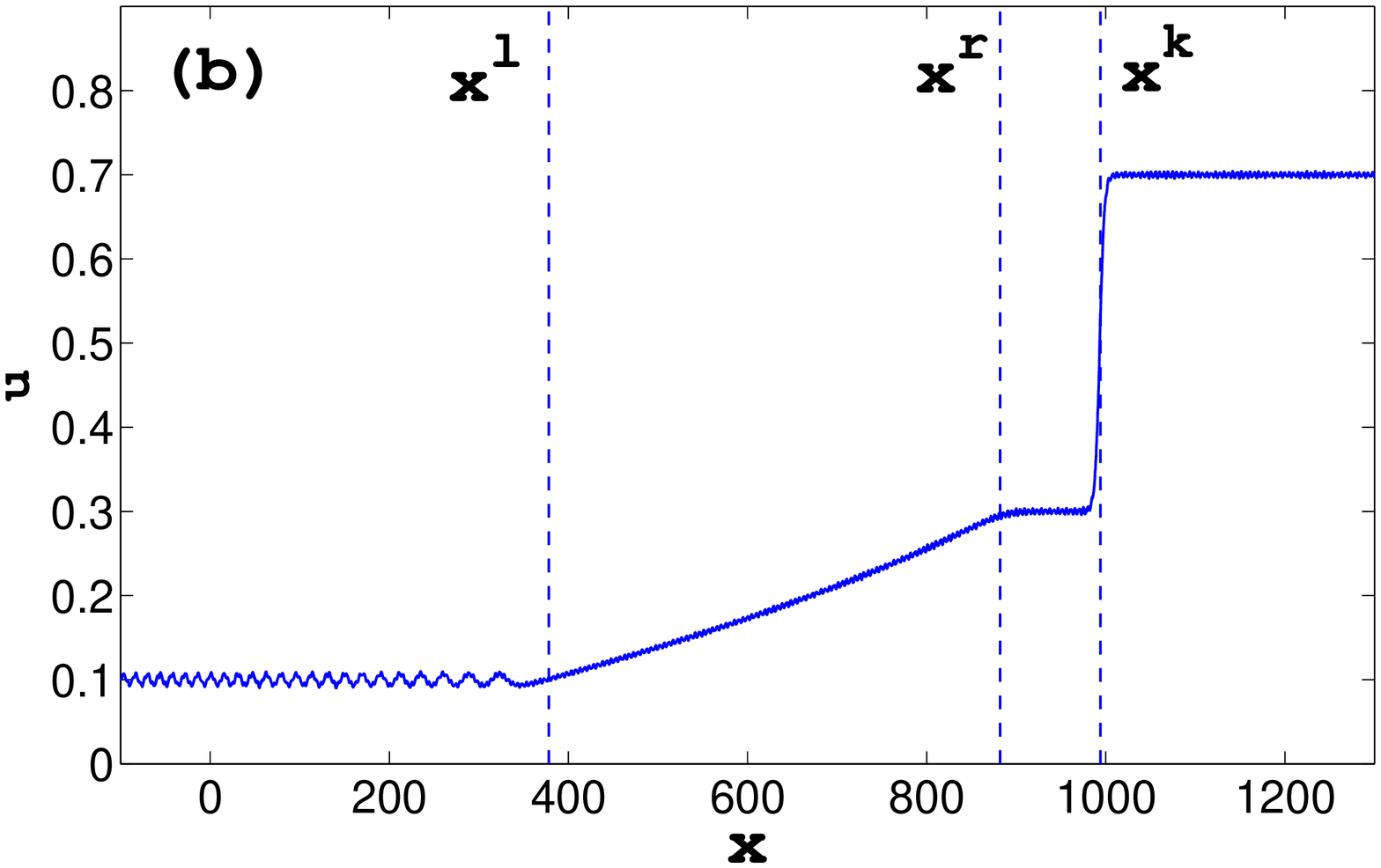}
\end{array}$
\end{center}
\caption{(Color online) Evolution of an initial discontinuity for the Gardner equation with $\alpha=1$. Region 7: $\{u^-$ {\bf RW} $\rightarrow \ (u^*)$ $\leftarrow$ {\bf SB} $u^+\}$.
The initial step parameters are: $u^-=0.1$, $u^+=0.7$. (a) Analytical (dispersionless limit) solution in the form of a rarefaction wave connected to a solibore; (b) Numerical solution of the Gardner equation. Dashed lines in (b) correspond to the analytically found positions of the solibore $x^k$ and rarefaction wave boundaries $x^{l,r}$ Both plots correspond to $t=700$.}
\end{figure}

\begin{figure}[bt]
\begin{center}
\includegraphics[width=8cm,height=5cm,clip]{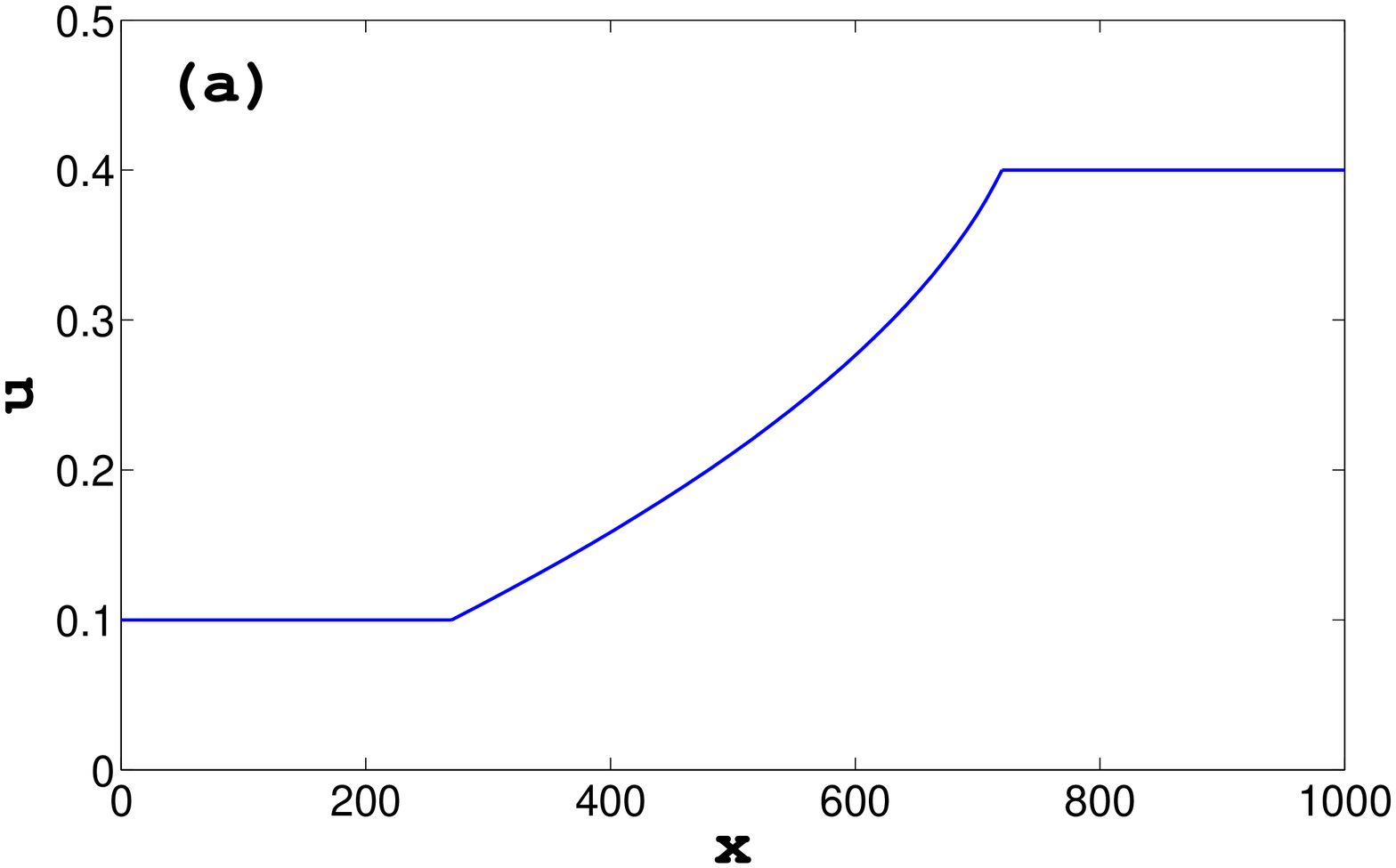}\\
\includegraphics[width=8cm,height=5cm,clip]{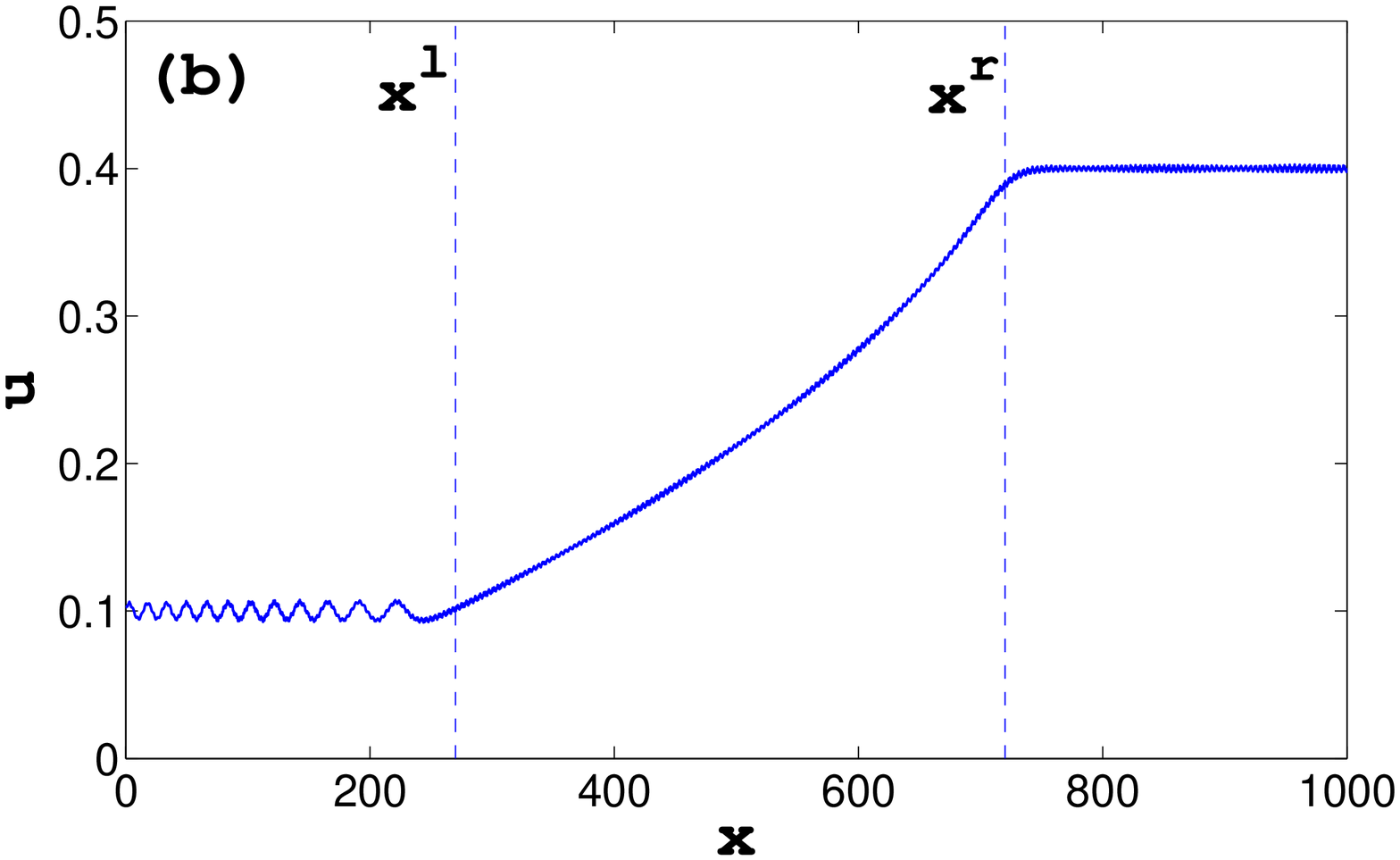}\\
\caption{(Color online) Evolution of an initial discontinuity for the Gardner equation with $\alpha=1$.
Region 8: $\{u^-$ {\bf RW $\rightarrow$ } $u^+\}$. The initial step parameters are: $u^-=0.1$, $u^+=0.4$. (a) Analytical (dispersionless limit) solution in the form of a rarefaction wave;
(b) Numerical solution of the Gardner equation. Dashed lines in (b) correspond to the analytically found positions $x^{l,r}$ of the rarefaction wave boundaries.
Both plots correspond to $t=500$.}
\end{center}
\end{figure}

\subsection{Classification for $\alpha<0$}
\begin{figure}[h!]
\begin{center}
\includegraphics[width=8cm]{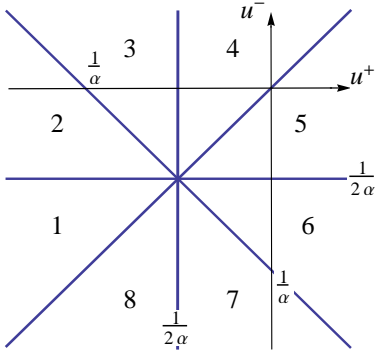}
\caption{(Color online) Parametric map of solutions of the step problem for the Gardner equation with
$\alpha<0$. The resolution diagrams corresponding to each of the cases on the plane of the initial step parameters
$u^-$ and $u^+$ are the following: Region 1: $\{u^-  \leftarrow {\bf RW} \ u^+\}$;
Region 2: $\{u^- \ {\bf TB} \rightarrow \ (u^*)  \leftarrow  {\bf RW} \ u^+\}$;
Region 3:  $\{u^- \ ({\bf TB}|{\bf UB}) \rightarrow  \ u^+\}$;
Region 4: $\{u^-$ {\bf  UB } $\rightarrow$ $u^+ \}$; Region 5: $\{u^-$ {\bf  RW } $\rightarrow$ $u^+ \}$;
Region 6: $\{u^- \ \leftarrow{\bf TB}  \ (u^*)  \ {\bf RW} \rightarrow \ u^+\}$;
Region 7: $\{u^- \leftarrow  \ ({\bf TB} | {\bf UB}) \ u^+ \}$;
Region 8:  $\{u^- \leftarrow {\bf UB} \ u^+\}$. In all relevant cases the intermediate state $u^*=1/\alpha -u^-$. }
\label{negalplane}
\end{center}
\end{figure}
Now we present the parametric map of solutions of the step problem for the Gardner equation with $\alpha <0$.
The most significant change in the structure of solutions compared to the case $\alpha>0$ is that the composite
solutions now contain trigonometric undular bores rather than solibores. The plane $(u^+, u^-)$ of the initial step
parameters is again split into 8 regions (see Fig.~\ref{negalplane}). The lines separating different regions are:
\begin{itemize}
\item $u^-=u^+$ separates the regions of pure undular bores and pure rarefaction waves;
\item $u^-=1/\alpha - u^+$   corresponds to the steps resolving into single trigonometric bore solutions and separates
the regions of composite solutions of different types: undular bore + trigonometric undular bore and rarefaction
wave + trigonometric bore;
\item $u^-=1/2\al$   separates regions of pure (undular bore) and composite (undular bore + trigonometric bore) solutions.
\end{itemize}

Let us now describe in some detail the  wave structures corresponding to different regions in Fig.~\ref{negalplane}.
\begin{figure}[h!]
\begin{center}
\includegraphics[width=8cm]{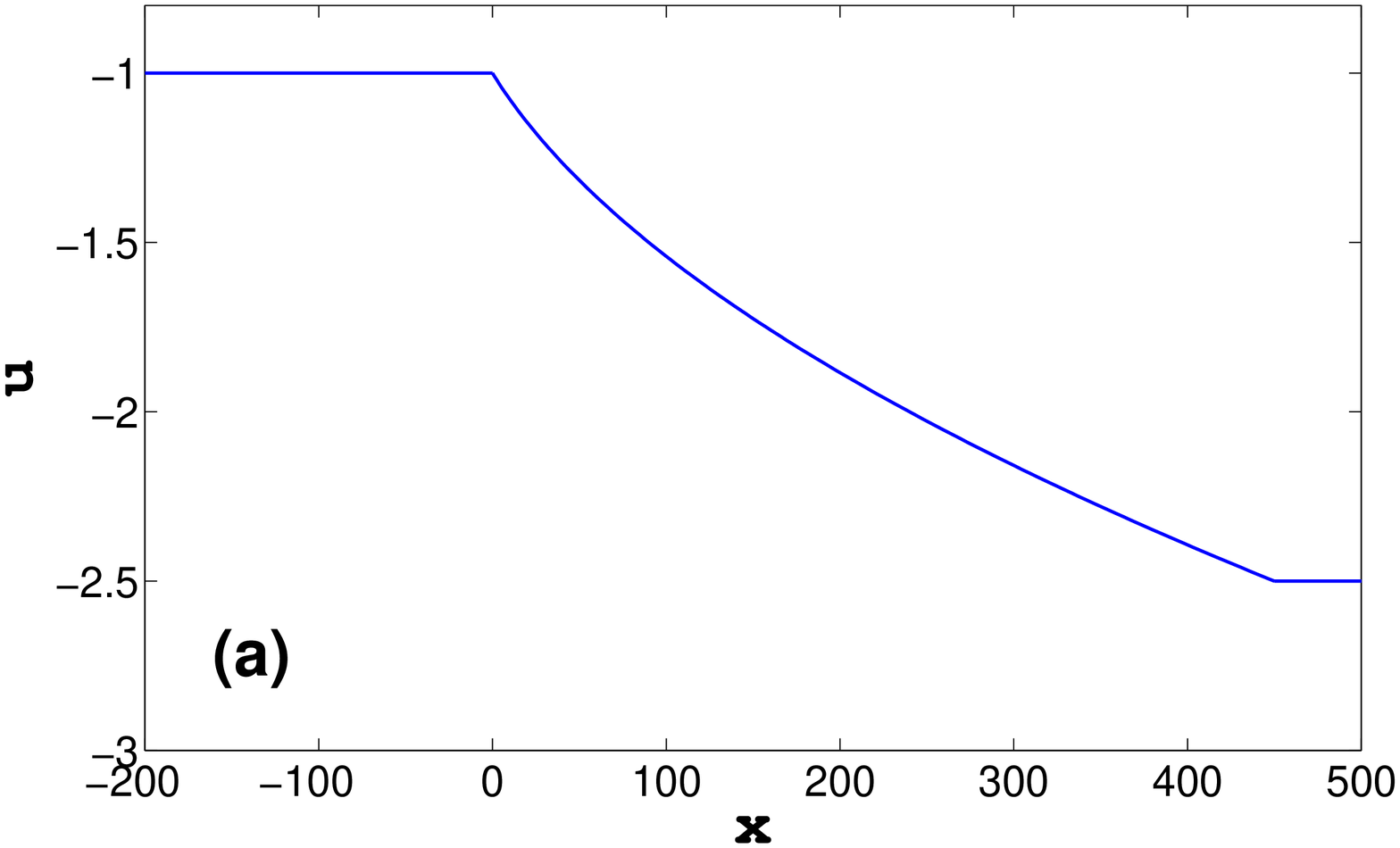}
\includegraphics[width=8cm]{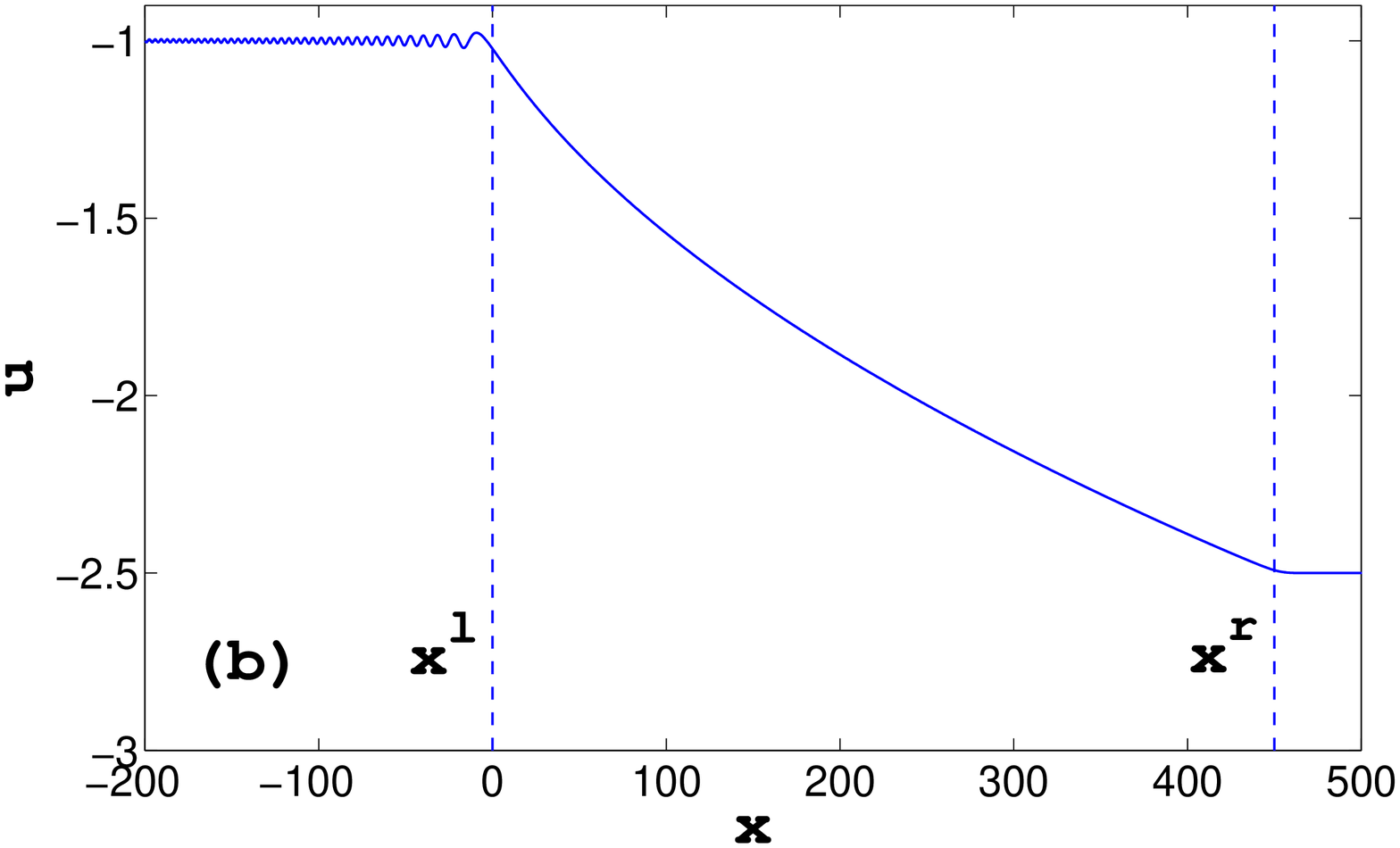}
\caption{(Color online) Evolution of an initial discontinuity in the Gardner equation with $\alpha=-1$:
Region 1: $\{u^-$ {\bf $\leftarrow$ RW } $u^+ \}$. The initial step parameters are: $u^-=-1$,  $u^+=-2.5$.  (a) Analytical (dispersionless limit) solution; (b) Numerical solution of the Gardner equation. Dashed lines in (b) correspond to the locations of the rarefaction wave edges found analytically. Both plots are made for $t=20$.
}
\label{fig17}
\end{center}
\end{figure}

\bigskip

{\bf Region 1}.  $u^+<u^-<\frac1{2\al}$, \ $\{u^-$ {\bf $\leftarrow$ RW } $u^+ \}$.

\medskip

Both values $u^-$ and $u^+$ lie in the domain where the function $w(u)=u(1-\alpha u)$ is monotonically decreasing so
there is one-to one correspondence between the  dispersionless limits of the Gardner and the KdV equations.
This suggests that the Region 1 initial discontinuity  is resolved by a single ``reversed'' simple rarefaction wave.
The rarefaction wave is described by the solution (\ref{revrare}) with $u^l=u^-$ and $u^r=u^+$. A typical
solution for Region 1 is shown in Fig.~\ref{region1}. The plot for the corresponding analytical solutions is
almost identical to that shown in Fig.~\ref{region1} with the exception for the small-amplitude wave train
at the left corner of the rarefaction wave so we do not present it here.
The analytically found boundaries of the rarefaction wave are shown by the dashed lines.

\medskip

{\bf Region 2}. $ u^+<\frac1{2\al}<u^-<\frac1{\al}-u^+$, \ $\{u^- \ {\bf TB} \rightarrow \ (u^*)  \leftarrow  {\bf RW} \ u^+\}$

\medskip
\begin{figure}[bt]
\begin{center}
\includegraphics[height=5cm,,width=9cm]{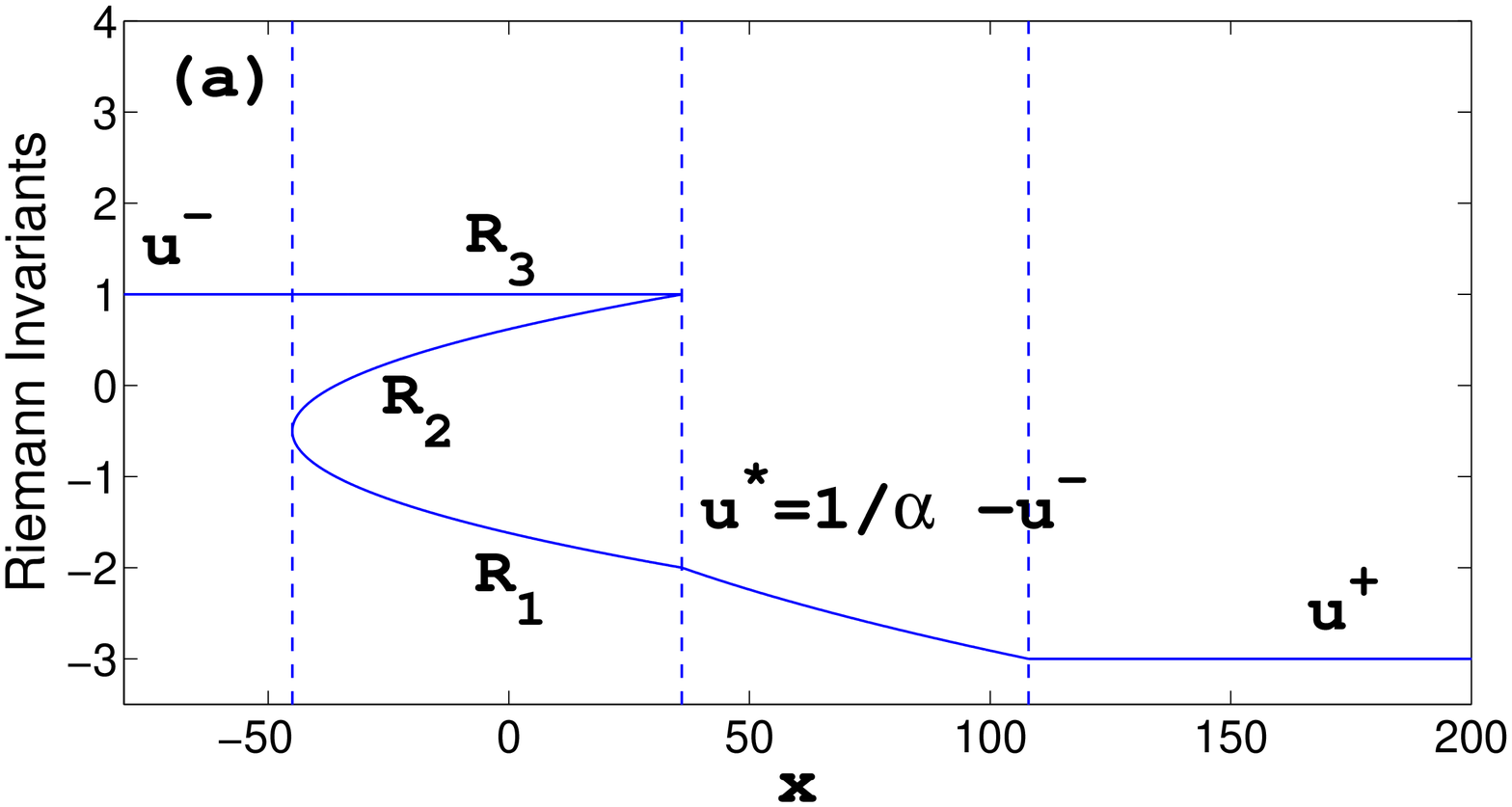}
\includegraphics[height=5cm,,width=9cm]{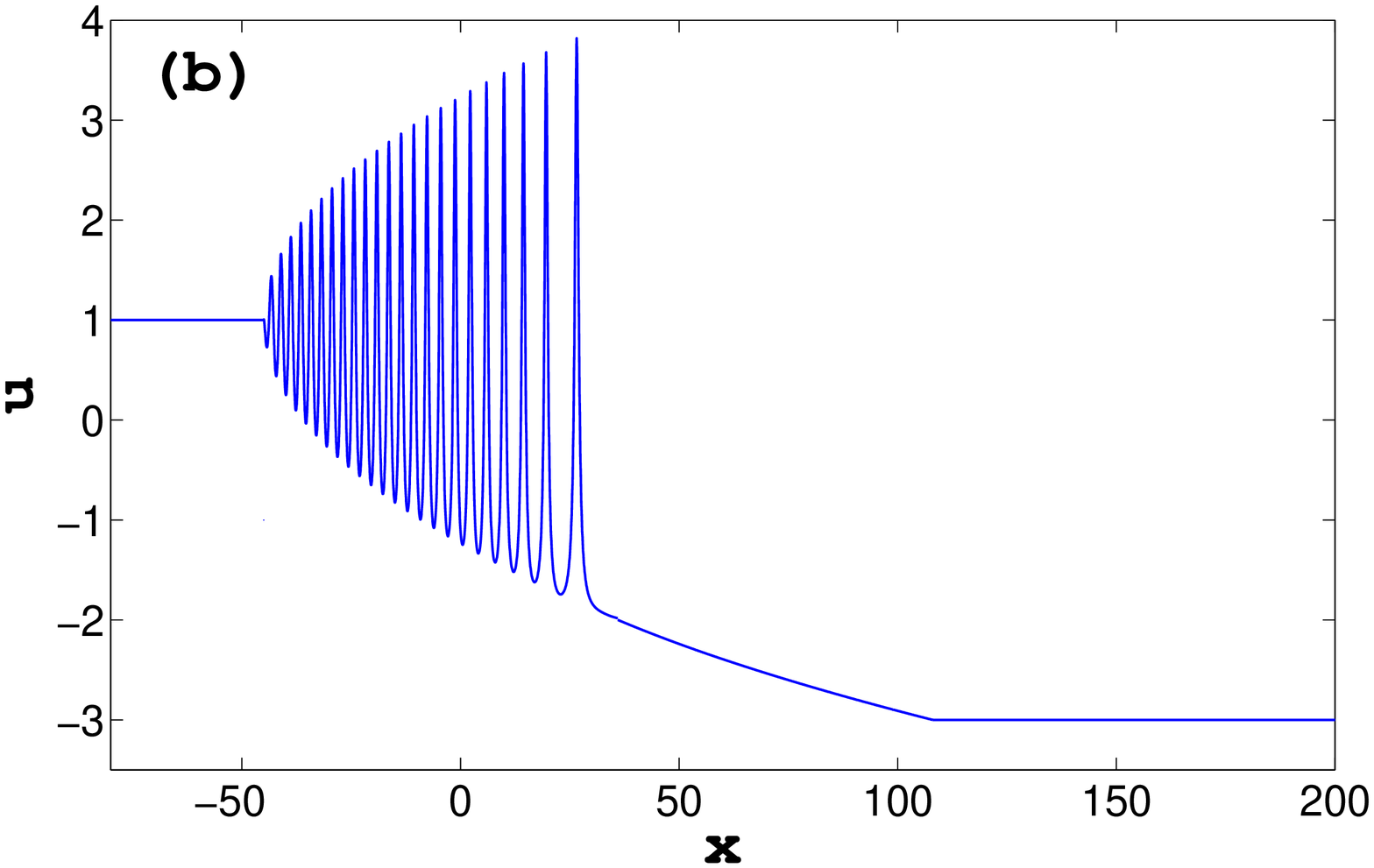}
\includegraphics[height=5cm,,width=9cm]{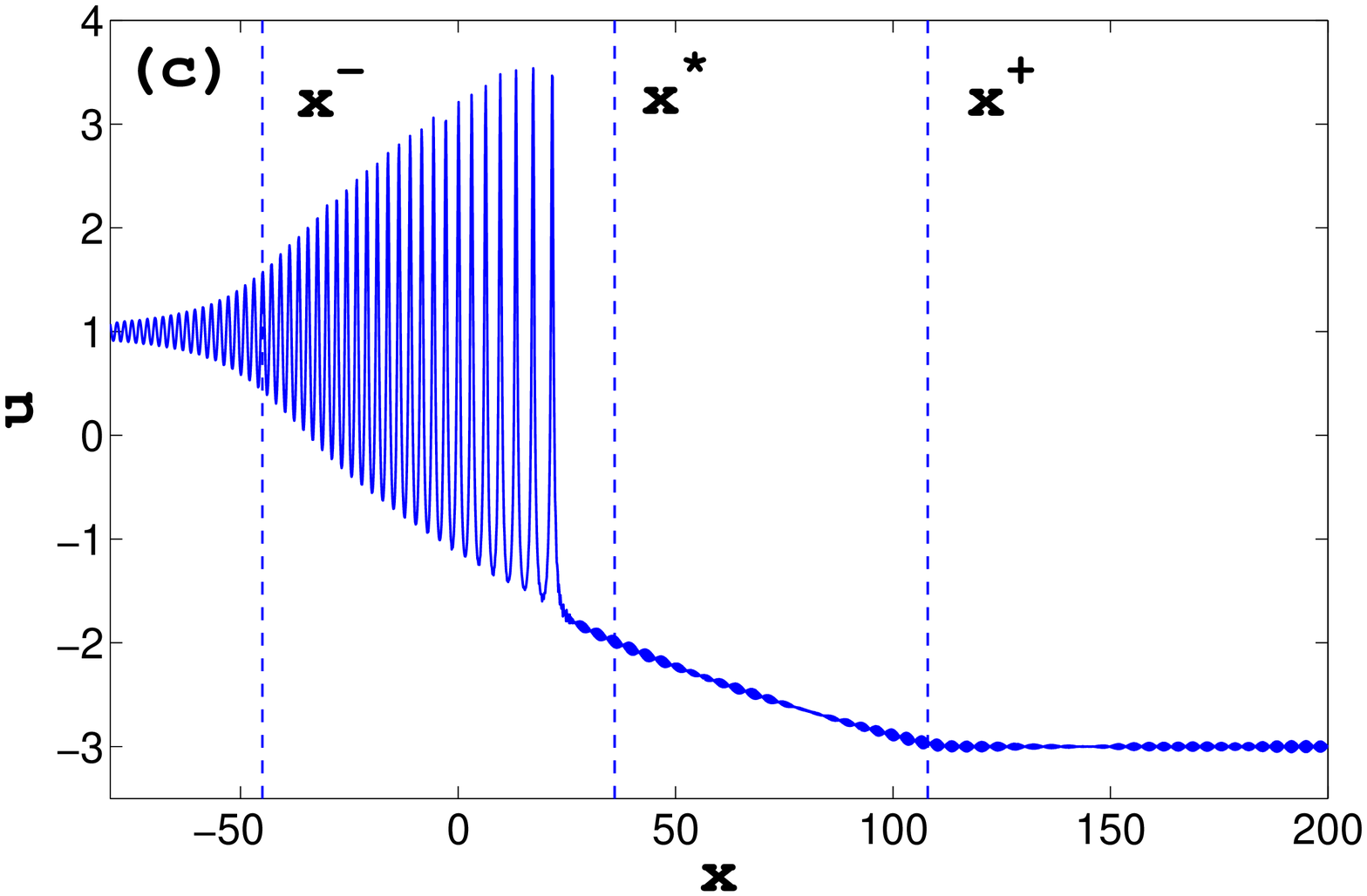}
\caption{(Color online) Evolution of an initial discontinuity for the Gardner equation with $\alpha=-1$ for $t=3$.
Region 2. $\{u^- \ {\bf TB} \rightarrow \ (u^*)  \leftarrow  {\bf RW} \ u^+\}$. The initial step parameters:
$u^-=1$, $u^+= -3$.
(a) Riemann invariants $R_1$, $R_2$ and $R_3$. (b) Analytical (modulation theory) solution.
(c) Numerical solution of the Gardner equation. Dashed lines on the numerical plot correspond to the
analytically found boundaries between different parts of the wave pattern.}
\label{reg2neg}
\end{center}
\end{figure}

\begin{figure}[bt]
\begin{center}
\includegraphics[height=5cm,width=9cm]{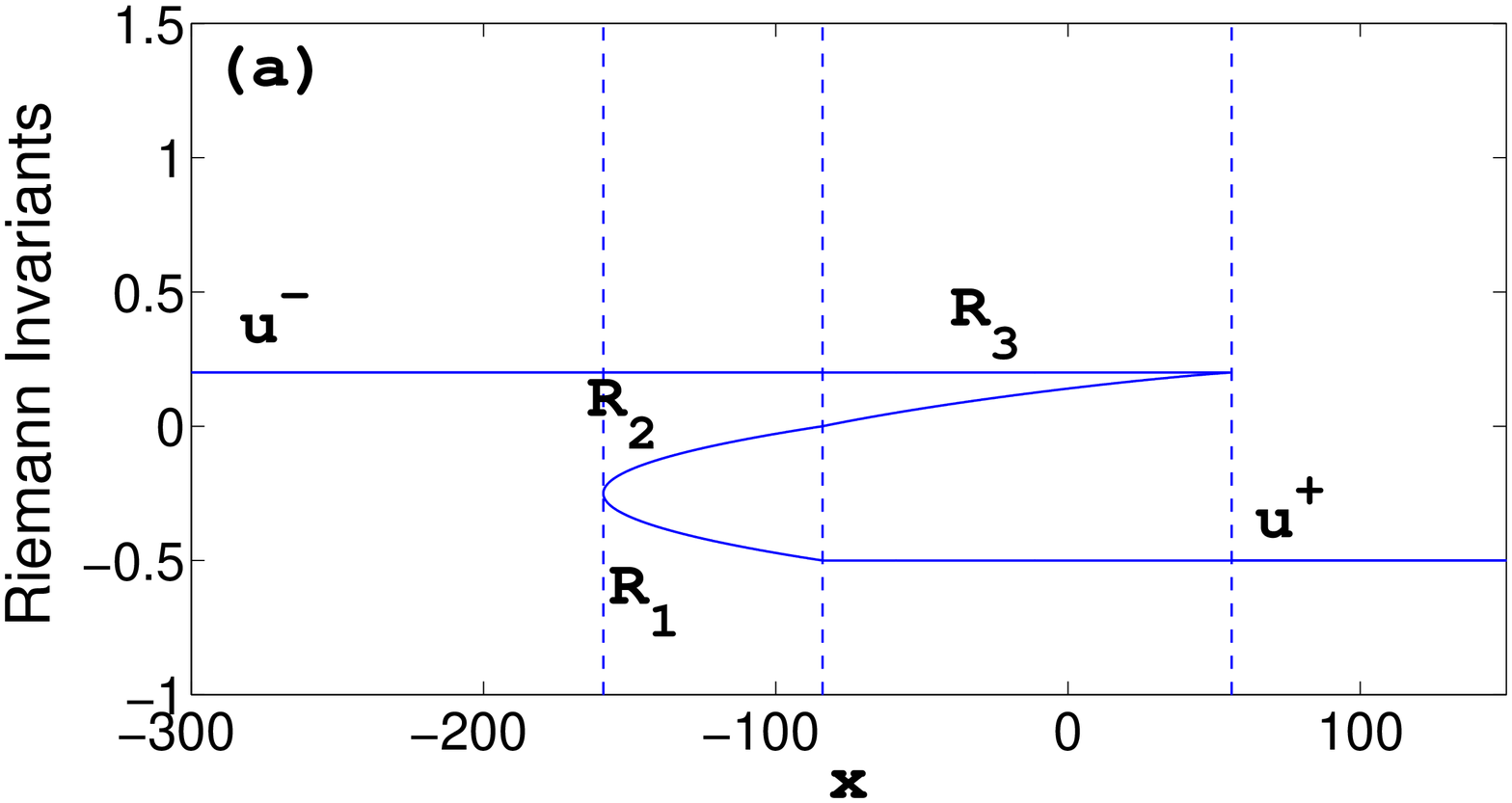}
\includegraphics[height=5cm,width=9cm]{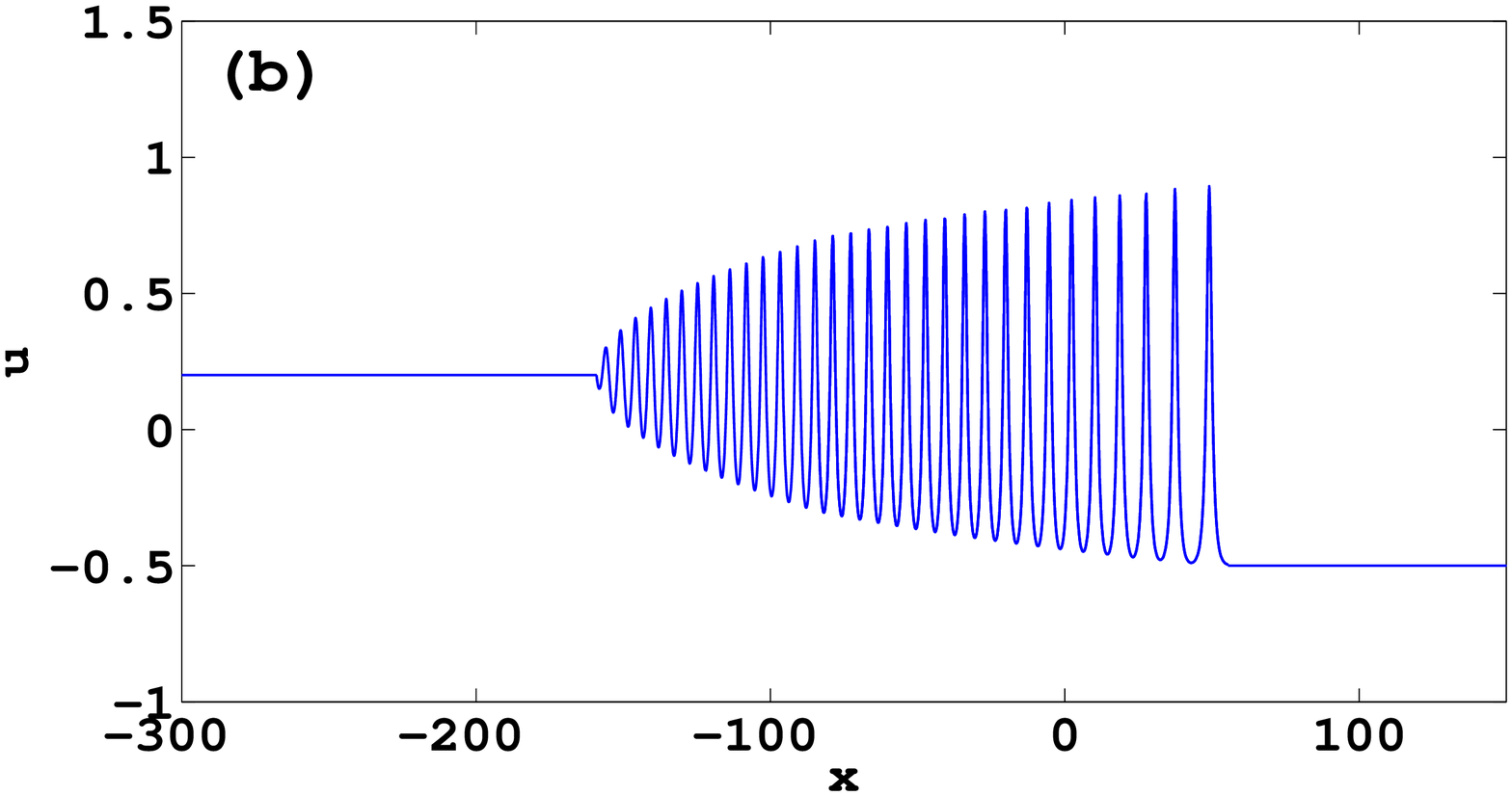}
\includegraphics[height=5cm,width=9cm]{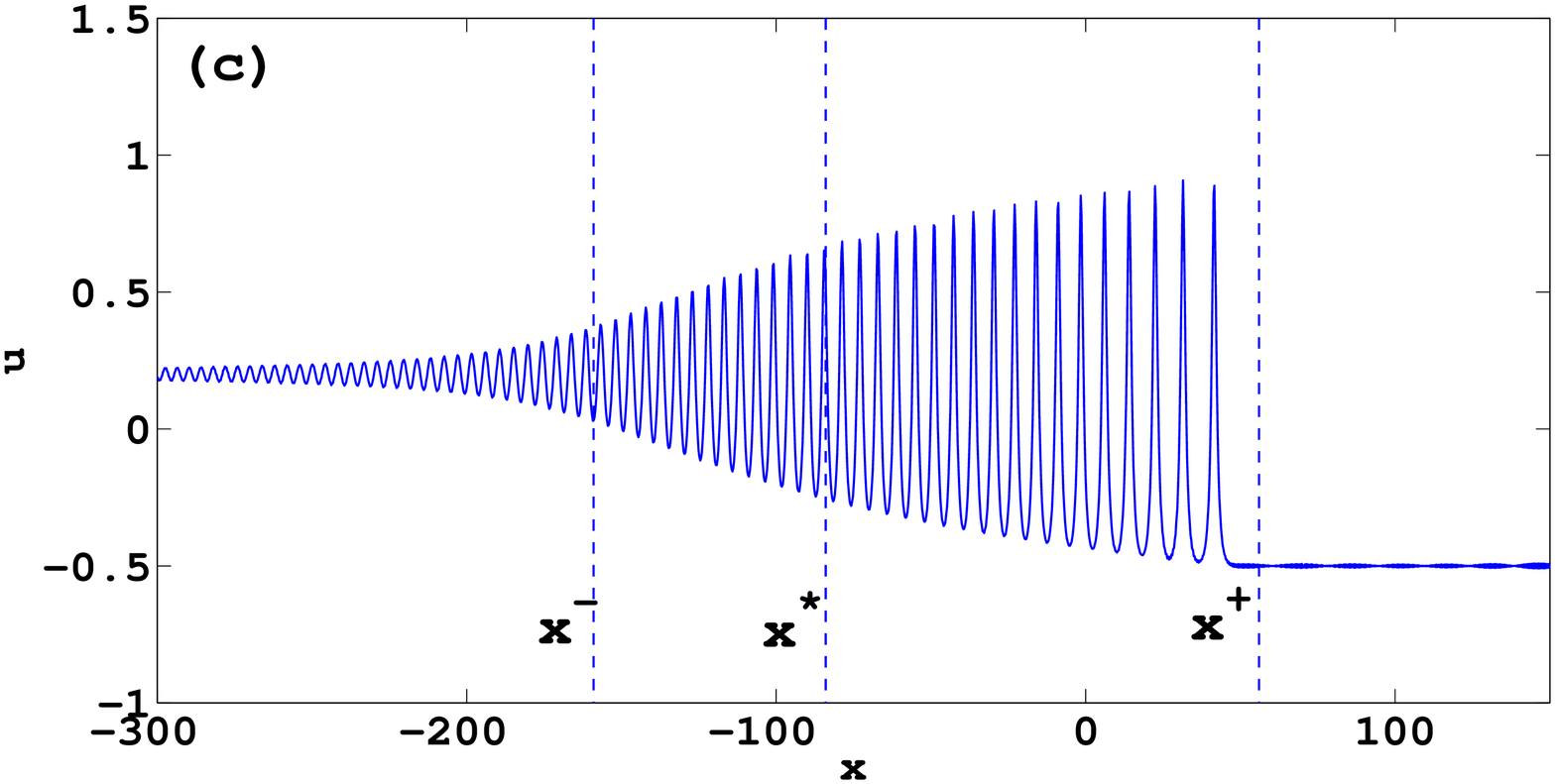}
\caption{(Color online) Evolution of an initial discontinuity for the Gardner equation with $\alpha=-2$. Region 3:  $\{u^- \ ({\bf TB} | {\bf UB}) \rightarrow  \ u^+\}$.
The initial step parameters are: $u^-=0.2$, $u^+=-0.5$. The plots are shown for $t=50$.
(a): Riemann invariants $R_1$, $R_2$ and $R_3$. (b): Analytical (modulation theory) solution.
(c): Numerical solution of the Gardner equation.  Dashed lines on the numerical plot correspond to the
analytically found boundaries between different parts of the wave pattern.}
\label{reg3neg}
\end{center}
\end{figure}
Since $u^+$ and $u^-$ now lie in different regions of monotonicity of the function $w(u)=u(1-\alpha u)$, the Region 1
solution in the form of a single reversed rarefaction wave is not able to provide the necessary continuous matching between
the given states. Instead, one needs to introduce a trigonometric undular bore joining the left constant state $u^-$ with rarefaction wave
at the level $u^*=1/\alpha - u^-$ (see Fig.~\ref{reg2neg}). The intermediate state $u^*$ is found from the condition $w(u^*)=w(u^-)$ (cf. condition $w(u^*)=w(u^+)$ for $\alpha>0$).
The modulation description of the relevant (normal) trigonometric bore was constructed in Section IVA in terms of the Riemann invariants $R_3 \ge R_2 \ge R_1$ (\ref{eq5a}).

The speeds $s^-$ and $s^*$ of the trailing and the leading edges of the trigonometric bore, and the speed of the
leading edge of the rarefaction wave $s^+$ are:
\begin{equation}\label{eq5e}
\begin{split}
    &s^-=\frac3{\al}-6u^-(1-\al u^-),\\
    &s^*=6u^*(1-\al u^*)=6u^-(1-\al u^-),\\
    &s^+=6u^+(1-\al u^+).
    \end{split}
\end{equation}
The analytical and numerical solutions along with the plot for the Riemann invariants $R_1, R_2, R_3$, are
shown in Fig.~\ref{reg2neg}.

\bigskip

At $u^-=1/\al-u^+> u^+$ the rarefaction wave disappears and one obtains a single normal trigonometric bore as
a result of the step evolution The relevant analytical description was presented in Section IVA.

\bigskip

{\bf Region 3}. $  \frac1{\al}-u^-<u^+<\frac1{2\al}$, \ $\{u^- \ ({\bf TB} |{\bf UB}) \rightarrow  \ u^+\}$

\medskip

In the Region 3 we get a composite undular bore consisting of normal trigonometric and cnoidal parts matching at
the point of the trailing edge $x^*=s^*t$ of the cnoidal bore.
The modulation solution for the entire composite bore is conveniently described in terms of the Riemann
invariants $R_3 \ge R_2 \ge R_1$ (\ref{eq5a}).  The corresponding analytical and numerical plots
are shown in Fig.~\ref{reg3neg}.

The characteristic speeds for this region are
\begin{equation}\label{eq7}
    \begin{split}
    &s^-=\frac3{\al}-6u^-(1-\al u^-),\\
    &s^*=12u^+(1-\al u^+)-6u^-(1-\al u^-),\\
    &s^+=2u^+(1-\al u^+)+4u^-(1-\al u^-).
    \end{split}
\end{equation}

We note that the boundary $x^*=s^*t$ between the trigonometric bore and the cnoidal bore parts in the composite bore solution  can be naturally defined only in the framework of the averaged (Whitham) equations, where it represents a characteristic separating two regions with qualitatively different behaviour  of the modulation solution (the modulation solution has a weak discontinuity at $x=x^*$ -- see Fig.~\ref{reg3neg}a). Due to the asymptotic nature of the modulation equations (the phase is washed out), this separating line cannot be consistently identified on the level of the genuine (rapidly oscillating) solution of the governing equation. The situation here is similar to that with the definition of the trailing edge of a standard undular bore: the trailing edge cannot be identified with a particular point in the bore but is rather associated with the linear group velocity characteristic of the modulation equations, hence the already mentioned noticeable difference in the behaviour of the asymptotic (modulation theory) solution and that of the full numerical solution near the trailing edge. (cf Figs.~\ref{reg3neg}b and  \ref{reg3neg}c).

\bigskip
{\bf Region 4.} \  $u^->u^+>\frac1{2\al}$,  $\{u^-$ {\bf  UB } $\rightarrow$ $u^+ \}$.

\begin{figure}[bt]
\begin{center}
\includegraphics[height=4cm,width=8cm]{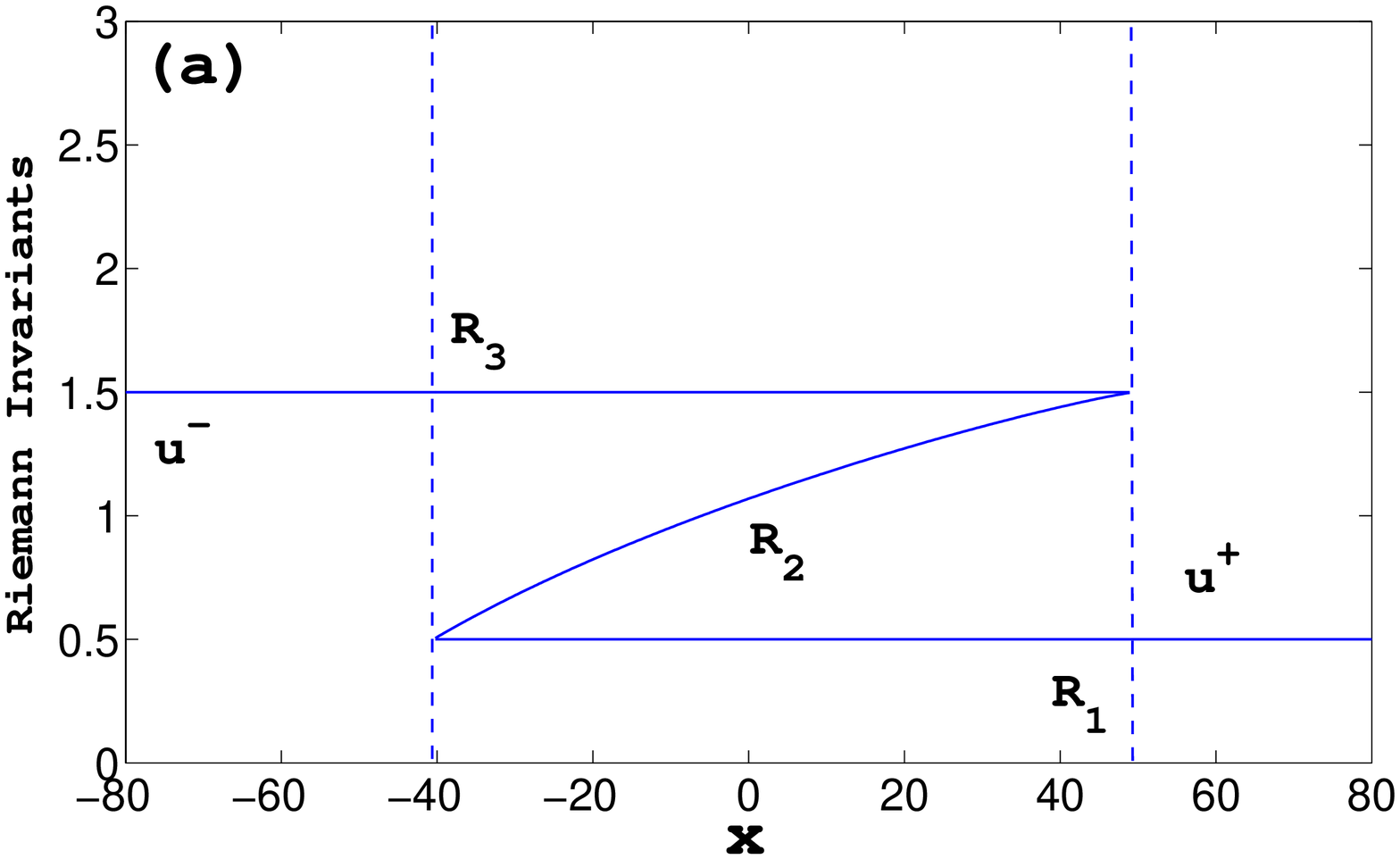}
\includegraphics[height=4cm,width=8cm]{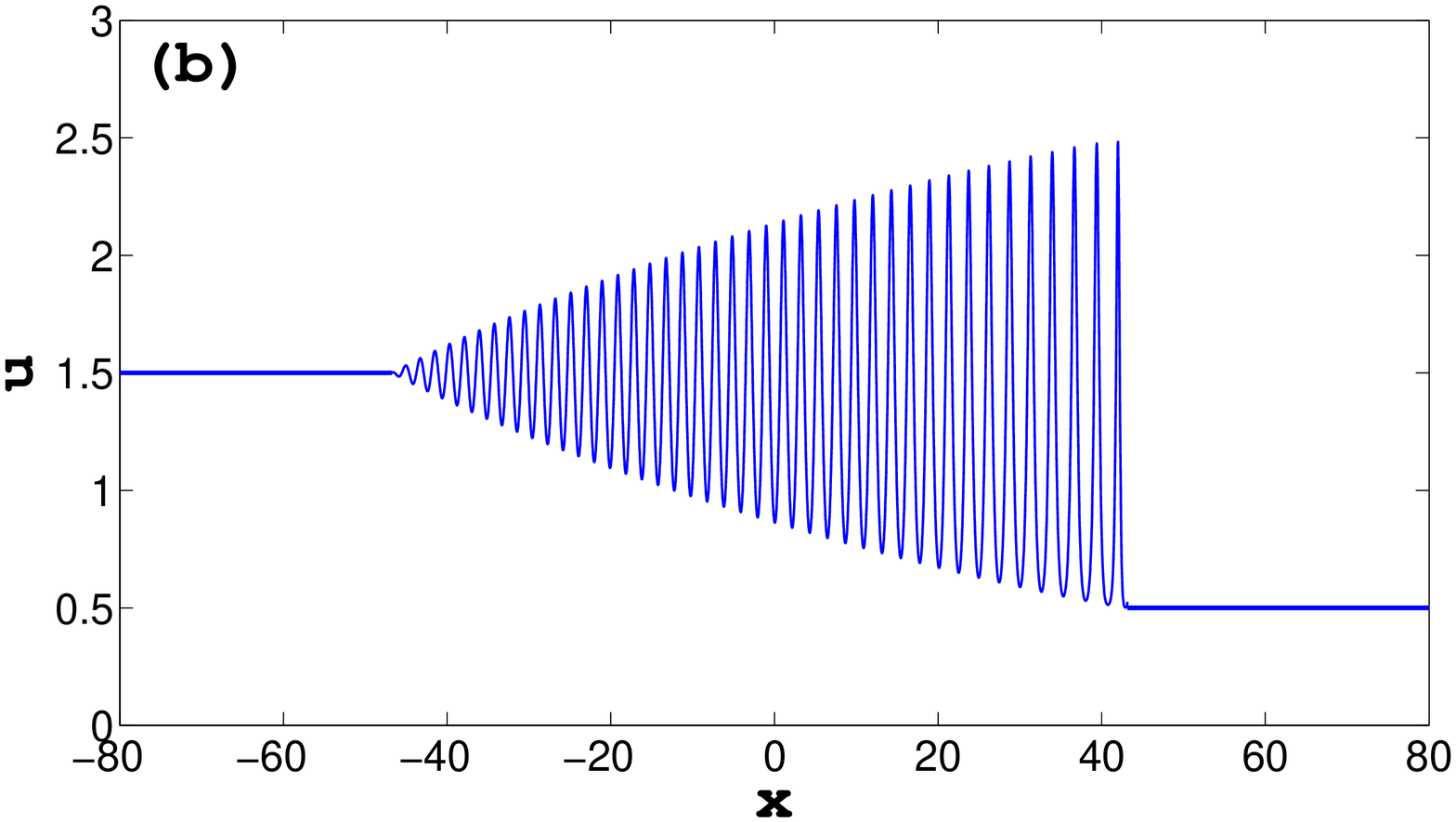}
\includegraphics[height=4cm,width=8cm]{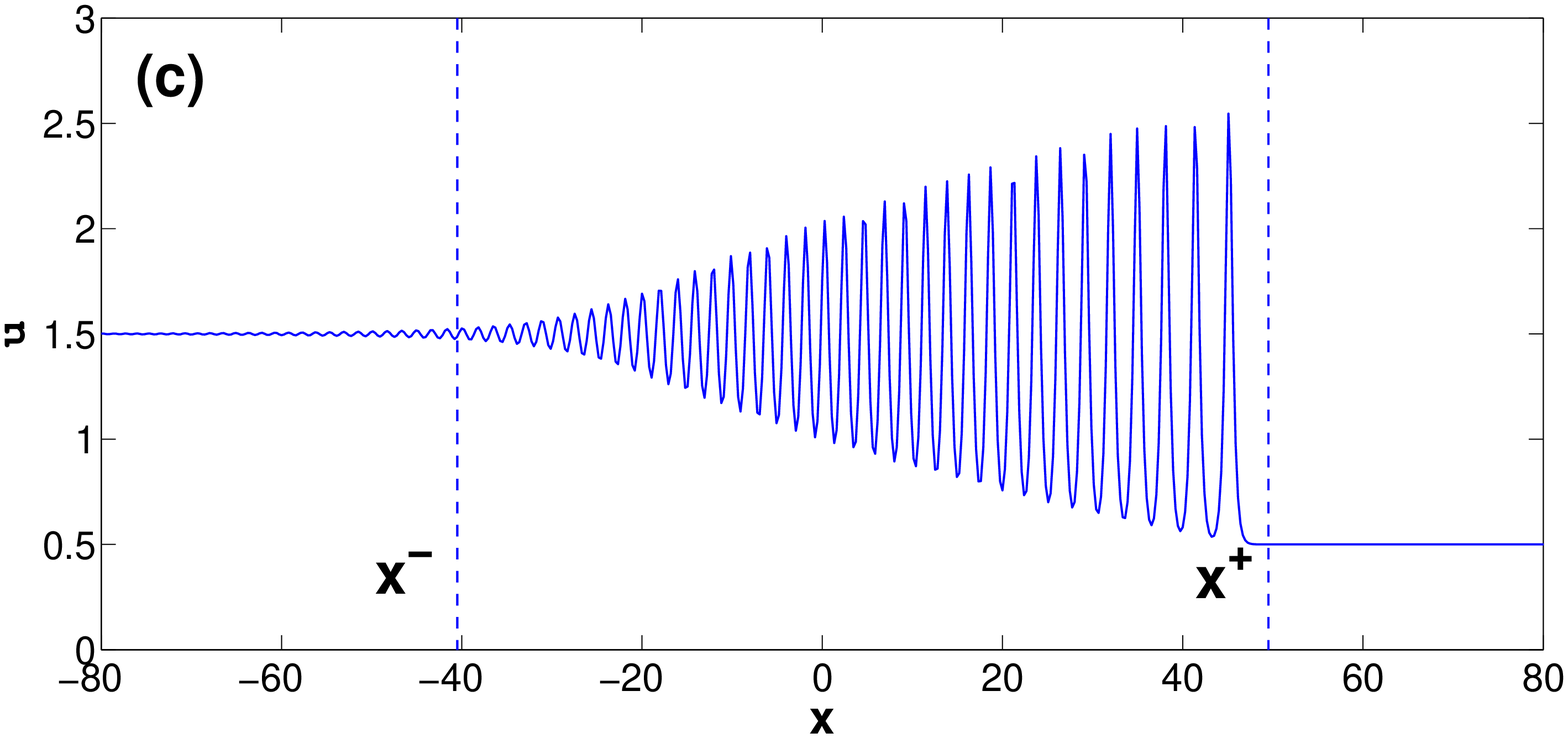}
\caption{(Color online) Evolution of an initial discontinuity for the Gardner equation with $\alpha=-1$. Region 4:  $\{u^-$ {\bf  UB } $\rightarrow$ $u^+ \}$.
The initial step parameters are $u^-=1.5$, $u^+=0.5$.
The plots are shown for $t=3$. (a) Riemann invariants $R_1$, $R_2$ and $R_3$. (b) Analytical (modulation theory) solution.
(c) Numerical solution of the Gardner equation.  Dashed lines on the numerical plot correspond to the
analytically found boundaries between different parts of the wave pattern.}
\label{reg4neg}
\end{center}
\end{figure}

\medskip
Both values $u^->u^+$ lie in the domain where the function $w(u)=u(1-\alpha u)$ increases, so the resolution occurs
via a single normal cnoidal undular bore; see Fig.~\ref{reg4neg}. To be consistent with other plots in this section
we present the Gurevich-Pitaevskii solution (\ref{GPconst}), (\ref{14-6}) for the undular bore in terms
of the Riemann invariants $\{R_j\}$ rather than $\{r_j\}$. The one-to-one correspondence between these two sets of the
Riemann invariants in Region 4 is given by relations (\ref{eq5a}).

The edge speeds are (see (\ref{gp-1}), (\ref{gp-2})):
\begin{equation}\label{eq11a}
\begin{split}
    &s^-=12u^-(1-\al u^-)-6u^+(1-\al u^+),\\
    &s^+=2u^-(1-\al u^-)+4u^+(1-\al u^+).
    \end{split}
\end{equation}

\medskip

{\bf Regions 5-8}. \quad  Similar to the classification for  $\alpha>0$ described in the previous section, the solutions for Regions 5 -- 8 can be obtained by applying the transformation (\ref{inv}) to their counterparts from the ``opposite'' regions of the parametric map in Fig.~16. Again, for the convenience of identification we present them below in the same format.

\medskip
{\bf Region 5.} \ $\frac1{2\al}<u^-<u^+$. $\{u^-$ {\bf  RW } $\rightarrow$ $u^+ \}$.

\medskip

\begin{figure}[bt]
\begin{center}
\includegraphics[width=8cm]{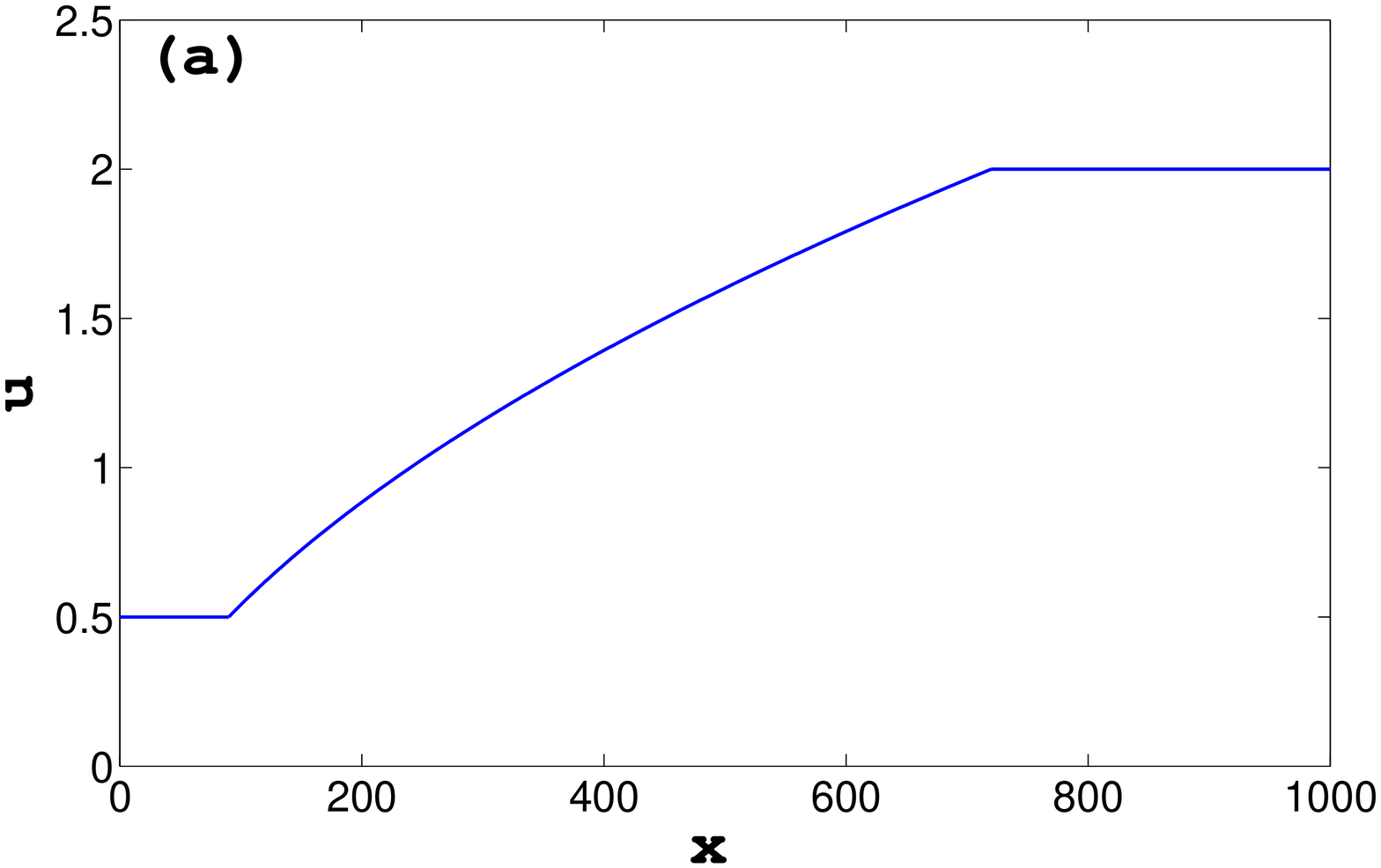}
\includegraphics[width=8cm]{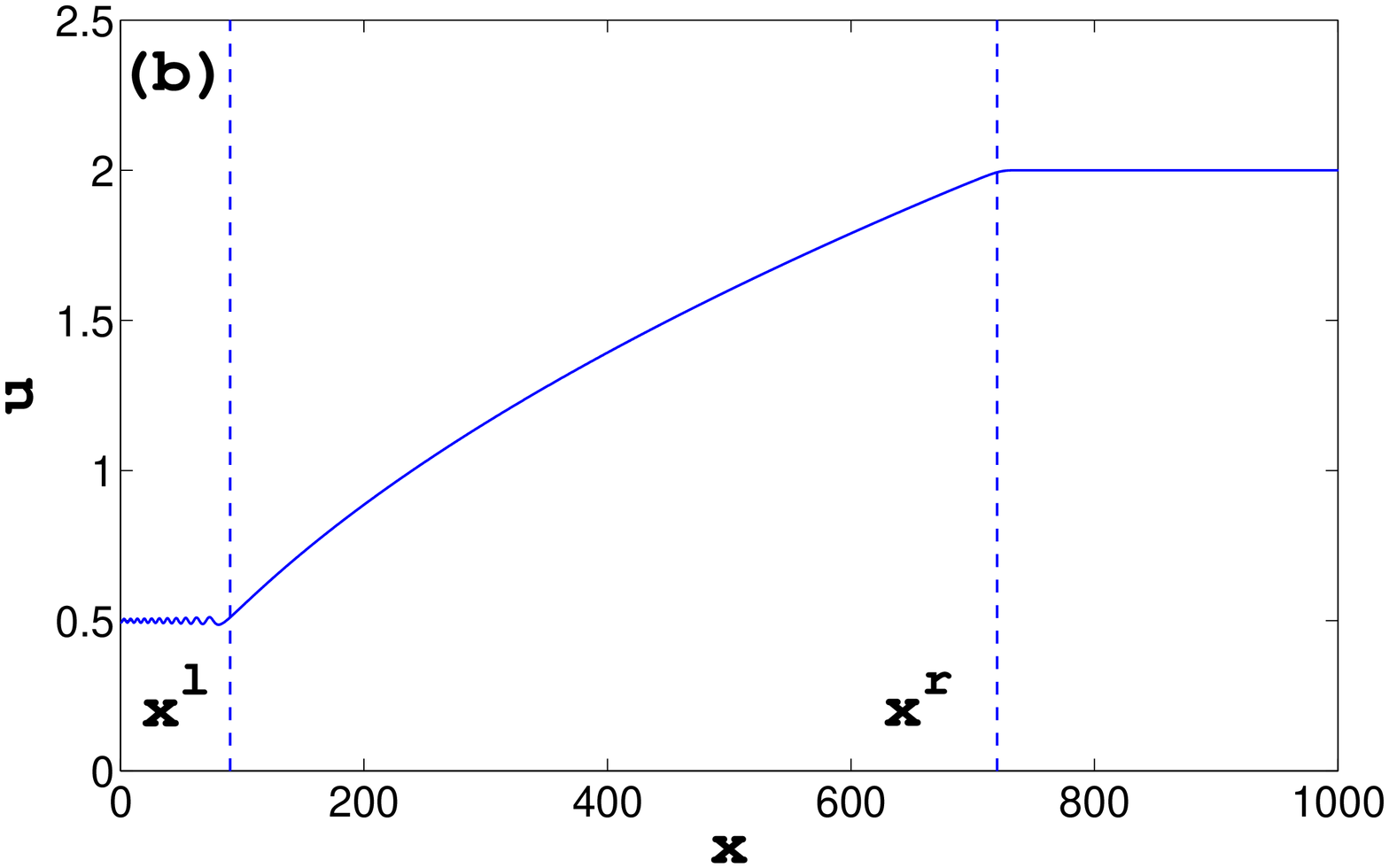}
\caption{(Color online) Evolution of an initial discontinuity in the Gardner equation with $\alpha=-1$.
Region 5: $\{u^-$ {\bf  RW } $\rightarrow$ $u^+ \}$. The initial discontinuity parameters are:
$u^-=0.5$,  $u^+=2$; (a) Analytical (dispersionless limit) solution; (b) Numerical solution of the Gardner equation.    Dashed lines correspond to the analytically found locations of the rarefaction wave edges. The plots are shown for $t=20$.}
\label{reg5neg}
\end{center}
\end{figure}

Both values $u^- < u^+$ lie now in the  domain where the function $w(u)=u(1-\alpha u)$ decreases, so the resolution occurs
via a single normal rarefaction wave described by (\ref{normrare})
with $u^l=u^-$, $u^r=u^+$.

The edge speeds are
\begin{equation}\label{eq13-2}
        s^-=6u^-(1-\al u^-),\quad
    s^+=6u^+(1-\al u^+).
\end{equation}
The corresponding numerical solution is shown in Fig.~\ref{reg5neg} along with the boundaries of the analytical
RW solution marked by dashed lines.


\bigskip

{\bf Region 6}. $ \frac1{\al}-u^+<u^-<\frac1{2\al}<u^+$, \  $\{u^- \ \leftarrow{\bf TB}  \ (u^*)  \ {\bf RW} \rightarrow \ u^+\}$.

\begin{figure}[bt]
\begin{center}
\includegraphics[height=5cm,width=9cm]{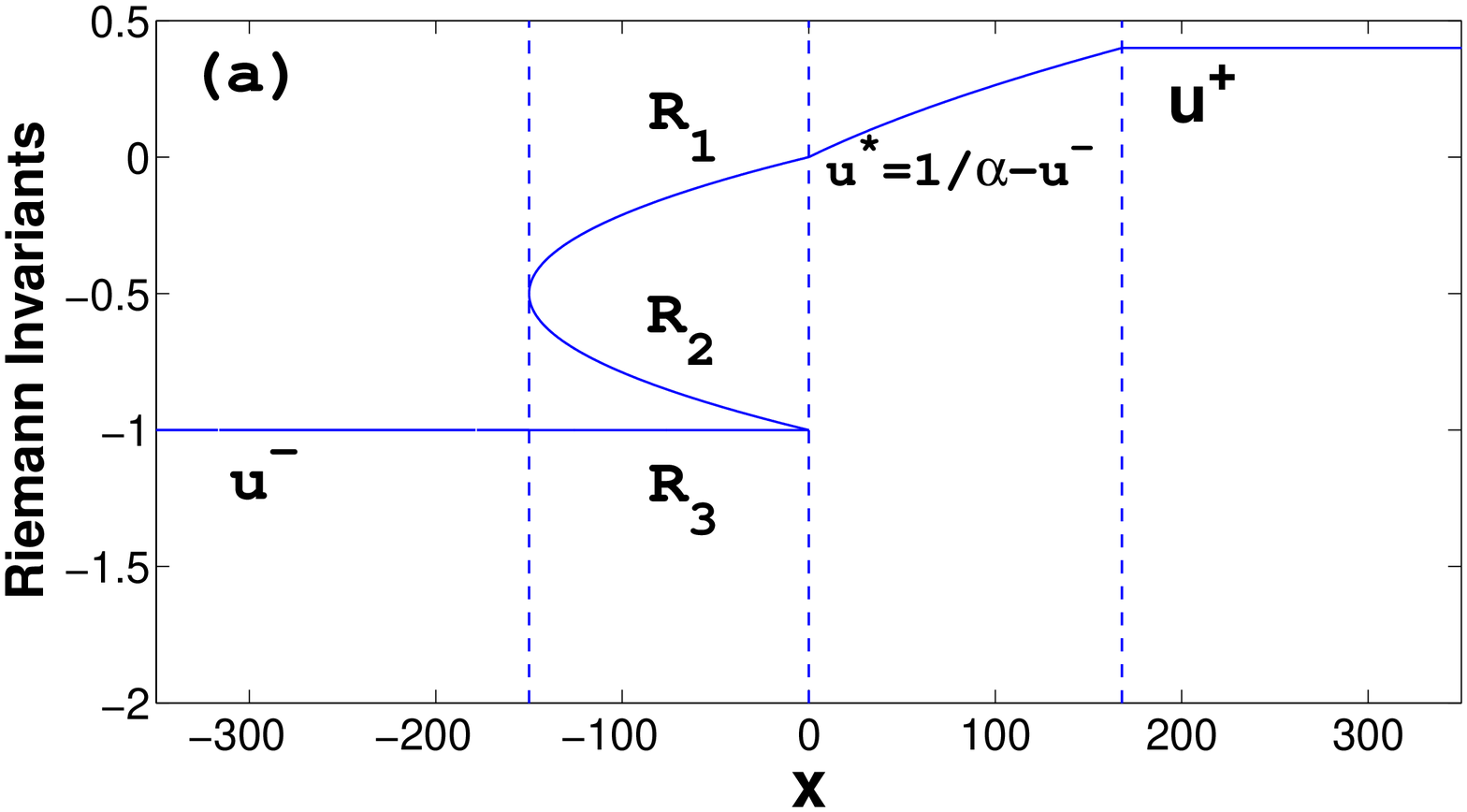}
\includegraphics[height=5cm,width=9cm]{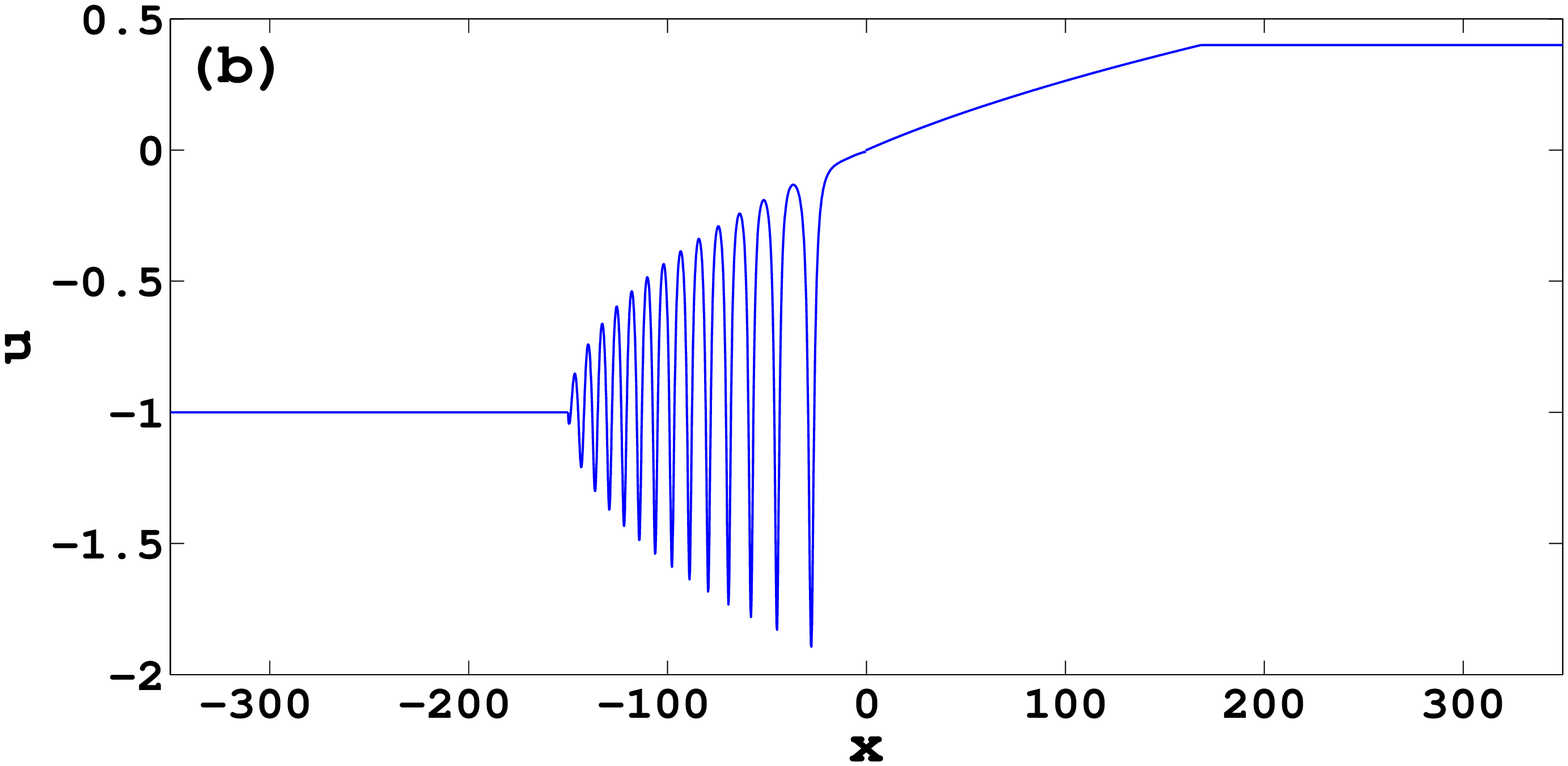}
\includegraphics[height=5cm,width=9cm]{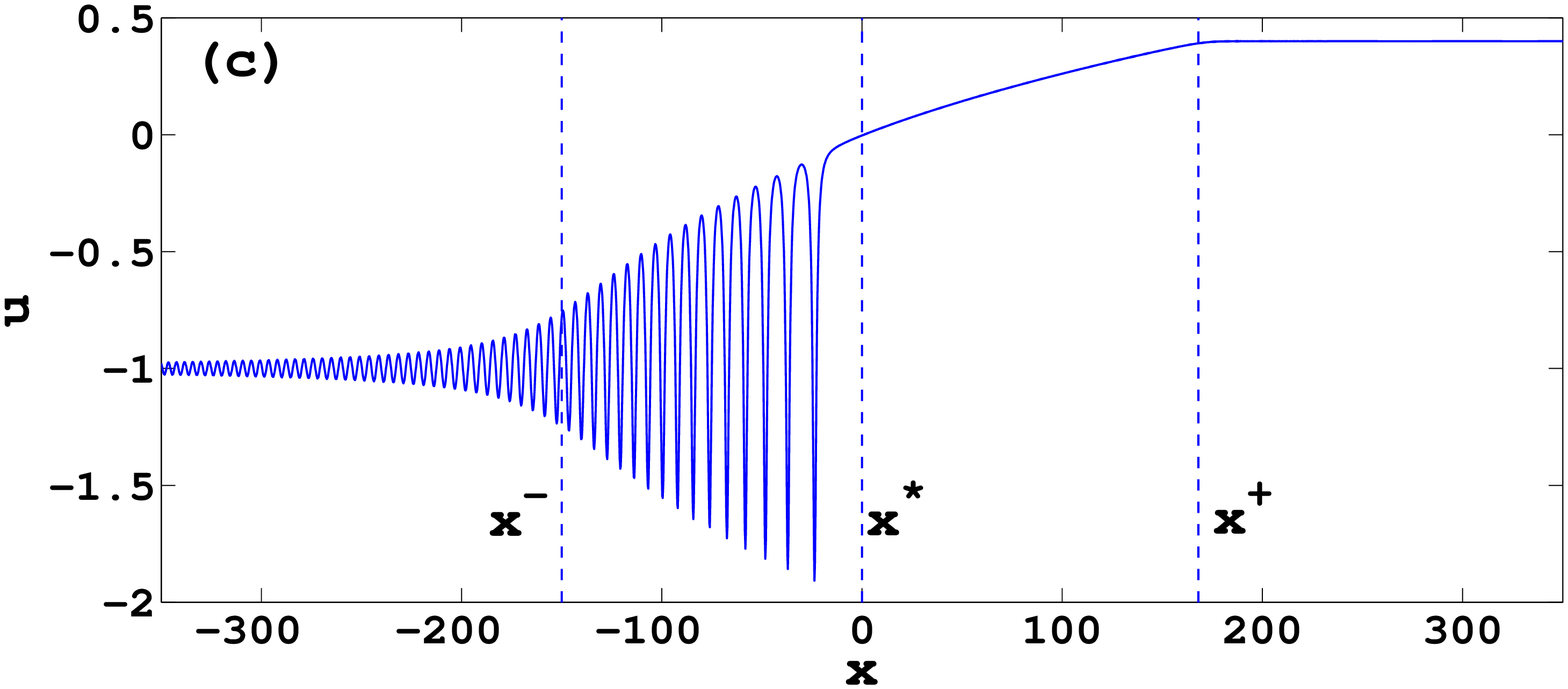}
\caption{(Color online) Evolution of an initial discontinuity for the Gardner equation with $\alpha=-1$.
The initial step parameters: $u^-=-1$, $u^+= 0.4$ (Region 6. $\{u^- \ \leftarrow{\bf TB}  \ (u^*)  \ {\bf RW} \rightarrow \ u^+\}$).
(a) Riemann invariants $R_1$, $R_2$ and $R_3$. (b) Analytical (modulation theory) solution.
(c) Numerical solution of the Gardner equation. $t=50$. Dashed lines on the numerical plot correspond to
the analytically found boundaries between different parts of the wave pattern.
}
\label{reg6neg}
\end{center}
\end{figure}

\medskip
The resolution pattern corresponding to this region is similar to that in Region 2 but now the resolution
occurs via the combination of the reversed trigonometric bore and normal rarefaction wave joined at the level
at the level $u^*=1/\alpha - u^-$.  The modulation description of such a bore is constructed in Section IVA
in terms of the Riemann invariants $R_1 \ge R_2 \ge R_3$ (\ref{eq5b}).
The relevant plots  are shown in Fig.~\ref{reg6neg}.

The edge speeds are equal to
\begin{equation}\label{eq14-2}
    \begin{split}
    &s^-=\frac3{\al}-6u^-(1-\al u^-), \\
    &s^*=6u^-(1-\al u^-),\\
    &s^+= 6u^+(1-\al u^+).
    \end{split}
\end{equation}

The part of the  line $u^-=1/\alpha - u^+$, where $u^-<u^+$, separating regions 6 and 7 corresponds
to a pure reversed trigonometric bore described in Section~IVA.

\bigskip

{\bf Region 7.}  $ u^-<\frac1{2\al}<u^+<\frac1{\al}-u^-$. $\{u^- \leftarrow  \ ({\bf TB} | {\bf UB}) \ u^+ \}$

\begin{figure}[bt]
\begin{center}
\includegraphics[height=5cm,width=9cm]{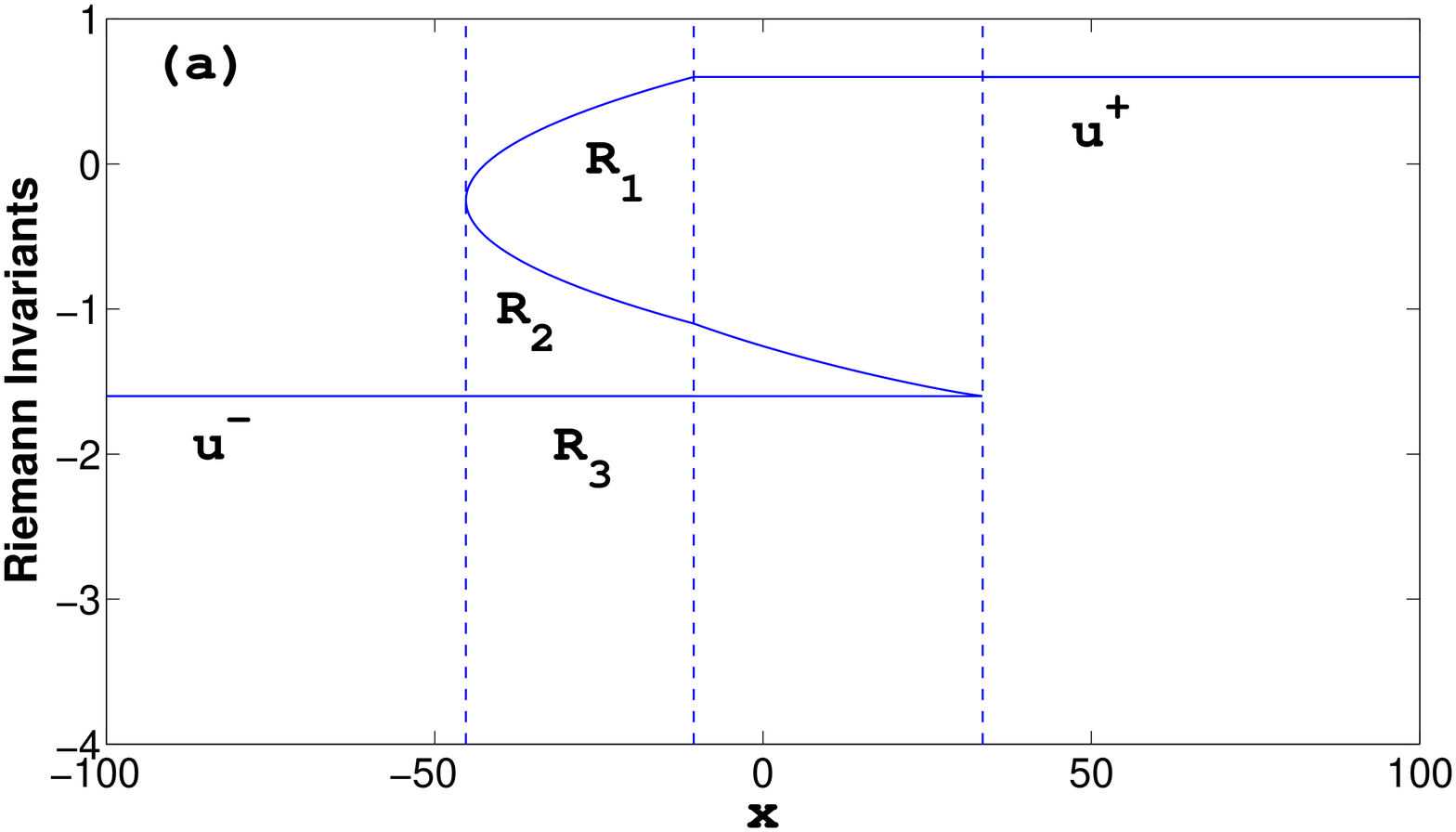}\hspace{0.5cm}
\includegraphics[height=5cm,width=9cm]{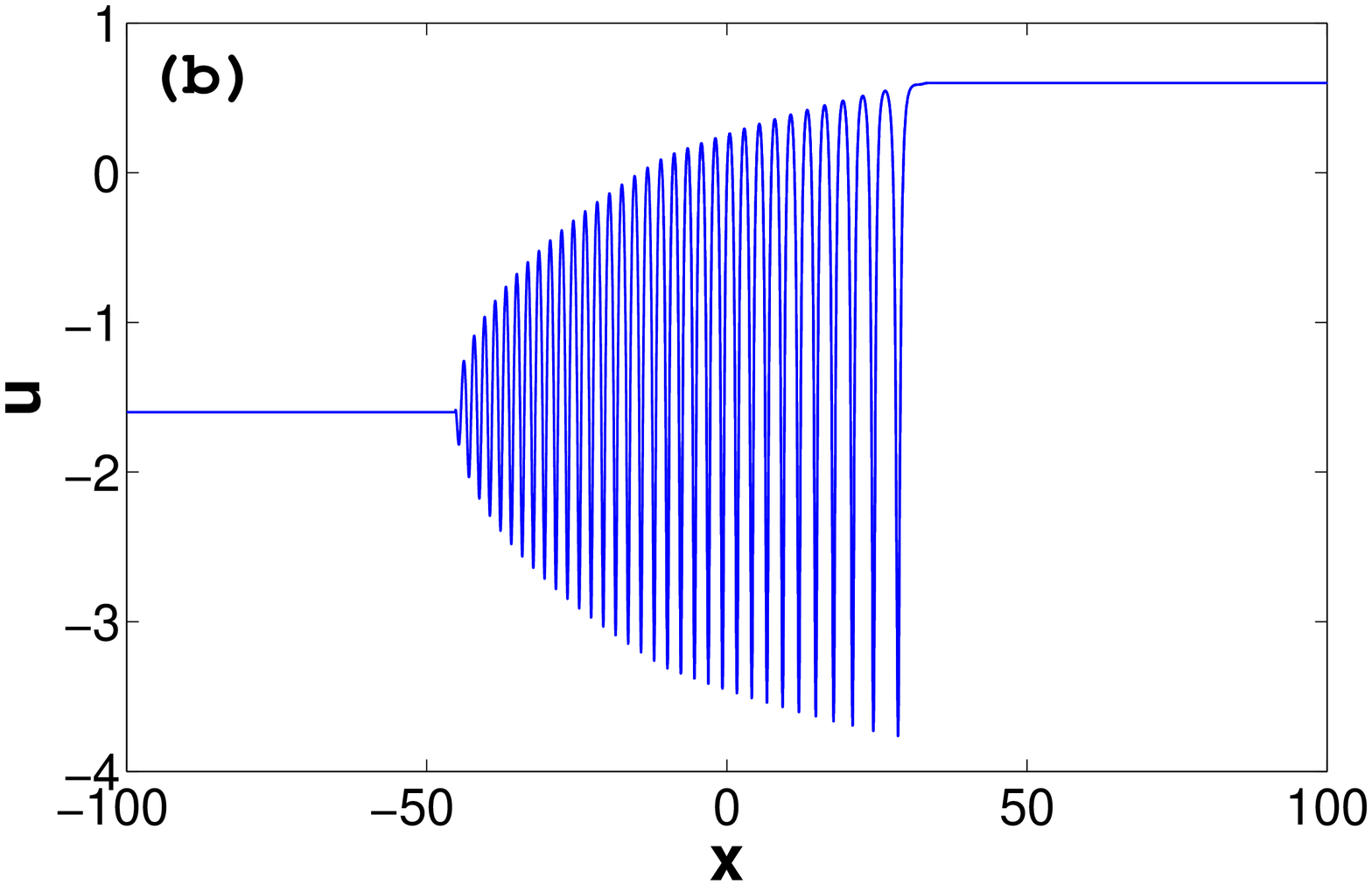}\\
\includegraphics[height=5cm,width=9cm]{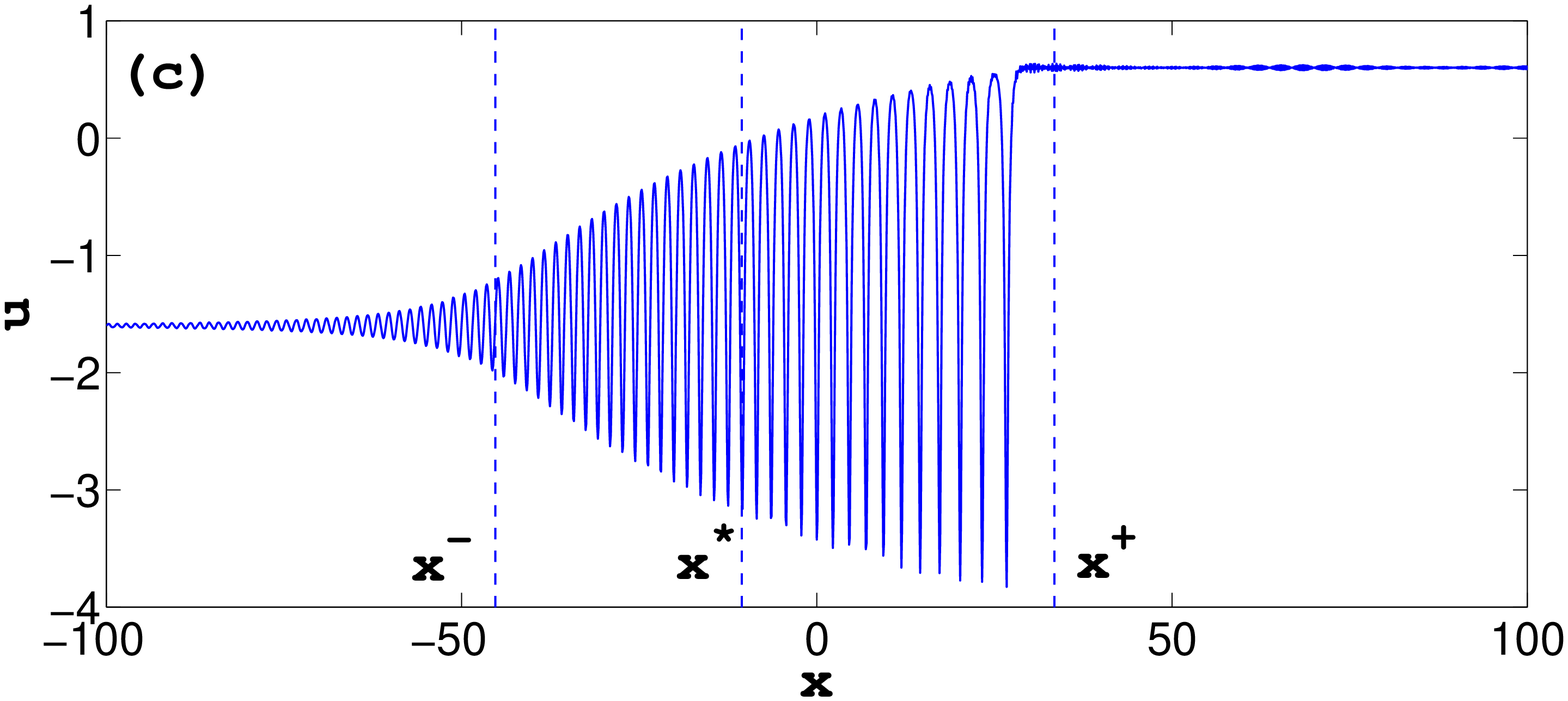}
\caption{(Color online) Evolution of an initial discontinuity for the Gardner equation with $\alpha=-2$. Region 7.  $\{u^- \leftarrow  \ ({\bf TB} | {\bf UB}) \ u^+ \}$).
The initial step parameters are: $u^-=-1.6$, $u^+=0.6$.
The plots are shown for $t=2$. (a): Riemann invariants $R_1$, $R_2$ and $R_3$.
(b): Analytical (modulation theory) solution; (c): Numerical solution of the Gardner equation.
Dashed lines on the numerical plot correspond to the analytically found boundaries between different parts of the wave pattern.}
\label{reg7neg}
\end{center}
\end{figure}

\medskip
Region 7 corresponds to the formation of a composite reversed trigonometric-undular bore (cf. Region 3 for the counterpart
normal resolution pattern).  The corresponding plots are shown in  Fig.~\ref{reg7neg}.

\begin{figure}[bt]
\begin{center}
\includegraphics[height=5cm,width=9cm]{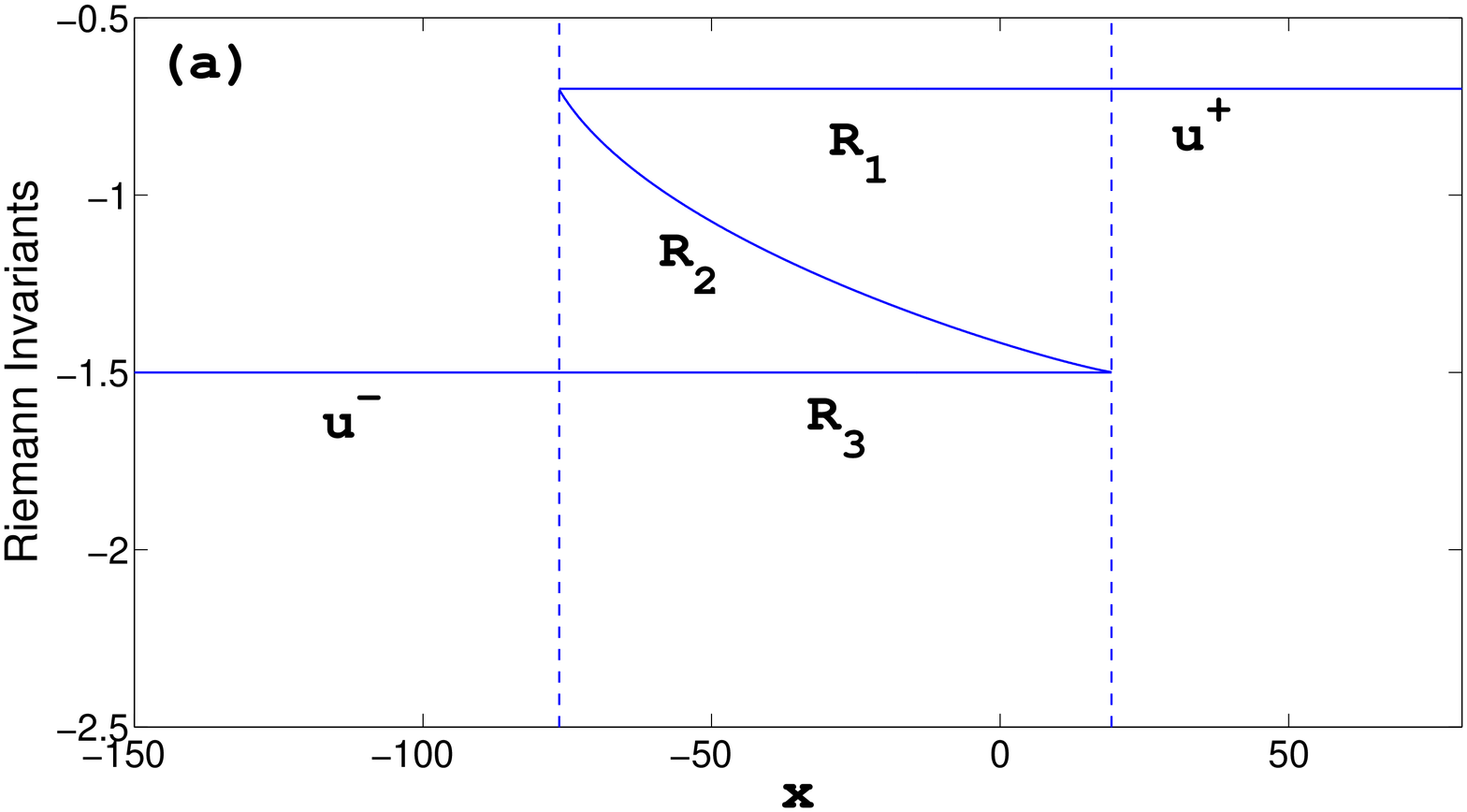}\hspace{0.5cm}
\includegraphics[height=5cm,width=9cm]{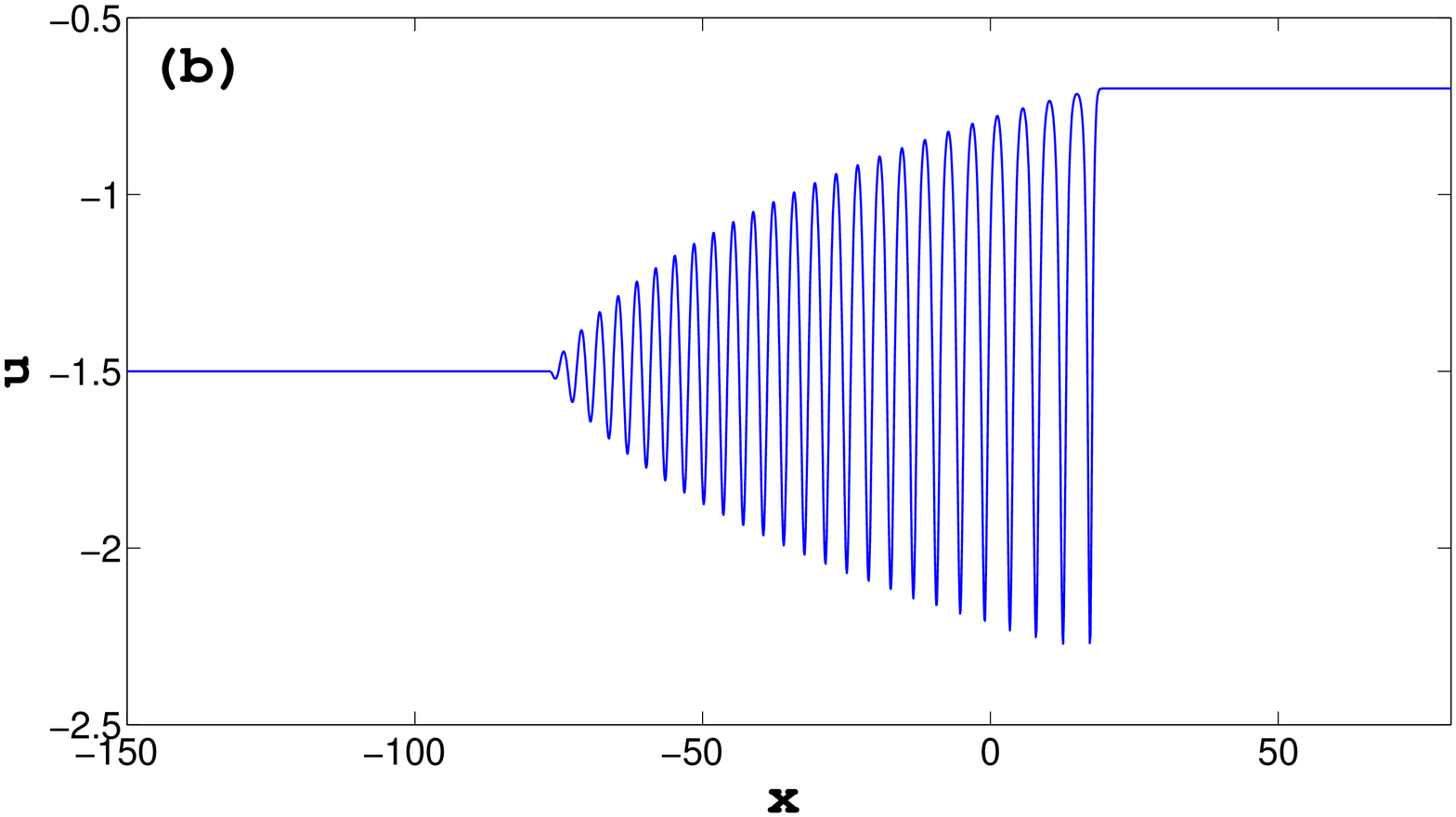}\hspace{0.5cm}
\includegraphics[height=5cm,width=9cm]{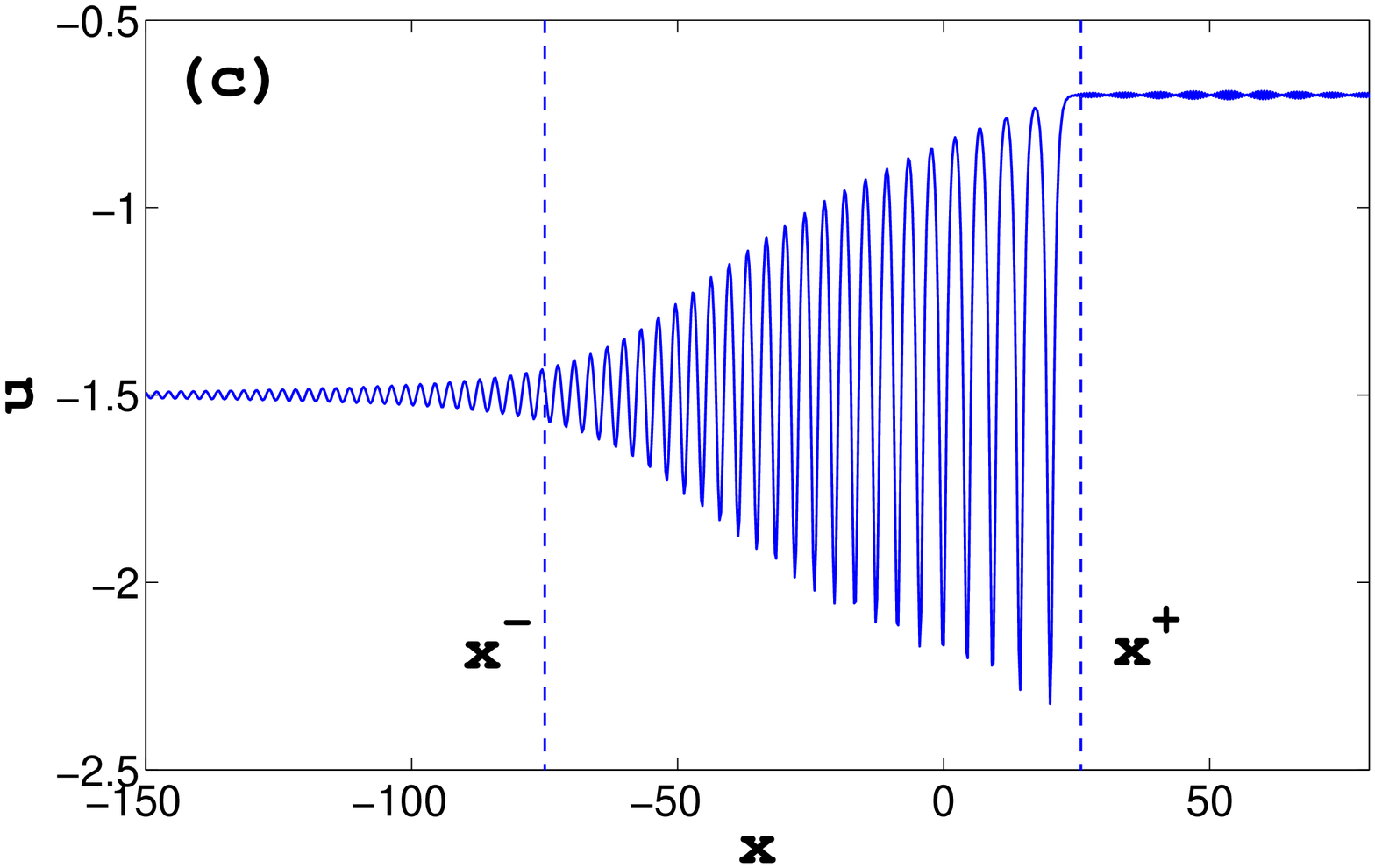}
\caption{(Color online) Evolution of an initial discontinuity for the Gardner equation with $\alpha=-1$. Region 8.  $\{u^- \leftarrow \ {\bf  UB } \ u^+ \}$.
The initial step parameters are $u^-=-1.5$, $u^+=-0.7$.  The plots are shown for $t=10$.
(a)Riemann invariants $R_1$, $R_2$ and $R_3$. (b) Analytical (modulation theory) solution.
(c) Numerical solution of the Gardner equation.  Dashed lines on the numerical plot correspond to the
analytically found boundaries between different parts of the wave pattern.}
\label{reg8neg}
\end{center}
\end{figure}
The edge speeds are
\begin{equation}\label{eq16-2}
    \begin{split}
    &s^-=\frac3{\al}-6u^+(1-\al u^+), \\
    &s^*=12u^+(1-\al u^+)-6u^-(1-\al u^-),\\
    &s^+= 2u^+(1-\al u^+)+4u^-(1-\al u^-).
    \end{split}
\end{equation}

\bigskip

{\bf Region 8}. $u^-<\frac1{\al}-u^+,\quad u^+<\frac1{2\al}$; $\{u^- \leftarrow {\bf  UB } u^+ \}$.

\medskip
This region corresponds to the to formation of a reversed cnoidal bore (cf. Region 4);
the plots are presented in Fig.~\ref{reg8neg}. Note that the Riemann invariants $R_1\ge R_2 \ge R_3$
(\ref{eq5b}) were used in the construction of the modulation solution for the reversed bore; as a matter of fact,
it is equivalent to the  Gurevich-Pitaevskii solution (\ref{GPconst}), (\ref{14-6})
in the original variables $r_3>r_2>r_1$ (see Fig.~3).

The edge speeds are
\begin{equation}\label{eq18}
\begin{split}
    &s^-=12u^+(1-\al u^+)-6u^-(1-\al u^-),\\
    &s^+=2u^+(1-\al u^+)+4u^-(1-\al u^-).
    \end{split}
\end{equation}

\section{Conclusions and Outlook}

We have constructed a full analytical description of the step problem for the Gardner equation (\ref{eq1}) for both signs of the coefficient $\alpha$ before the cubic term.
The complete classification of arising solutions for different parameters $u^+, u^-$ defining the initial step (\ref{step})  includes 16 possible
cases (8 for each sign of $\alpha$). Each sector on the $(u^+, u^-)$ plane of the parametric map of solutions
corresponds to a unique wave pattern representing one of the following: undular bore, a rarefaction wave, a solibore,
a trigonometric bore; or a combination of two of the above wave structures. The wave pattern arising in each case depends
on the position of the initial step parameters $u^-, u^+$ relative to each other and to the turning point $1/(2 \alpha)$ of the function
$6u(1-\alpha u)$ defining the characteristic speed of the dispersionless limit of the Gardner equation.
The analytical description of undular bores is made using the Whitham modulation theory. The observed rich phenomenology
of solutions arising in the step problem for the Gardner equation is due to the fact that the modulation
Whitham system associated with the Gardner equation, unlike that for the KdV equation, is neither strictly hyperbolic
nor genuinely nonlinear. Our analytical solutions are supported by numerical simulations.

One of the important applications of the  obtained  solutions is an analytical description  of transcritical flow in
a stratified fluid in the framework of the forced Gardner equation (cf. \cite{mh87,gcc-2002}). Other possible
applications include the consideration of the interaction of internal undular bores with variable topographies
(cf. \cite{egt12}) and the description of the perturbed modulation regimes for internal waves (e.g. due to the inclusion
of  weak dissipation). In the latter case, the description will require a perturbed modulation theory approach developed
in \cite{kamch04}.
The obtained classification will also provide a guidance for the similar classifications for fully nonlinear non-integrable
counterparts of the Gardner equation (such as Myatta-Choi-Camassa system \cite{cc99}) to which the analytic technique
of the undular bore description developed in \cite{el-2005} is applicable (see \cite{ep-11}).

\subsection*{Acknowledgements}

AMK thanks National Taiwan University and Taida Institute for Mathematical Sciences,
where this work was started, for kind hospitality.
T.L.H. acknowledges support of the National Science Council of Taiwan under Grant
No. NSC-100-2632-E-035-001-MY3. S.C.G. acknowledges
support from the National Center for Theoretical Science, Taiwan.

\end{document}